\documentclass[11pt,a4paper]{article}

\usepackage{amssymb}
\usepackage[dvips]{graphicx}
\usepackage{bm}

\begin{document}

\title{The all-loop perturbative derivation of the NSVZ $\beta$-function and the NSVZ scheme in the non-Abelian case by summing singular contributions}

\author{
K.V.Stepanyantz\\
{\small{\em Moscow State University,}}\\
{\small{\em Faculty of Physics, Department of Theoretical Physics,}}\\
{\small{\em 119991, Moscow, Russia}}}

\maketitle

\begin{abstract}
The perturbative all-loop derivation of the NSVZ $\beta$-function for ${\cal N}=1$ supersymmetric gauge theories regularized by higher covariant derivatives is finalized by calculating the sum of singularities produced by quantum superfields. These singularities originate from integrals of double total derivatives and determine all contributions to the $\beta$-function starting from the two-loop approximation. Their sum is expressed in terms of the anomalous dimensions of the quantum gauge superfield, of the Faddeev--Popov ghosts, and of the matter superfields. This allows obtaining the NSVZ equation in the form of a relation between the $\beta$-function and these anomalous dimensions for the renormalization group functions defined in terms of the bare couplings. It holds for an arbitrary renormalization prescription supplementing the higher covariant derivative regularization. For the renormalization group functions defined in terms of the renormalized couplings we prove that in all loops one of the NSVZ schemes is given by the HD+MSL prescription.
\end{abstract}

\unitlength=1cm

\section{Introduction}
\hspace*{\parindent}

Ultraviolet divergences in supersymmetric theories are resticted by some non-renormalization theorems. In particular, in ${\cal N}=1$ supersymmetric theories the superpotential does not receive divergent quantum corrections \cite{Grisaru:1979wc}. However, even in the case of ${\cal N}=1$ supersymmetry there are some other similar statements. For example, the triple gauge-ghost vertices (which have two external lines of the Faddeev--Popov ghosts and one external line of the quantum gauge superfield) are also finite in all orders \cite{Stepanyantz:2016gtk}. Moreover, according to Refs. \cite{Novikov:1983uc,Jones:1983ip,Novikov:1985rd,Shifman:1986zi}, the $\beta$-function of ${\cal N}=1$ supersymmetric gauge theories is related to the anomalous dimension of the matter superfields $(\gamma_\phi)_j{}^i$ by an equation which is usually called ``the exact NSVZ $\beta$-function''. For the theory with a simple gauge group $G$ and chiral matter superfields in its representation $R$ this equation is written as

\begin{equation}\label{NSVZ_Exact_Beta_Function}
\frac{\beta(\alpha,\lambda)}{\alpha^2} = - \frac{3 C_2 - T(R) + C(R)_i{}^j \big(\gamma_\phi\big)_j{}^i(\alpha,\lambda)/r}{2\pi(1- C_2\alpha/2\pi)}.
\end{equation}

\noindent
Note that we do not so far specify the definitions of the renormalization group functions (RGFs) and use the notations $r\equiv \dim\, G$,

\begin{equation}
\mbox{tr}(T^A T^B) \equiv T(R)\delta^{AB};\qquad  (T^A T^A)_i{}^j \equiv C(R)_i{}^j; \qquad f^{ACD} f^{BCD} \equiv C_2\delta^{AB},
\end{equation}

\noindent
where $T^A$ are the generators of the group $G$ in the representation $R$ such that $[T^A, T^B] = i f^{ABC} T^C$. Also we assume that the generators of the fundamental representation $t^A$ are normalized by the condition $\mbox{tr}(t^A t^B) = \delta^{AB}/2$.

Eq. (\ref{NSVZ_Exact_Beta_Function}) implies that the all-order $\beta$-function of the ${\cal N}=1$ supersymmetric Yang--Mills (SYM) theory without matter superfields is given by the geometric series. Moreover, if ${\cal N}=2$ supersymmetric gauge theories are considered as a special case of the ${\cal N}=1$ supersymmetric theories, then Eq. (\ref{NSVZ_Exact_Beta_Function}) leads to the finiteness beyond the one-loop approximation provided the quantization is made in ${\cal N}=2$ supersymmetric way \cite{Shifman:1999mv,Buchbinder:2014wra}. This implies that the NSVZ $\beta$-function is closely related to the ${\cal N}=2$ non-renormalization theorem derived in \cite{Grisaru:1982zh,Howe:1983sr,Buchbinder:1997ib}. However, for its derivation ${\cal N}=2$ supersymmetry should be manifest at all steps of calculating quantum corrections. This can be achieved with the help of the harmonic superspace \cite{Galperin:1984av,Galperin:2001uw} and the invariant regularization \cite{Buchbinder:2015eva}. The finiteness of ${\cal N}=4$ SYM theory \cite{Grisaru:1982zh,Howe:1983sr,Mandelstam:1982cb,Brink:1982pd} follows from the ${\cal N}=2$ non-renormalization  theorem and, therefore, from the NSVZ $\beta$-function.

Equations analogous to NSVZ are also known for the Adler $D$-function in ${\cal N}=1$ SQCD \cite{Shifman:2014cya,Shifman:2015doa} and for the renormalization of the gaugino mass in gauge theories with softly broken supersymmetry \cite{Hisano:1997ua,Jack:1997pa,Avdeev:1997vx}. Recently an NSVZ-like equation was constructed for the renormalization of the Fayet--Iliopoulos term in $D=2$, ${\cal N}=(2,0)$ supersymmetric theories \cite{Chen:2019eta}.

It is important that the NSVZ and NSVZ-like equations are valid only for certain renormalization prescriptions. In particular, for theories regularized by dimensional reduction \cite{Siegel:1979wq} supplemented by modified minimal subtraction \cite{Bardeen:1978yd} (this scheme is usually called $\overline{\mbox{DR}}$) Eq. (\ref{NSVZ_Exact_Beta_Function}) is not valid starting from the order $O(\alpha^2,\alpha\lambda^2,\lambda^4)$, which corresponds to the three-loop $\beta$-function and the two-loop anomalous dimension \cite{Jack:1996vg,Jack:1996cn,Jack:1998uj,Harlander:2006xq,Mihaila:2013wma}.\footnote{For the NSVZ-like equations the situation is the same, see Refs. \cite{Jack:1997pa} and \cite{Aleshin:2019yqj} for details.} However, the detailed analysis made in these papers demonstrated that by a specially tuned finite renormalization of the gauge coupling constant one can restore the NSVZ equation (\ref{NSVZ_Exact_Beta_Function}), at least, in the three- and four-loop approximations. It should be noted that the possibility of this tuning is non-trivial, because of the presence of various group invariants (like $C_2$, $C(R)_i{}^j$, etc.). If one considers only finite renormalizations polynomial in these invariants, then the NSVZ equation leads to some scheme-independent consequences \cite{Kataev:2013csa,Kataev:2014gxa}. This implies that, although the calculations made in the $\overline{\mbox{DR}}$-scheme do not produce the NSVZ equation, they nevertheless confirm that it is valid in a certain (NSVZ) subtraction scheme. Using the general equations describing how the NSVZ equation changes under finite renormalizations \cite{Kataev:2014gxa,Kutasov:2004xu}, it is possible to construct an infinite set of the NSVZ schemes \cite{Goriachuk:2018cac}.

For a long time it was unknown, how to construct an all-order renormalization prescription giving the NSVZ scheme. However, recently it was understood that the solution can be found in the case of using the higher covariant derivative method \cite{Slavnov:1971aw,Slavnov:1972sq} for regularizing supersymmetric theories. Unlike dimensional reduction \cite{Siegel:1980qs}, this regularization is mathematically consistent and can be formulated in a manifestly supersymmetric way in ${\cal N}=1$ superspace \cite{Krivoshchekov:1978xg,West:1985jx}. Although the calculations in higher derivative theories are rather complicated, some of them have been done in the last decades. For instance, a number of calculations of the one-loop effective potential for ${\cal N}=1$ higher derivative supersymmetric theories were made in \cite{Gomes:2009ev,Gama:2011ws,Gama:2013rsa,Gama:2014fca,BezerradeMello:2016bjn,Gama:2017ets,Gama:2020pte}. In theories {\it regularized} by higher derivatives quantum corrections are obtained in a similar way. Some higher order calculations made with this regularization (see, e.g., \cite{Stepanyantz:2012zz,Shakhmanov:2017soc,Kazantsev:2018nbl,Kuzmichev:2019ywn}) demonstrated that the NSVZ equation is valid for RGFs defined in terms of the bare couplings. (In the Abelian case this has been proved in all orders \cite{Stepanyantz:2011jy,Stepanyantz:2014ima}. Similar proofs of the NSVZ-like equations have also been constructed for the Adler $D$-function in ${\cal N}=1$ SQCD \cite{Shifman:2014cya,Shifman:2015doa} and for the renormalization of the photino mass in ${\cal N}=1$ SQED with softly broken supersymmetry \cite{Nartsev:2016nym}.) RGFs defined in terms of the bare couplings are scheme-independent for a fixed regularization \cite{Kataev:2013eta}, but depend on a regularization, so that Eq. (\ref{NSVZ_Exact_Beta_Function}) is valid for them for an arbitrary renormalization prescription supplementing the higher derivative regularization. The above mentioned calculations confirmed this in such an approximation where the dependence on a regularization is essential. Note that, according to \cite{Aleshin:2019yqj,Aleshin:2016rrr}, RGFs defined in terms of the bare couplings do not satisfy the NSVZ and NSVZ-like equations in the case of using dimensional reduction, again, for an arbitrary renormalization prescription supplementing it.

The important statement which allows constructing the NSVZ scheme is that RGFs defined in terms of the bare couplings and RGFs standardly defined in terms of the renormalized couplings up to the renaming of arguments coincide in the HD+MSL scheme \cite{Kataev:2013eta}, when the divergences of the theory regularized by Higher covariant Derivatives are removed by Minimal Subtractions of Logarithms \cite{Shakhmanov:2017wji,Stepanyantz:2017sqg}. This implies that the renormalization constants are constructed in such a way that they include only powers of $\ln\Lambda/\mu$, where $\Lambda$ is a regularization parameter, analogous to the ultraviolet cut-off, and $\mu$ is a renormalization point. All finite constants in this case, by definition, are set to 0. The coincidence of various definitions of RGFs implies that the HD+MSL scheme appears to be one of the NSVZ schemes. In the Abelian case this has been proved in all loops \cite{Kataev:2013eta}. Note that another all-loop NSVZ scheme in the Abelian case is the on-shell scheme \cite{Kataev:2019olb}. For the renormalization of the photino mass in softly broken ${\cal N}=1$ SQED and for the Adler $D$-function in ${\cal N}=1$ SQCD the NSVZ-like schemes are also given by the HD+MSL prescription, see Refs. \cite{Nartsev:2016mvn} and \cite{Kataev:2017qvk}, respectively. In the non-Abelian case there are strong indications \cite{Stepanyantz:2016gtk} that HD+MSL = NSVZ. Also there are some nontrivial explicit multiloop calculations confirming this fact \cite{Shakhmanov:2017soc,Kazantsev:2018nbl,Kuzmichev:2019ywn}, but, nevertheless, this statement has not yet been proved in all orders. This proof (started in Refs. \cite{Stepanyantz:2016gtk,Stepanyantz:2019ihw,Stepanyantz:2019lfm}) will be finalized in this paper.

The main observation used for the derivation of the NSVZ and NSVZ-like equations for RGFs defined in terms of the bare couplings in theories regularized by higher derivatives is that the loop integrals giving the $\beta$-function are integrals of double total derivatives with respect to loop momenta.\footnote{The total and double total derivative structure of the integrands has first been noted in calculating the lowest quantum corrections in Refs. \cite{Soloshenko:2003nc} and \cite{Smilga:2004zr}, respectively. For theories regularized by dimensional reduction such a structure is not valid \cite{Aleshin:2015qqc}.} Certainly, at least one of the loop integrals can be calculated analytically using equations like

\begin{equation}\label{Integral_Toy}
\int \frac{d^4Q}{(2\pi)^4} \frac{\partial^2}{\partial Q_\mu \partial Q^\mu}\Big(\frac{f(Q^2)}{Q^2}\Big) = \frac{1}{4\pi^2} f(0),
\end{equation}

\noindent
where $f(Q^2)$ is a nonsingular function of the Euclidean momentum $Q_\mu$ which rapidly tends to 0 at infinity. Note that the result does not vanish due to the nontrivial surface integration over the small sphere $S^3_\varepsilon$ surrounding the point $Q_\mu =0$. If we consider an $L$-loop contribution to the $\beta$-function, then Eq. (\ref{Integral_Toy}) allows calculating one of the loop integrals, so that the resulting expression will contain only $(L-1)$ loop integrations. Therefore, it is possible to suggest that the result is a certain $(L-1)$-loop quantum correction. According to \cite{Stepanyantz:2011jy,Stepanyantz:2014ima}, in the Abelian case it is really proportional to the $(L-1)$-loop contribution to anomalous dimension of the matter superfields, so that the Abelian NSVZ equation \cite{Vainshtein:1986ja,Shifman:1985fi}

\begin{equation}
\frac{\beta(\alpha_0)}{\alpha_0^2} = \frac{N_f}{\pi}\Big(1-\gamma(\alpha_0)\Big),
\end{equation}

\noindent
where $N_f$ is a number of flavors, is naturally produced for RGFs defined in terms of the bare couplings.

In the non-Abelian case the situation is more complicated, because Eq. (\ref{NSVZ_Exact_Beta_Function}) relates the $\beta$-function to the anomalous dimension of the matter superfields {\it in all previous orders} due to the gauge coupling constant dependent denominator in the right hand side. However, according to Ref. \cite{Stepanyantz:2016gtk}, using the all-loop finiteness of triple gauge-ghost vertices the NSVZ equation (\ref{NSVZ_Exact_Beta_Function}) can be presented in an equivalent form

\begin{equation}\label{NSVZ_Equivalent_Form}
\frac{\beta(\alpha,\lambda)}{\alpha^2} = - \frac{1}{2\pi}\Big(3 C_2 - T(R) - 2C_2 \gamma_c(\alpha,\lambda) - 2C_2 \gamma_V(\alpha,\lambda) + \frac{1}{r} C(R)_i{}^j \big(\gamma_\phi\big)_j{}^i(\alpha,\lambda)\Big),
\end{equation}

\noindent
which does not contain the coupling constant dependent denominator in the right hand side similarly to the Abelian case. (Note that in Eq. (\ref{NSVZ_Equivalent_Form}) we again do not specify the definitions of RGFs and omit some other possible arguments.) Similarly to the Abelian case, Eq. (\ref{NSVZ_Equivalent_Form}) relates an $L$-loop contribution {\it only} to $(L-1)$-loop contributions to the anomalous dimensions of the quantum gauge superfield, of the Faddeev--Popov ghosts, and of the matter superfields, denoted by $\gamma_V$, $\gamma_c$, and $(\gamma_\phi)_j{}^i$, respectively. This fact has a simple graphical interpretation analogous to the ${\cal N}=1$ SQED case considered in \cite{Smilga:2004zr,Kazantsev:2014yna}. Namely, given a supergraph without external lines, the corresponding superdiagrams contributing to the $\beta$-function are constructed by attaching two external lines of the background gauge superfield in all possible ways, while the superdiagrams contributing to the anomalous dimensions are obtained by all possible cuts of internal lines. This qualitative picture is related to the structure of loop integrals giving the $\beta$-function defined in terms of the bare couplings. According to \cite{Stepanyantz:2019ihw} they are integrals of double total derivatives in all orders in agreement with numerous calculations made with the higher covariant derivative regularization, see, e.g., \cite{Shakhmanov:2017soc,Kazantsev:2018nbl,Pimenov:2009hv,Stepanyantz:2011bz,Stepanyantz:2019lyo}. Each cut of a certain propagator corresponds to taking an integral of a double total derivative analogous to (\ref{Integral_Toy}). The sums of singularities generated by the cuts of various propagators produce the corresponding anomalous dimensions in Eq. (\ref{NSVZ_Equivalent_Form}) even at the level of loop integrals.\footnote{The superdiagrams contributing to $\gamma_V$ can be compared with a certain part of the $\beta$-function only if their sum is transversal. For the sum of {\it all} superdiagrams this is always true due to the Slavnov--Taylor identities \cite{Taylor:1971ff,Slavnov:1972fg}.} In the lowest orders this was explicitly demonstrated in \cite{Shakhmanov:2017soc,Kazantsev:2018nbl,Kuzmichev:2019ywn,Shakhmanov:2017wji,Stepanyantz:2019lyo}. In all loops the sums of singularities obtained by cutting the matter and Faddeev--Popov ghost propagators have been found in \cite{Stepanyantz:2019lfm} using the method based on the Schwinger--Dyson effective superdiagrams proposed in \cite{Stepanyantz:2004sg}. These sums coincide with the terms containing $(\gamma_\phi)_j{}^i$ and $\gamma_c$ defined in terms of the bare couplings, respectively. Therefore, to complete the derivation of the NSVZ equation, one should find a sum of all singularities produced by cuts of the gauge propagators and demonstrate that it gives the term containing $\gamma_V$ in Eq. (\ref{NSVZ_Equivalent_Form}).\footnote{The term $-(3C_2-T(R))/2\pi$ corresponds to the one-loop contribution to the $\beta$-function, which should be calculated separately. With the higher covariant derivative regularization this has been done in \cite{Aleshin:2016yvj}.} This is a purpose of the present paper.

The paper is organized as follows. Section \ref{Section_N=1_Gauge_Theories} describes the quantization of ${\cal N}=1$ supersymmetric gauge theories regularized by higher covariant derivatives. In section \ref{Section_Perturbative_Derivation} we rewrite the NSVZ $\beta$-function in the form of a relation between some two-point Green functions, which will be proved below. The proof is based on the method for constructing the integrals of double total derivatives giving the function $\beta/\alpha_0^2$ which was proposed in \cite{Stepanyantz:2019ihw}, see also \cite{Kuzmichev:2019ywn}. This method is described in section \ref{Subsection_Idea}. It is slightly modified in section \ref{Subsection_Graphs}. Using this modification the sums of singularities produced by the matter superfields, by the Faddeev--Popov ghosts, and by the quantum gauge superfield are calculated exactly in all loops in section \ref{Section_Singularities}. In particular, in subsection \ref{Subsection_Gauge_Singularities} we find the sum of singularities produced by the quantum gauge superfield propagators, which is needed for finalizing the proof of the NSVZ $\beta$-function. Combining the results we derive Eq. (\ref{NSVZ_Equivalent_Form}) for RGFs defined in terms of the bare couplings. In the next section \ref{Section_NSVZ_Scheme} we demonstrate that the HD+MSL prescription really gives one of the NSVZ schemes in all orders for RGFs defined in terms of the renormalized couplings. The last section \ref{Section_Explicit_Calculation} is devoted to the explicit perturbative verification of the results in the lowest orders of the perturbation theory. Doing the corresponding calculations we pay especial attention to checking the modification of the method for constructing integrals giving the function $\beta/\alpha_0^2$ proposed in section \ref{Subsection_Graphs}.

\section{${\cal N}=1$ renormalizable supersymmetric gauge theories regularized by higher covariant derivatives}
\hspace*{\parindent}\label{Section_N=1_Gauge_Theories}

We will consider a general renormalizable ${\cal N}=1$ SYM theory with a simple gauge group $G$ and chiral matter superfields $\phi_i$ in a representation $R$. In the massless limit in terms of ${\cal N}=1$ superfields \cite{Gates:1983nr,West:1990tg,Buchbinder:1998qv} this theory is described by the action

\begin{eqnarray}\label{Action_Under_Consideration}
S = \frac{1}{2e_0^2}\, \mbox{Re}\,\mbox{tr} \int d^6x\, W^a W_a + \frac{1}{4} \int d^8x\, \phi^{*i} (e^{2{\cal F}(V)} e^{2\bm{V}})_i{}^j \phi_j
+ \Big(\frac{1}{6} \lambda_0^{ijk} \int d^6x\, \phi_i \phi_j \phi_k + \mbox{c.c.}\Big),
\end{eqnarray}

\noindent
where for the superspace integration measures we use the brief notations

\begin{equation}
d^8x \equiv d^4x\, d^4\theta_x;\qquad d^6x \equiv d^4x\, d^2\theta_x;\qquad  d^6\bar x \equiv d^4x\, d^2\bar\theta_x.
\end{equation}

\noindent
In Eq.~(\ref{Action_Under_Consideration}) $\bm{V}$ is the Hermitian background gauge superfield and $V$ is the quantum gauge superfield, which satisfies the constraint $V^+ = e^{-2\bm{V}} V e^{2\bm{V}}$. Note that in the first term of Eq.~(\ref{Action_Under_Consideration}) the quantum and background gauge superfields are expanded in the generators of the fundamental representation as $V = e_0 V^A t^A$ and $\bm{V} = e_0 \bm{V}^A t^A$, while in the second term $V = e_0 V^A T^A$ and $\bm{V} = e_0 \bm{V}^A T^A$, where $T^A$ are the generators of the gauge group in the representation $R$.

The gauge and Yukawa couplings are denoted by $e_0$ and $\lambda_0^{ijk}$, respectively, where the subscript $0$ points the bare values. The function ${\cal F}(V) = e_0 {\cal F}(V)^A t^A$ is needed, because the quantum gauge superfield is renormalized in the non-linear way \cite{Piguet:1981fb,Piguet:1981hh,Tyutin:1983rg}. In the lowest approximation this function was found in Refs. \cite{Juer:1982fb,Juer:1982mp} and is written as

\begin{equation}\label{Nonlinear_Function_F}
{\cal F}(V)^A = V^A + y_0\, e_0^2\, G^{ABCD}\, V^B V^C V^D + \ldots,
\end{equation}

\noindent
where $y_0$ is the first bare constant in an infinite set of parameters describing the nonlinearity, and $G^{ABCD} \equiv \big(f^{AKL} f^{BLM} f^{CMN} f^{DNK} + \mbox{permutations of $B$, $C$, and $D$}\big)/6$. The necessity of introducing the function ${\cal F}(V)$ was also confirmed by the calculation of Ref. \cite{Kazantsev:2018kjx}, where it was demonstrated that the renormalization group equations are satisfied only if the renormalization of the parameter $y$ is taken into account.

The gauge superfield strength is described by the chiral superfield

\begin{equation}\label{W_a_Definition}
W_a \equiv \frac{1}{8} \bar D^2 \Big[ e^{-2\bm{V}} e^{-2{\cal F}(V)}\, D_a \Big(e^{2{\cal F}(V)}e^{2\bm{V}}\Big)\Big],
\end{equation}

\noindent
which is a right spinor with respect to the Lorentz group.

If the Yukawa couplings satisfy the equation

\begin{equation}\label{Yukawa_Constraint}
\lambda_0^{ijm} (T^A)_m{}^k + \lambda_0^{imk} (T^A)_m{}^j + \lambda_0^{mjk} (T^A)_m{}^i = 0,
\end{equation}

\noindent
the theory (\ref{Action_Under_Consideration}) is invariant under the background gauge transformations

\begin{equation}\label{Background_Gauge_Invariance_Original}
e^{2\bm{V}} \to e^{-A^+} e^{2\bm{V}} e^{-A};\qquad V \to  e^{-A^+} V e^{A^+};\qquad \phi_i \to (e^A)_i{}^j \phi_j
\end{equation}

\noindent
parameterized by the chiral superfield $A$ which takes values in the Lie algebra of the gauge group $G$. The action (\ref{Action_Under_Consideration}) is also invariant under the quantum gauge transformations, but this invariance is broken by the gauge fixing procedure.

The regularization is implemented by inserting into the classical action the higher derivative regulator functions $R$ and $F$. They should rapidly increase at infinity and satisfy the condition $R(0)=F(0)=1$. Then the action of the regularized theory can be written as

\begin{eqnarray}\label{Action_Regularized_Without_G}
&& S_{\mbox{\scriptsize reg}} = \frac{1}{2 e_0^2}\,\mbox{Re}\, \mbox{tr} \int d^6x\, W^a \Big[e^{-2\bm{V}} e^{-2{\cal F}(V)}\,  R\Big(-\frac{\bar\nabla^2 \nabla^2}{16\Lambda^2}\Big)\, e^{2{\cal F}(V)}e^{2\bm{V}}\Big]_{Adj} W_a \qquad\nonumber\\
&& + \frac{1}{4} \int d^8x\, \phi^{*i} \Big[\, F\Big(-\frac{\bar\nabla^2 \nabla^2}{16\Lambda^2}\Big) e^{2{\cal F}(V)}e^{2\bm{V}}\Big]_i{}^j \phi_j
+ \Big(\frac{1}{6} \lambda_0^{ijk} \int d^6x\, \phi_i \phi_j \phi_k + \mbox{c.c.} \Big),\qquad
\end{eqnarray}

\noindent
where for $f(x) = 1 + f_1 x + f_2 x^2 +\ldots$

\begin{equation}
f(X)_{Adj} Y \equiv Y + f_1 [X, Y] + f_2 [X,[X,Y]] + \ldots,
\end{equation}

\noindent
and the explicit expressions for the covariant derivatives have the form

\begin{equation}\label{Covariant_Derivative_Definition}
\nabla_a = D_a;\qquad \bar\nabla_{\dot a} = e^{2{\cal F}(V)} e^{2\bm{V}} \bar D_{\dot a} e^{-2\bm{V}} e^{-2{\cal F}(V)}.
\end{equation}

\noindent
The modification of the action $S \to S_{\mbox{\scriptsize reg}}$ allows regularizing all divergences beyond the one-loop approximation \cite{Faddeev:1980be}.

The generating functional for the regularized theory should also include a gauge fixing term, ghost actions, sources, and Pauli--Villars determinants needed for removing the remaining one-loop divergences \cite{Slavnov:1977zf},

\begin{eqnarray}\label{Z_Functional_Generating}
Z = \int D\mu\, \mbox{Det}(PV,M_\varphi)^{-1}\mbox{Det}(PV,M)^c \exp\Big\{i\Big(S_{\mbox{\scriptsize reg}}
+ S_{\mbox{\scriptsize gf}} + S_{\mbox{\scriptsize FP}} + S_{\mbox{\scriptsize NK}} + S_{\mbox{\scriptsize sources}}\Big)\Big\}.
\end{eqnarray}

\noindent
The source term is given by the expression

\begin{equation}
S_{\mbox{\scriptsize sources}} = \int d^8x\, J^A V^A + \Big(\int d^6x\, \Big(j^i \phi_i + j_c^A c^A + \bar j_c^A \bar c^A\Big) + \mbox{c.c.}\Big),
\end{equation}

\noindent
where $J^A$ and $j^i$ are the real and chiral superfields, respectively. The anticommuting chiral superfields $j_c^A$, and $\bar j_c^A$ are the sources for the anticommuting chiral Faddeev--Popov ghost and antighost superfields denoted by $c^A$ and $\bar c^A$, respectively.

We will use the gauge fixing term

\begin{equation}\label{Term_For_Fixing_Gauge}
S_{\mbox{\scriptsize gf}} = -\frac{1}{16\xi_0 e_0^2}\, \mbox{tr} \int d^8x\,  \bm{\nabla}^2 V K\Big(-\frac{\bm{\bar\nabla}^2 \bm{\nabla}^2}{16\Lambda^2}\Big)_{Adj} \bm{\bar\nabla}^2 V
\end{equation}

\noindent
invariant under the background gauge transformations (\ref{Background_Gauge_Invariance_Original}) due to the presence of the background covariant derivatives

\begin{equation}\label{Background_Values}
\bm{\nabla}_a \equiv D_a;\qquad\quad \bm{\bar\nabla}_{\dot a} \equiv e^{2\bm{V}} \bar D_{\dot a} e^{-2\bm{V}}.
\end{equation}

\noindent
The corresponding actions for the Faddeev--Popov and Nielsen--Kallosh ghosts ($S_{\mbox{\scriptsize FP}}$ and $S_{\mbox{\scriptsize NK}}$, respectively) can be found in Refs. \cite{Stepanyantz:2019ihw,Stepanyantz:2019lfm}. In this paper we will not use the explicit expressions for them. The bare gauge parameter $\xi_0$ and the parameters present in the function ${\cal F}(V)$ (i.e., $y_0$, etc.) can conveniently be included into a single infinite set $Y_0 \equiv (\xi_0,\,y_0,\ldots)$.

Following Refs. \cite{Aleshin:2016yvj,Kazantsev:2017fdc}, for regularizing the one-loop divergences we use two sets of the Pauli--Villars superfields. The first one consists of three commuting chiral superfields $\varphi_a$ in the adjoint representation of the gauge group with the mass $M_\varphi = a_\varphi\Lambda$. They cancel the divergences coming from the gauge and ghost loops. The corresponding determinant present in Eq. (\ref{Z_Functional_Generating}) is denoted by $ \mbox{Det}(PV,M_\varphi)$. The second set giving the determinant $\mbox{Det}(PV,M)$ consists of the commuting chiral superfields $\Phi_i$ in a certain representation $R_{\mbox{\scriptsize PV}}$ that admits the gauge invariant mass term such that $M^{ij} M_{jk} = M^2 \delta^i_k$ with $M = a\Lambda$.\footnote{For simple gauge groups it is possible to choose $R_{\mbox{\scriptsize PV}} = Adj$.} These superfields cancel the one-loop divergences produced by the matter superfields $\phi_i$ if in Eq. (\ref{Z_Functional_Generating}) the degree of the corresponding determinant is $c=T(R)/T(R_{\mbox{\scriptsize PV}})$. Again we will not use explicit expressions of the Pauli--Villars determinants which can also be found in Refs. \cite{Stepanyantz:2019ihw,Stepanyantz:2019lfm}. It should be only mentioned that the constants $a_\varphi$ and $a$ must be independent of couplings.

Starting from the generating functional for the connected Green functions $W \equiv -i\ln Z$, one can construct the effective action

\begin{equation}
\Gamma[\bm{V}, V, \phi_i,\ldots] \equiv W - S_{\mbox{\scriptsize sources}}\Big|_{\mbox{\scriptsize sources}\, \to\, \mbox{\scriptsize fields}},
\end{equation}

\noindent
where it is necessary to express the sources $J^A, j^i,\ldots$ in terms of quantum superfields $V^A,\phi_i,\ldots$ using the equations

\begin{equation}\label{Field_Definitions}
V^A \equiv \frac{\delta W}{\delta J^A},\qquad \phi_i \equiv \frac{\delta W}{\delta j_i},\qquad  \mbox{etc}.
\end{equation}

\section{The NSVZ equation as a relation between two-point Green functions}
\hspace*{\parindent}\label{Section_Perturbative_Derivation}

The NSVZ equation (\ref{NSVZ_Equivalent_Form}) for RGFs defined in terms of the bare couplings in the case of using the higher covariant derivative regularization can be written as a certain equation relating two-point Green functions of the theory. The terms in the effective action corresponding to these Green functions can be presented in the form

\begin{eqnarray}\label{Gamma2_Background_V}
&&\hspace*{-12mm}  \Gamma^{(2)}_{\bm{V}} = - \frac{1}{8\pi} \mbox{tr} \int \frac{d^4p}{(2\pi)^4}\, d^4\theta\, \bm{V}(-p,\theta) \partial^2 \Pi_{1/2} \bm{V}(p,\theta)\, d^{-1}(\alpha_0, \lambda_0, Y_0, \Lambda/p);\\
\label{Gamma2_Quantum_V}
&&\hspace*{-12mm} \Gamma^{(2)}_V - S_{\mbox{\scriptsize gf}}^{(2)} = -\frac{1}{4} \int \frac{d^4q}{(2\pi)^4}\, d^4\theta\, V^A(-q,\theta) \partial^2\Pi_{1/2} V^A(q,\theta)\, G_V(\alpha_0, \lambda_0, Y_0, \Lambda/q);\\
\label{Gamma2_Phi}
&&\hspace*{-12mm} \Gamma^{(2)}_\phi = \frac{1}{4}\int \frac{d^4q}{(2\pi)^4}\, d^4\theta\, \phi^{*i}(-q,\theta) \phi_j(q,\theta) \big(G_\phi\big)_i{}^j(\alpha_0, \lambda_0, Y_0, \Lambda/q);\\
\label{Gamma2_C}
&&\hspace*{-12mm} \Gamma^{(2)}_c = \frac{1}{4}\int \frac{d^4q}{(2\pi)^4}\, d^4\theta\, \Big(\bar c^{+A}(-q,\theta) c^{A}(q,\theta) - \bar c^A(-q,\theta) c^{+A}(q,\theta)\Big) G_c(\alpha_0, \lambda_0, Y_0, \Lambda/q).
\end{eqnarray}

\noindent
where $\partial^2\Pi_{1/2} \equiv - D^a \bar D^2 D_a/8$ is the supersymmetric transversal projection operator.

Writing Eq. (\ref{Gamma2_Background_V}) we took into account that the two-point Green function of the background gauge superfield is transversal due to the manifest background gauge invariance of the effective action. Similarly, in Eq. (\ref{Gamma2_Quantum_V}) we used the fact that quantum corrections to the two-point Green function of the quantum gauge superfield are also transversal due to the Slavnov--Taylor identity \cite{Taylor:1971ff,Slavnov:1972fg}.

In our notation the renormalization constants are introduced with the help of the equations\footnote{There is an infinite set of the renormalization constants for parameters of the nonlinear renormalization. Here we explicitly write only the first one ($Z_y$).}

\begin{eqnarray}
&& Z_\alpha \equiv \frac{\alpha}{\alpha_0};\qquad\quad V^A \equiv Z_V  (V_R)^A;\qquad\quad  Z_y \equiv \frac{y_0}{y};\qquad\quad \ldots \nonumber\\
&& Z_\xi \equiv \frac{\xi}{\xi_0};\qquad \bar c^A c^B = Z_c  (\bar c_R)^A (c_R)^B;\qquad \phi_i = \big(\sqrt{Z_\phi}\big)_i{}^j \big(\phi_R\big)_j,\qquad
\end{eqnarray}

\noindent
where the subscript $R$ stands for the renormalized superfields. The renormalization constants are constructed from the requirement that the functions $d^{-1}$, $Z_V^2 G_V$, $(Z_\phi G_\phi)_i{}^j$, and $Z_c G_c$ expressed in terms of the renormalized couplings $\alpha$, $\lambda^{ijk}$, and $Y$ should be finite in the limit $\Lambda\to\infty$. Note that due to the non-renormalization of the superpotential \cite{Grisaru:1979wc} the renormalized Yukawa couplings are given by

\begin{equation}\label{Lambda_Renormalization}
\lambda^{ijk} = \big(\sqrt{Z_\phi}\big)_m{}^i \big(\sqrt{Z_\phi}\big)_n{}^j \big(\sqrt{Z_\phi}\big)_p{}^k \lambda_0^{mnp}.
\end{equation}

\noindent
We will always assume that no finite constants corresponding to finite renormalizations appear in this equation, so that Eq. (\ref{Lambda_Renormalization}) partially fixes the subtraction scheme. Similarly, due to the all-loop finiteness of the triple gauge-ghost vertices \cite{Stepanyantz:2016gtk}\footnote{Similar statements were also known for some theories formulated in terms of the usual fields in the Landau gauge $\xi_0=0$ \cite{Dudal:2002pq,Capri:2014jqa}.} it is possible to choose a subtraction scheme in which the renormalization constants satisfy the condition

\begin{equation}\label{ZZZ_Constraint}
Z_\alpha^{-1/2} Z_c Z_V =1.
\end{equation}

\noindent
Note that this equation is compatible with minimal subtractions of logarithms, because in the HD+MSL scheme all renormalization constants $Z_\alpha$, $Z_c$, and $Z_V$ are equal to 1 for $\mu=\Lambda$.

The equation (\ref{ZZZ_Constraint}) allows to present the NSVZ equation in an equivalent form relating the $\beta$-function to the anomalous dimensions of the quantum gauge superfield, the Faddeev--Popov ghosts, and the matter superfields,

\begin{eqnarray}\label{NSVZ_Equivalent_Form_Bare}
&& \frac{\beta(\alpha_0,\lambda_0,Y_0)}{\alpha_0^2} = - \frac{1}{2\pi}\Big(3 C_2 - T(R) - 2C_2 \gamma_c(\alpha_0,\lambda_0,Y_0)\nonumber\\
&&\qquad\qquad\qquad\qquad\qquad\ \ - 2C_2 \gamma_V(\alpha_0,\lambda_0,Y_0) + \frac{1}{r} C(R)_i{}^j \big(\gamma_\phi\big)_j{}^i(\alpha_0,\lambda_0,Y_0)\Big),\qquad
\end{eqnarray}

\noindent
where we take into account that RGFs (at least, $\gamma_V$ and $\gamma_c$) may in general depend on $Y_0$. RGFs defined in terms of the bare couplings entering this equation can be related to the corresponding two-point Green functions by the equations

\begin{eqnarray}\label{Beta_Bare_Definition}
&& \frac{\beta(\alpha_0,\lambda_0,Y_0)}{\alpha_0^2} = -\frac{d}{d\ln\Lambda}\Big(\frac{1}{\alpha_0}\Big)\bigg|_{\alpha,\lambda,Y=\mbox{\scriptsize const}}
= \frac{d}{d\ln\Lambda} \Big(d^{-1} - \alpha_0^{-1}\Big)\bigg|_{\alpha,\lambda,Y=\mbox{\scriptsize const};\, p\to 0};\qquad\\
\label{Gamma_V_Bare_Definition}
&& \gamma_V(\alpha_0,\lambda_0,Y_0) = - \frac{d\ln Z_V}{d\ln\Lambda}\bigg|_{\alpha,\lambda,Y = \mbox{\scriptsize const}} = \frac{1}{2}\,\frac{d\ln G_V}{d\ln\Lambda}\bigg|_{\alpha,\lambda,Y = \mbox{\scriptsize const};\ q\to 0};\\
\label{Gamma_C_Bare_Definition}
&& \gamma_c(\alpha_0,\lambda_0,Y_0) = - \frac{d\ln Z_c}{d\ln\Lambda}\bigg|_{\alpha,\lambda,Y = \mbox{\scriptsize const}} = \frac{d\ln G_c}{d\ln\Lambda}\bigg|_{\alpha,\lambda,Y = \mbox{\scriptsize const};\ q\to 0};\\
\label{Gamma_Phi_Bare_Definition}
&& \big(\gamma_\phi\big)_i{}^j(\alpha_0,\lambda_0,Y_0) = - \frac{d\big(\ln Z_\phi\big)_i{}^j}{d\ln\Lambda}\bigg|_{\alpha,\lambda,Y = \mbox{\scriptsize const}} = \frac{d\big(\ln G_\phi\big)_i{}^j}{d\ln\Lambda}\bigg|_{\alpha,\lambda,Y = \mbox{\scriptsize const};\ q\to 0}.
\end{eqnarray}

\noindent
That is why Eq. (\ref{NSVZ_Equivalent_Form_Bare}) can also be rewritten as an equation relating these Green functions

\begin{eqnarray}\label{NSVZ_For_Green_Functions_With_1Loop}
&&  \left.\frac{d}{d\ln\Lambda} \Big(d^{-1} -\alpha_0^{-1}\Big)\right|_{\alpha,\lambda,Y = \mbox{\scriptsize const};\ p\to 0} = -\frac{1}{2\pi} \Big(3C_2-T(R)\Big) + \frac{1}{2\pi}\, \frac{d}{d\ln\Lambda} \Big( 2C_2 \ln G_c  \qquad\nonumber\\
&& + C_2 \ln G_V - \frac{1}{r} C(R)_i{}^j \big(\ln G_\phi\big)_j{}^i\Big)\bigg|_{\alpha,\lambda,Y = \mbox{\scriptsize const};\ q\to 0},\qquad
\end{eqnarray}

\noindent
which was first proposed in Ref. \cite{Stepanyantz:2016gtk}. In this paper we will prove it in all orders of the perturbation theory.

\section{Integrals of total derivatives}
\label{Section_Integrals}

\subsection{How to construct and calculate integrals of total derivatives}
\hspace*{\parindent}\label{Subsection_Idea}

The key observation needed for deriving the NSVZ $\beta$-function is the factorization of loop integrals which give the function $\beta(\alpha_0,\lambda_0,Y_0)/\alpha_0^2$ into integrals of double total derivatives in the case of using the higher covariant derivative regularization. The all-loop proof of this fact was done in Ref. \cite{Stepanyantz:2019ihw}. The ideas used in this proof allowed constructing a method for obtaining these integrals in each order of the perturbation theory. For this purpose one should calculate only specially modified vacuum supergraphs\footnote{The word ``vacuum'' means that they do not contain any external legs.} instead of a much larger number of superdiagrams with two external legs of the background gauge superfield. Some higher-order calculations made with the help of this method in Refs. \cite{Kuzmichev:2019ywn,Stepanyantz:2019lyo,Aleshin:2020gec} demonstrated that it works correctly and reproduces all known results. Moreover, the structure of quantum corrections which were first obtained by this method is in excellent agreement with some general theorems. Say, the $\beta$-function in the considered theories appeared to be gauge independent and satisfies the NSVZ equation in the HD+MSL scheme. Here using this method we will find an exact all-order expression for the function

\begin{equation}\label{Delta_Beta}
\frac{\beta(\alpha_0,\lambda_0,Y_0)}{\alpha_0^2} - \frac{\beta_{\mbox{\scriptsize 1-loop}}(\alpha_0)}{\alpha_0^2},
\end{equation}

\noindent
where

\begin{equation}\label{Beta_1Loop}
\beta_{\mbox{\scriptsize 1-loop}}(\alpha_0) = -\frac{\alpha_0^2}{2\pi}\left(3C_2-T(R)\right)
\end{equation}

\noindent
is the one-loop $\beta$-function defined in terms of the bare couplings. (For the higher covariant derivative regularization considered in this paper it was calculated in Ref. \cite{Aleshin:2016yvj}.)

Now, let us formulate the algorithm for constructing contributions to the function (\ref{Delta_Beta}) following Refs. \cite{Kuzmichev:2019ywn,Stepanyantz:2019ihw}.

1. As a starting point we consider a vacuum supergraph with $L$ loops and construct an expression for it using the superspace Feynman rules.

2. Next, one needs to find a point with the integration over $d^4\theta$ (or convert the integration over $d^2\theta$ into the integration over $d^4\theta$) and insert the expression $\theta^4 (v^B)^2$ to the corresponding point of this supergraph. Here $v^B$ denotes a function slowly decreasing at a very large scale $R\to \infty$. (Note that without this insertion any vacuum supergraph vanishes due to the integration over the anticommuting variables $\theta$.)

3. We calculate the supergraph modified by the above insertion and omit terms suppressed by powers of $1/(R\Lambda)$. As a result we obtain an expression proportional to

\begin{equation}
{\cal V}_4 \equiv \int d^4x\, \big(v^B\big)^2 \to \infty.
\end{equation}

4. At the next step it is necessary to choose $L$ propagators with independent momenta denoted by $Q_i^\mu$, where $i=1,\ldots,L$. (In our notation the capital letters denote Euclidean momenta which appear after the Wick rotation.) Let the gauge group indices corresponding to the beginnings and endings of these propagators be $a_i$ and $b_i$, respectively. Then the product of the marked propagators contains the factor $\prod_{i=1}^L \delta_{a_i}^{b_i}$.

5. The product $\prod_{i=1}^L \delta_{a_i}^{b_i}$ coming from the marked propagators in the integrand of the (Euclidean) loop integral should formally be replaced by the operator

\begin{equation}\label{Replacement}
\sum_{k,l=1}^L  \Big(\prod\limits_{i\ne k,l} \delta_{a_i}^{b_i}\Big)\, (T^A)_{a_k}{}^{b_k} (T^A)_{a_l}{}^{b_l} \frac{\partial^2}{\partial Q^\mu_k \partial Q^\mu_l}.
\end{equation}

6. At the last step, it is necessary to apply the operator

\begin{equation}\label{Operator}
- \frac{2\pi}{r{\cal V}_4}\cdot \frac{d}{d\ln\Lambda}
\end{equation}

\noindent
to the resulting expression, where the derivative with respect to $\ln\Lambda$ should be calculated at fixed vales of the renormalized couplings prior to the integration over loop momenta.

After the above described procedure we obtain a contribution  to the function (\ref{Delta_Beta}) corresponding to the considered supergraph. It is produced by the sum of all two-point superdiagrams which are constructed from this supergraph by attaching two external lines of the background gauge superfield $\bm{V}$ in all possible ways. By construction, the result is given by a certain integral of double total derivatives with respect to the loop momenta.

The integrals of total derivatives do not vanish due to the singularities of the integrands, which appear when two momentum derivatives act on an inverse squared momentum,

\begin{equation}\label{Delta_Singularity}
\frac{\partial^2}{\partial Q_\mu \partial Q_\mu} \Big(\frac{1}{Q^2}\Big) = -4\pi^2 \delta^4(Q).
\end{equation}

\noindent
We will also need a modification of this identity which is obtained when the derivatives are taken with respect to the different momenta,

\begin{equation}\label{Delta_Singularity_General}
\frac{\partial^2}{\partial Q_{\mu,1} \partial Q_{\mu,2}} \Big(\frac{1}{(a_1 Q_{\mu,1}+ a_2 Q_{\mu,2}+ Q_{\mu,3})^2}\Big) = - 4\pi^2 a_1 a_2\, \delta^4\Big(a_1 Q_{\mu,1}+ a_2 Q_{\mu,2}+ Q_{\mu,3}\Big),
\end{equation}

\noindent
where $a_1$ and $a_2$ are some constants.

Note that in calculating the integrals of double total derivatives we should take these singular contributions with the opposite sign. Really, if $f(Q^2)$ is a non-singular function which rapidly tends to 0 at infinity, then

\begin{equation}\label{Integral_And_Singularity}
\int \frac{d^4Q}{(2\pi)^4}\, \frac{\partial^2}{\partial Q_\mu \partial Q_\mu}\Big(\frac{f(Q^2)}{Q^2}\Big) = \frac{1}{4\pi^2} f(0) = \int  \frac{d^4Q}{(2\pi)^4}\, f(Q^2)\cdot 4\pi^2 \delta^4(Q).
\end{equation}

Note that terms in which double total derivatives act on $Q^{-4}$ are not well-defined and cannot appear in the final expression for the function (\ref{Delta_Beta}), although they can be present in expressions for separate supergraphs. This statement has been confirmed by some explicit two- and three-loop calculations in Refs. \cite{Kazantsev:2018nbl,Stepanyantz:2019lyo}.

\subsection{Graphs and total derivatives}
\hspace*{\parindent}\label{Subsection_Graphs}

Let consider a vacuum supergraph with $L$ loops, $V$ vertices, and $P$ internal lines and set directions of all internal lines in an arbitrary way. These directions will be pointed by arrows. Also we denote momenta of all internal lines by certain letters. In our conventions an incoming momentum has the sign ``minus'', while an outcoming one has the sign ``plus''.

Let us construct a $(V-1)\times P$ matrix $M$ corresponding to the supergraph under consideration. For this purpose we numerate the vertices in an arbitrary order and write the energy-momentum conservation laws in all vertices, except for the last one. (Evidently, the last equation is a linear combination of the others.) The resulting system of equations can be written in the form

\begin{equation}
\left(
\begin{array}{cccc}
M_{1,1} & M_{1,2} & \ldots & M_{1,P}\\
M_{2,1} & M_{2,2} & \ldots & M_{2,P}\\
\vdots & \vdots & \ddots & \vdots\\
M_{V-1,1} & M_{V-1,2} & \ldots & M_{V-1,P}
\end{array}
\right)
\left(
\begin{array}{c}
Q_{\mu,1}\\
Q_{\mu,2}\\
\vdots\\
Q_{\mu,P}
\end{array}
\right) =0.
\end{equation}

\noindent
The matrix with the elements $M_{i,J}$, where $i=1,\ldots,V-1$ and $J=1,\ldots,P$, is the required matrix matched to the considered vacuum supergraph. From the above equations, which can briefly be written as

\begin{equation}\label{Conservation_Laws}
\sum\limits_{J=1}^P M_{i,J} Q_{\mu,J} = 0,
\end{equation}

\noindent
we conclude that only $P-V+1$ momenta are independent. According to the topological identity $L=P-V+1$, this implies that (as well known) there are $L$ independent momenta $Q_{\mu,\alpha}$, $\alpha=1,\ldots, L$ in the considered supergraph, and the others can be expressed in terms of them,

\begin{equation}\label{Dependent_Momenta}
Q_{\mu,I} = \sum\limits_{\alpha=1}^L N_{I,\alpha} Q_{\mu,\alpha}.
\end{equation}

\begin{figure}[h]
\begin{picture}(0,3)
\put(4.05,0.2){\includegraphics[scale=0.12]{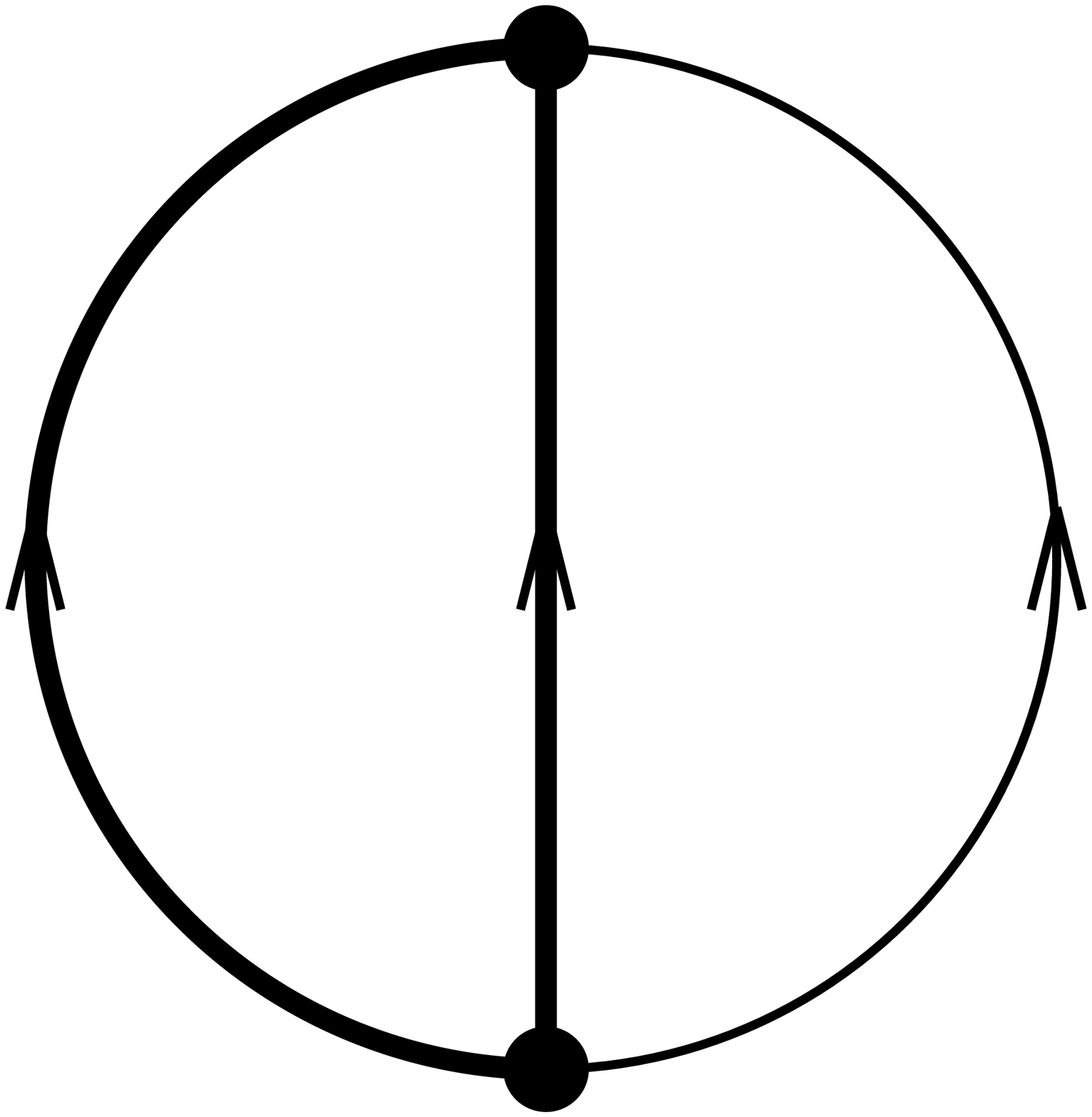}}
\put(5.0,2.6){2} \put(5.05,-0.2){1}
\put(3.3,1.1){$Q_{\mu,1}$} \put(4.3,1.3){$Q_{\mu,2}$} \put(6.3,1.1){$Q_{\mu,3}$}
\put(10.05,0.2){\includegraphics[scale=0.12]{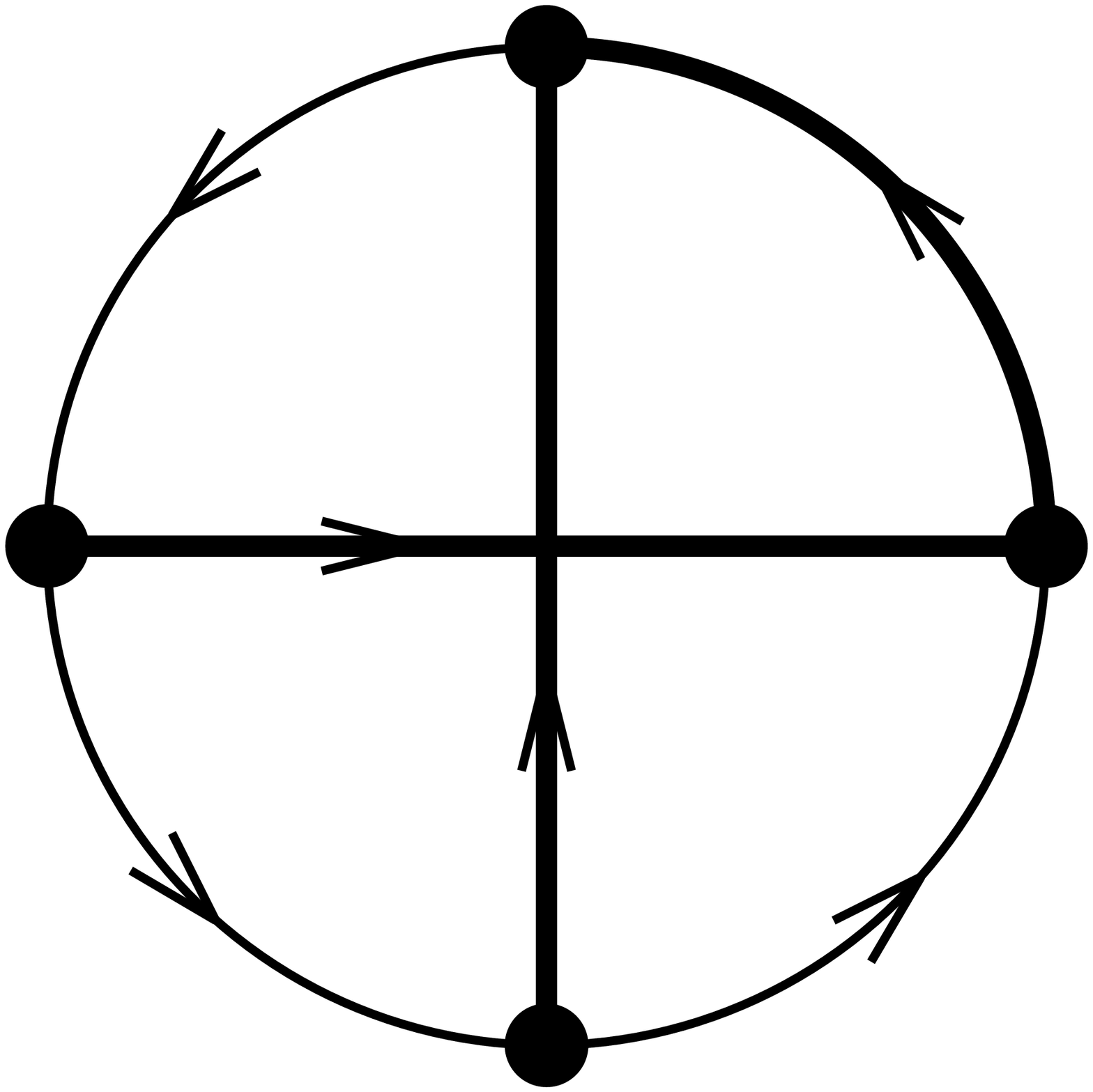}}
\put(11.02,2.6){1} \put(11.05,-0.2){3} \put(9.7,1.2){2} \put(12.35,1.2){4}
\put(12.0,2.1){$Q_{\mu,3}$} \put(11.25,0.8){$Q_{\mu,2}$} \put(10.3,1.5){$Q_{\mu,1}$} \put(9.6,2.1){$Q_{\mu,4}$} \put(9.7,0.3){$Q_{\mu,5}$} \put(12.0,0.3){$Q_{\mu,6}$}
\end{picture}
\caption{Examples of vacuum supergraphs, for which we construct the matrices $M$. The bold lines denote propagators the momenta of which are considered as independent.}\label{Figure_Examples}
\end{figure}

As a simple example we can consider a supergraph presented in Fig. \ref{Figure_Examples} on the left. In this case the equation (\ref{Conservation_Laws}), the matrix $M$, and the equation (\ref{Dependent_Momenta}) read as

\begin{equation}
\Big(1\ 1 \ 1\Big) \left(
\begin{array}{c}
Q_{\mu,1}\\
Q_{\mu,2}\\
Q_{\mu,3}
\end{array}
\right) = 0;\qquad M = \Big(1\ 1 \ 1\Big);\qquad \left(
\begin{array}{c}
Q_{\mu,1}\\
Q_{\mu,2}\\
Q_{\mu,3}
\end{array}
\right) = \left(
\begin{array}{c}
1\\
0\\
-1
\end{array}
\right) Q_{\mu,1} + \left(
\begin{array}{c}
0\\
1\\
-1
\end{array}
\right) Q_{\mu,2}.
\end{equation}

\noindent
For a more complicated three-loop graph presented in Fig. \ref{Figure_Examples} on the right as another example, the matrix $M$ takes the form

\begin{equation}
M = \left(
\begin{array}{cccccc}
0 & -1 & -1 & 1 & 0 & 0\\
1 & 0 & 0 & -1 & 1 & 0\\
0 & 1 & 0 & 0 & -1 & 1
\end{array}
\right).
\end{equation}

\noindent
Evidently, in this case there are three independent momenta. For example, it is possible to choose $Q_{\mu,1}$, $Q_{\mu,2}$, and $Q_{\mu,3}$ as independent variables. The corresponding propagators are denoted in Fig. \ref{Figure_Examples} by the bold lines. Then the equation (\ref{Dependent_Momenta}) takes the form

\begin{equation}
\left(
\begin{array}{c}
Q_{\mu,1} \\ Q_{\mu,2} \\ Q_{\mu,3} \\ Q_{\mu,4} \\ Q_{\mu,5} \\ Q_{\mu,6}
\end{array}
\right) = \left(
\begin{array}{c}
1 \\ 0 \\ 0 \\ 0 \\ -1 \\ -1
\end{array}
\right) Q_{\mu,1} + \left(
\begin{array}{c}
0 \\ 1 \\ 0 \\ 1 \\ 1 \\ 0
\end{array}
\right) Q_{\mu,2} + \left(
\begin{array}{c}
0 \\ 0 \\ 1 \\ 1 \\ 1 \\ 1
\end{array}
\right) Q_{\mu,3}.
\end{equation}

An important observation made in Ref. \cite{Stepanyantz:2019ihw} is that the equations which reflect the gauge invariance of vertices are very similar to Eq. (\ref{Conservation_Laws}). Really, let us consider a vertex with $n$ outcoming lines corresponding to the term

\begin{equation}
\int d^8x\, \hat V^{I_1 I_2 \ldots I_n}\, \varphi_{I_1} \varphi_{I_2} \ldots \varphi_{I_n},
\end{equation}

\noindent
where $\hat V$ is an operator acting on the product of various superfields of the theory denoted by $\varphi_I$. Then the energy-momentum conservation in the considered vertex is expressed by the equation

\begin{equation}\label{Conservation_In_Vertex}
Q_{\mu,1} + Q_{\mu,2} + \ldots + Q_{\mu,n} = 0,
\end{equation}

\noindent
which is very similar to the equation which follows from the gauge invariance of theory

\begin{equation}\label{Global_Gauge_Invariance_In_Vertex}
(T^A)_{K}{}^{I_1}\, \hat V^{K I_2 \ldots I_n} + (T^A)_{K}{}^{I_2}\, \hat V^{I_1 K \ldots I_n}  + \ldots + (T^A)_{K}{}^{I_n}\, \hat V^{I_1 I_2 \ldots K} = 0,
\end{equation}

\noindent
where $(T^A)_I{}^J$ are generators of the gauge group in a relevant representation.

For the incoming lines the signs of the corresponding momenta in equations like (\ref{Conservation_In_Vertex}) should be changed. Similarly, in equations like (\ref{Global_Gauge_Invariance_In_Vertex}) one should replace the generators by the transposed generators of the conjugated representation taken with an opposite sign, although this replacement does not change the equation, because

\begin{equation}
\big(T^A_{R}\big)_K{}^I \ \to\  - \big(T^A_{\bar R}\big){}^I{}\vphantom{\big)}_K = \big(T^A_R\big)_K{}^I.
\end{equation}

Let us consider a vacuum supergraph and replace a $\delta$-symbol $\delta_{a_J}^{b_J}$ coming from the propagator with a number $J$ by a certain matrix $A_{a_J}{}^{b_J}$. We will denote the resulting modified supergraph by $[A]_J$. If we replace the $i$-th vertex operator $\hat V^{I_1 I_2\ldots I_n}$ in a certain vacuum supergraph by the left hand side of Eq. (\ref{Global_Gauge_Invariance_In_Vertex}), then, using this notation, we obtain the equation

\begin{equation}
\sum\limits_{J=1}^P M_{i,J} \big[\,T^A\big]_J = 0.
\end{equation}

\noindent
The system of these equations is analogous to Eq. (\ref{Conservation_Laws}) and contains the same matrix $M_{i,J}$. Evidently, the solution of these equations can be written in a form similar to Eq. (\ref{Dependent_Momenta}),

\begin{equation}\label{TA_Solution}
\big[\,T^A\big]_I = \sum\limits_{\alpha=1}^L N_{I,\alpha} \big[\,T^A\big]_\alpha,
\end{equation}

\noindent
where $N_{I,\alpha}$ are exactly the same coefficients as in Eq. (\ref{Dependent_Momenta}).

According to the algorithm described in section \ref{Subsection_Idea} we need to insert $\theta^4 (v^B)^2$ into the vacuum supergraph and make the replacement described in the item 5. We will denote supergraphs obtained in this way as

\begin{equation}
\sum\limits_{\alpha,\beta=1}^L \Big[\,\theta^4 (v^B)^2; \frac{\partial^2}{\partial Q_{\mu,\alpha} \partial Q_{\mu,\beta}} \big[\, T^A\big]_\alpha \big[\, T^A\big]_\beta \Big].
\end{equation}

\noindent
This means that $\theta^4 (v^B)^2$ is inserted into an arbitrary point of the supergraph, two $\delta$-symbols are replaced by the generators, and the double total derivatives are applied to the integrand of the loop integral.

Then the sum of all perturbative contributions to the function (\ref{Delta_Beta}) can be presented in the form

\begin{equation}\label{Supergraph_Sum}
\frac{\beta(\alpha_0,\lambda_0,Y_0)}{\alpha_0^2} - \frac{\beta_{\mbox{\scriptsize 1-loop}}(\alpha_0)}{\alpha_0^2} = -\frac{2\pi}{r{\cal V}_4} \frac{d}{d\ln\Lambda} \hspace*{-2mm} \sum\limits_{\stackrel{\mbox{\scriptsize vacuum}}{\mbox{\scriptsize supergraphs}}} \sum\limits_{\alpha,\beta=1}^L \Big[\,\theta^4 (v^B)^2;\, \frac{\partial^2}{\partial Q_{\mu,\alpha} \partial Q_{\mu,\beta}} \big[\, T^A\big]_\alpha \big[\, T^A\big]_\beta \Big].
\end{equation}

\noindent
By construction, this expression is an integral of double total derivatives. However, it does not vanish because the double total derivatives produce singular contributions acting on $1/Q_I^2$. It is important to note that expressions for separate supergraphs (presented in the form of scalar integrals) can contain not only $1/Q_I^2$, but also $1/(Q_I^2)^n \equiv 1/Q_I^{2n}$ with $n\ge 2$ if $Q_I$ is a momentum of a quantum gauge superfield propagator. Acting on $1/Q_I^{2n}$ with $n\ge 2$ the double total derivatives produce an expression which is not well-defined due to infrared divergences. However, we know that the left hand side of Eq. (\ref{Supergraph_Sum}) is well-defined. Therefore, all bad terms coming from the $1/Q_I^{2n}$ singularities should cancel each other. The calculations in the lowest orders \cite{Kazantsev:2018nbl,Stepanyantz:2019lyo} exactly confirm this statement.

\begin{figure}[h]
\begin{picture}(0,3)
\put(6.8,0){\includegraphics[scale=0.15]{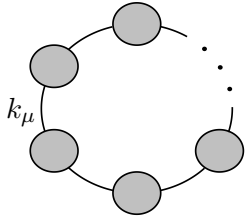}}
\put(6.5,1.3){$k_\mu$}
\end{picture}
\caption{The structure of vacuum supergraphs with coinciding momenta (which are equal to $k_\mu$). The gray circles denote 1PI subdiagrams.}\label{Figure_Graph_With_Coinciding_Momenta}
\end{figure}

Because all bad terms should cancel each other, it is necessary to consider only the $1/Q_I^2$ singularities. Then in the expression for a supergraph we need to consider only the product of all {\it different} inverse squared momenta multiplied by a nonsingular function $f$,

\begin{equation}\label{Product_Of_Momenta}
\frac{f(Q_1,Q_2,\ldots, Q_{P'})}{Q_1^2 Q_2^2 \ldots Q_{P'}^2} \equiv f(Q_1,Q_2,\ldots, Q_{P'}) \prod\limits_{I=1}^{P'}\vphantom{\prod}' \frac{1}{Q_I^2},
\end{equation}

\noindent
where $Q_{\mu,I} \ne Q_{\mu,J}$ for $I\ne J$. The prime indicates that the product includes only different momenta, a number of them being denoted by $P'$ (which is evidently less or equal to $P$). This is essential for supergraphs which contain coinciding momenta. The structure of such supergraphs is illustrated in Fig. \ref{Figure_Graph_With_Coinciding_Momenta}. In this figure the gray circles denote 1PI subdiagrams, which are connected by propagators with coinciding momenta. By construction, the expression (\ref{Product_Of_Momenta}) should include only one such momentum.

$\delta$-singularities appear when double total derivatives act on various factors in the product (\ref{Product_Of_Momenta}). Note that, according to Eq. (\ref{Delta_Singularity_General}), such singularities can appear even if derivatives with respect to different momenta act on the same inverse squared momentum,

\begin{equation}
\frac{\partial^2}{\partial Q_{\mu,\alpha} \partial Q_{\mu,\beta}} \Big(\frac{1}{Q_I^2}\Big) = - 4\pi^2 N_{I,\alpha} N_{I,\beta}\, \delta^4(Q_I),
\end{equation}

\noindent
where we also took Eq. (\ref{Dependent_Momenta}) into account. This implies that acting on the product (\ref{Product_Of_Momenta}) the double total derivatives give the singular contribution

\begin{equation}\label{Singularity}
\frac{\partial^2}{\partial Q_{\mu,\alpha} \partial Q_{\mu,\beta}} \prod\limits_{I=1}^{P'}\vphantom{\prod}' \frac{1}{Q_I^2}\ \to\ - 4\pi^2 \sum\limits_{I=1}^{P'}\vphantom{\sum}' N_{I,\alpha} N_{I,\beta}\, \delta^4(Q_I) \prod\limits_{J\ne I}\vphantom{\prod}' \frac{1}{Q_J^2}.
\end{equation}

\noindent
Taking into account that the bad terms containing $1/Q_I^{2n}$ with $n\ge 2$ should cancel each other, we can rewrite the expression (\ref{Supergraph_Sum}) as a sum of supergraphs in which they are omitted.  According to Eq. (\ref{Singularity}), in the remaining terms we replace one of $1/Q_I^2$ by $4\pi^2\delta^4(Q_I)$ and sum up all expressions obtained after these substitutions. (The sign is ``plus'', because the singular contribution of the form (\ref{Delta_Singularity}) should be taken with the minus sign, see Eq. (\ref{Integral_And_Singularity}).) The result can be presented as

\begin{eqnarray}
&& -\frac{2\pi}{r{\cal V}_4} \frac{d}{d\ln\Lambda} \hspace*{-2mm} \sum\limits_{\stackrel{\mbox{\scriptsize vacuum}}{\mbox{\scriptsize supergraphs}}} \sum\limits_{I=1}^{P'} \sum\limits_{\alpha,\beta=1}^L N_{I,\alpha} N_{I,\beta}\nonumber\\
&&\qquad\qquad\qquad\qquad \times \Big[\,\theta^4 (v^B)^2;\, \big[\, T^A\big]_\alpha \big[\, T^A\big]_\beta;\, \frac{1}{Q^{2n}}\bigg|_{n\ge 2} \to 0;\, \frac{1}{Q_I^2} \to 4\pi^2\delta^4(Q_I) \Big].\qquad
\end{eqnarray}

\noindent
Next, using Eq. (\ref{TA_Solution}) we calculate the sums over the indices $\alpha$ and $\beta$,

\begin{equation}\label{New_Expression_For_Delta_Beta}
-\frac{2\pi}{r{\cal V}_4} \frac{d}{d\ln\Lambda} \hspace*{-2mm} \sum\limits_{\stackrel{\mbox{\scriptsize vacuum}}{\mbox{\scriptsize supergraphs}}} \sum\limits_{I=1}^{P'} \Big[\,\theta^4 (v^B)^2;\, \big[\, T^A T^A\big]_I;\, \frac{1}{Q^{2n}}\bigg|_{n\ge 2} \to 0;\, \frac{1}{Q_I^2} \to 4\pi^2\delta^4(Q_I) \Big].
\end{equation}

\noindent
Note that for the gauge and ghost propagators $(T^A T^A)_{BC} = C_2\delta_{BC}$, while for the matter propagators $(T^A T^A)_i{}^j = C(R)_i{}^j$.

Basing on Eq. (\ref{New_Expression_For_Delta_Beta}) it is possible to modify the step 5 of the algorithm described in section \ref{Subsection_Idea}. Note that in what follows we formulate this step before the Wick rotation.

5. We construct a set of supergraphs in which one of the propagators is marked. If the original supergraph contains some propagators with the same momentum, then one can mark only one propagator with this momentum. After calculating the $D$-algebra, one should omit all terms proportional to $1/q_I^{2n}$ with $n\ge 2$ in scalar integrals and for any marked propagator with the momentum $q_\mu$ in the remaining terms make the replacement

\begin{eqnarray}\label{Replacement_Modified}
&& \frac{1}{q^2}\quad\ \to\ \ C_2\, \frac{\partial^2}{\partial q_\mu \partial q^\mu} \Big(\frac{1}{q^2}\Big)\qquad\qquad \mbox{for gauge and ghost propagators};\nonumber\\
&& \delta_i^j\, \frac{1}{q^2}\ \to\ \ C(R)_i{}^j\, \frac{\partial^2}{\partial q_\mu \partial q^\mu} \Big(\frac{1}{q^2}\Big)\qquad \mbox{for matter propagators}.
\end{eqnarray}

\noindent
After these replacements it is necessary to sum up the expressions for all supergraphs in the constructed set.

Note that this prescription is the same for all vacuum supergraphs, unlike the one described in section \ref{Subsection_Idea}. Really, the form of the operator (\ref{Replacement}) depends on the topology of the supergraph, while the replacement (\ref{Replacement_Modified}) does not. That is why it is possible, first, to find a sum of all vacuum supergraphs and, only after this, make these replacements. Below we will demonstrate that the sum of vacuum supergraphs really does not contain any bad terms which include $1/Q_I^{2n}$ with $n\ge 2$. This confirms the above argumentation, although the explicit mechanism of cancelling the bad contributions to the function (\ref{Delta_Beta}) should be analyzed separately.

In the next section using the modification of the algorithm described above we will find the sum of all contributions to the function (\ref{Delta_Beta}) which come from $\delta$-singularities of the form (\ref{Delta_Singularity_General}). Various perturbative verifications of the above described modification will be made in section \ref{Section_Explicit_Calculation}.

\section{The all-loop sum of singular contributions}
\label{Section_Singularities}

\subsection{The starting point and main idea}
\hspace*{\parindent}\label{Subsection_Main_Idea}

In this section we will calculate the sum of all singular contributions. They appear when the double total derivatives act on matter, ghost and gauge propagators and effectively cut the corresponding internal line. In \cite{Stepanyantz:2019lfm} the sums of singularities produced by the cuts of matter and ghost propagators have been calculated in all orders. However, the method which is used in this paper is different. That is why it is expedient to recalculate the sums of matter and ghost singularities and verify that both approaches give the same result. Moreover, in this section we will find the all-loop expression for a sum of singularities produced by the cuts of quantum gauge superfield propagators. The method of summation generalizes the one proposed in Ref. \cite{Stepanyantz:2011jy} for the Abelian case.

First, we introduce the auxiliary action

\begin{equation}
S_g \equiv S_{\mbox{\scriptsize total}} +\Delta S_g
\end{equation}

\noindent
depending on a real constant $g$, where $S_{\mbox{\scriptsize total}} \equiv S_{\mbox{\scriptsize reg}} + S_{\mbox{\scriptsize gf}} + S_{\mbox{\scriptsize FP}} + S_{\mbox{\scriptsize NK}}$ and

\begin{eqnarray}
&& \Delta S_g \equiv \frac{1}{4} \Big(\frac{1}{g}-1\Big) \int d^8x\,\Big[ - V^A R(\partial^2/\Lambda^2)\, \partial^2 \Pi_{1/2} V^A \nonumber\\
&&\qquad\qquad\qquad -\frac{1}{8\xi_0} D^2 V^A K(\partial^2/\Lambda^2) \bar D^2 V^A + \bar c^{+A} c^A -\bar c^A c^{+A} + \phi^{*i} F(\partial^2/\Lambda^2)\phi_i\Big].\qquad
\end{eqnarray}

\noindent
For $g=1$ the action $S_g$ evidently coincides with $S_{\mbox{\scriptsize total}}$. Moreover, all vertices generated by $S_g$ coincide with the ones produced by $S_{\mbox{\scriptsize total}}$. However, the parts of $S_g$ quadratic in the quantum gauge superfield, the Faddeev--Popov ghosts, and the matter superfields differ from the corresponding quadratic parts of $S_{\mbox{\scriptsize total}}$ by the factor $1/g$. This implies that, in comparison with the original theory, all propagators of the above mentioned quantum superfields obtained from the generating functional

\begin{equation}
Z_g \equiv \int D\mu\, \mbox{Det}(PV, M_\varphi)^{-1}\, \mbox{Det}(PV, M)^c \exp\Big(i S_g + i S_{\mbox{\scriptsize sources}}\Big),
\end{equation}

\noindent
with $c=T(R)/T(R_{\mbox{\scriptsize PV}})$, acquire the factor $g$. Note that the propagators of the massive Pauli--Villars superfields and of the Nielsen--Kallosh ghosts remain the same. We need not modify them, because the Pauli--Villars propagators do not produce singularities, and the Nielsen--Kallosh ghosts are essential only in the one-loop approximation which is considered separately. Certainly, the functional $Z_g$ is treated formally, and we do not care about regularizing divergences in the corresponding supergraphs.

Now, let us find the derivative of the functional $\Gamma_g$ (defined as a Legendre transform of $W_g = - i\ln Z_g$) with respect to the parameter $g$ at vanishing superfields and $g=1$. It is convenient to present the result in the form

\begin{eqnarray}\label{Gamma_Derivative}
&&\hspace*{-3mm} \frac{\partial \Gamma_g}{\partial g}\bigg|_{g=1;\, \mbox{\scriptsize fields=0}} = \Big\langle\frac{\partial S_g}{\partial g}\Big\rangle\Big|_{g=1,\, \mbox{\scriptsize fields=0}}
= \Big\langle \frac{1}{4} \int d^8x\, \bigg\{ V^A\Big[ \partial^2\Pi_{1/2} R(\partial^2/\Lambda^2) + \frac{1}{16\xi_0}\Big(\bar D^2 D^2 \qquad\nonumber\\
&&\hspace*{-3mm} + D^2 \bar D^2\Big) K(\partial^2/\Lambda^2) \Big]V^A - \phi^{*i} F(\partial^2/\Lambda^2) \phi_i - \bar c^{+A} c^A + \bar c^A c^{+A} \bigg\} \Big\rangle\bigg|_{g=1;\, \mbox{\scriptsize fields=0}}.\qquad
\end{eqnarray}

\noindent
The angular brackets entering this equation are defined in the standard way

\begin{equation}
\langle A \rangle \equiv \frac{1}{Z}\int D\mu\, A\, \mbox{Det}(PV, M_\varphi)^{-1}\, \mbox{Det}(PV, M)^c \exp\Big(i S_{\mbox{\scriptsize total}} + i S_{\mbox{\scriptsize sources}}\Big),
\end{equation}

\noindent
where the sources should be expressed in terms of fields using Eq. (\ref{Field_Definitions}).

Evidently, the expression (\ref{Gamma_Derivative}) can be rewritten in terms of the inverse two-point Green functions of the quantum superfields,

\begin{eqnarray}\label{Gamma_Derivative_In_Terms_Of_Inverse_Functions}
&& \frac{\partial \Gamma_g}{\partial g}\bigg|_{g=1;\, \mbox{\scriptsize fields=0}}
= \frac{i}{4}\int d^8x\,\bigg\{\Big[ \partial^2\Pi_{1/2} R(\partial^2/\Lambda^2) + \frac{1}{16\xi_0}\Big(\bar D^2 D^2 + D^2 \bar D^2\Big) K(\partial^2/\Lambda^2) \Big]_x
\nonumber\\
&& \times \Big(\frac{\delta^2\Gamma}{\delta V^A_{x}\delta V^A_y}\Big)^{-1} - F(\partial^2/\Lambda^2)_x\Big(\frac{\delta^2\Gamma}{\delta\phi^{*i}_{\,,x}\delta\phi_{i,y}}\Big)^{-1}
- \Big(\frac{\delta^2\Gamma}{\delta c^A_{x}\,\delta \bar c^{+A}_{y}}\Big)^{-1}
- \Big(\frac{\delta^2\Gamma}{\delta \bar c^A_{x}\,\delta c^{+A}_{y}}\Big)^{-1}
\bigg\}\bigg|_{y=x}.\qquad
\end{eqnarray}

\noindent
(Certainly, in the last expression we also assume that $g=1$ and the (super)fields are set to 0.)

The original effective action $\Gamma$ at vanishing fields is given by the sum of vacuum supergraphs. So far we do not calculate these supergraphs and work only with the formal expressions constructed with the help of supersymmetric Feynman rules. The derivative (\ref{Gamma_Derivative}) is contributed by the same vacuum supergraphs, but each of them is multiplied by the number of propagators. Really, as we have already mentioned, each propagator of a quantum superfield is proportional to the parameter $g$, while the vertices do not contain $g$. Evidently, the first and second terms in Eq. (\ref{Gamma_Derivative_In_Terms_Of_Inverse_Functions}) are contributed by vacuum supergraphs multiplied by the number of gauge and matter propagators, respectively. The last two terms give a sum of vacuum supergraphs multiplied by the number of ghost propagators.

Note that if a supergraph has some coinciding momenta as in Fig. \ref{Figure_Graph_With_Coinciding_Momenta}, all propagators with the coinciding momenta should be taken into account when calculating the number of propagators. The contribution of these graphs to $\partial\Gamma_g/\partial g$ is given by a relevant term in Eq. (\ref{Gamma_Derivative_In_Terms_Of_Inverse_Functions}) and is proportional to a certain inverse Green function. If a usual two-point Green function is proportional to the function

\begin{equation}
G \equiv 1+ \Delta G
\end{equation}

\noindent
depending on the momentum, then the inverse Green function will include

\begin{equation}\label{Inverse_G}
G^{-1} = 1 + \sum\limits_{n=1}^\infty (-1)^n \big(\Delta G\big)^n.
\end{equation}

\noindent
Evidently, a term containing $(\Delta G)^n$ corresponds to supergraphs of the structure presented in Fig. \ref{Figure_Graph_With_Coinciding_Momenta} with $n$ coinciding momenta $k_\mu$. The gray circles in this supergraph certainly give various $\Delta G$.

As we discussed in the previous section, for constructing the $\beta$-function one should cut only one propagator with the momentum $k_\mu$ in a supergraph which have the structure presented in Fig. \ref{Figure_Graph_With_Coinciding_Momenta}. (This cut is made by $\delta^4(k)$ which appears after the replacement (\ref{Replacement_Modified}).) From the other side, the corresponding contribution to the expression (\ref{Gamma_Derivative}) obtained by counting the propagators with the momentum $k_\mu$ is proportional to $n$. Therefore, for calculating the contribution to the $\beta$-function it is necessary to replace $G^{-1}$ by

\begin{equation}
\sum\limits_{n=1}^\infty \frac{(-1)^{n}}{n} \big(\Delta G\big)^n = -\ln G.
\end{equation}

\noindent
(The first term in Eq. (\ref{Inverse_G}) corresponds to the one-loop approximation, which is considered separately.) After this we proceed according to the algorithm described in section \ref{Section_Integrals}. Details of this calculation will be considered below, separately for the cuts of gauge, ghost, and matter propagators.

\subsection{The all-loop sum of matter singularities}
\hspace*{\parindent}\label{Subsection_Matter_Singularities}

Let us start with calculating a contribution to the function (\ref{Delta_Beta}) coming from the cuts of matter propagators. According to Eq. (\ref{Gamma_Derivative_In_Terms_Of_Inverse_Functions}), the corresponding contribution to $\partial\Gamma_g/\partial g$ is given by the expression

\begin{equation}\label{Matter_Part}
-\frac{i}{4}\int d^8x\, F(\partial^2/\Lambda^2)_x \Big(\frac{\delta^2\Gamma}{\delta\phi^{*i}_{\,,x}\delta\phi_{i,y}}\Big)^{-1}\bigg|_{y=x;\ \mbox{\scriptsize fields}=0},
\end{equation}

\noindent
which encodes the sum of vacuum supergraphs with a marked matter propagator. (By definition, the marking of a propagator does not change the expression for a supergraph, but supergraphs in which marked propagators are different are considered as different.) The two-point Green function of the matter superfields obtained by differentiating Eq. (\ref{Gamma2_Phi}) with respect to $\phi$ and $\phi^*$ reads as

\begin{equation}\label{Matter_Green_Function}
\frac{\delta^2\Gamma}{\delta\phi^{*i}_{\,,x}\, \delta\phi_{j,y}}\bigg|_{\mbox{\scriptsize fields}=0} = \frac{1}{16} D_x^2 \bar D_y^2 \big(G_\phi\big)_i{}^j \delta^8_{xy},
\end{equation}

\noindent
where

\begin{equation}
\delta^8_{xy} \equiv \delta^4(x-y)\, \delta^4(\theta_x-\theta_y).
\end{equation}

The function $(G_\phi)_i{}^j$ present in Eq. (\ref{Matter_Green_Function}) is normalized in such a way that in the tree approximation it is equal to $\delta_i^j F(\partial^2/\Lambda^2)$. (In the limit $\Lambda\to\infty$ this expression gives $\delta_i^j$.) Therefore, it can be presented as

\begin{equation}
\big(G_\phi\big)_i{}^j = \delta_i^j F + \big(\Delta G_\phi\big)_i{}^j,
\end{equation}

\noindent
where the sum of quantum corrections is denoted by $(\Delta G_\phi)_i{}^j$. The function inverse to (\ref{Matter_Green_Function}) entering Eq. (\ref{Matter_Part}) is given by the expression

\begin{eqnarray}\label{Inverse_Matter_Green_Function}
&& \Big(\frac{\delta^2\Gamma}{\delta\phi^{*i}_{\,,x}\, \delta\phi_{j,y}}\Big)^{-1} = -\frac{D_x^2 \bar D_y^2}{4\partial^2} \big(G_\phi^{-1}\big)_j{}^i \delta^8_{xy} = -4 \Big(\frac{D_x^2 \bar D_x^2}{16\partial^2 F} \delta_j^i + \frac{D_x^2 \bar D_x^2}{16\partial^2 F}\, \big(\Delta G_\phi\big)_j{}^i\, \frac{D_x^2 \bar D_x^2}{16\partial^2 F}\qquad\vphantom{\Bigg(}\nonumber\\
&& + \frac{D_x^2 \bar D_x^2}{16\partial^2 F}\, \big(\Delta G_\phi\big)_j{}^k\, \frac{D_x^2 \bar D_x^2}{16\partial^2 F}\, \big(\Delta G_\phi\big)_k{}^i\, \frac{D_x^2 \bar D_x^2}{16\partial^2 F} +\ldots \Big) \delta^8_{xy}. \vphantom{\Bigg(}
\end{eqnarray}

\noindent
This can easily be verified using the first of the identities

\begin{equation}\label{Chiral_Identities}
\bar D^2 D^2 \bar D^2 = -16 \bar D^2\partial^2;\qquad  D^2 \bar D^2 D^2 = -16 D^2\partial^2.
\end{equation}

To obtain an expression which encodes the sum of vacuum supergraphs in which only one of propagators with coinciding momenta is marked, one should multiply terms with $n$ insertions of $\Delta G_\phi$ by $1/n$. (As we have already mentioned, the first term, which corresponds to the one-loop approximation, should be omitted.) After this we obtain the function

\begin{eqnarray}
&& -4 \Big(\frac{D_x^2 \bar D_x^2}{16\partial^2 F} \big(\Delta G_\phi\big)_j{}^i \frac{D_x^2 \bar D_x^2}{16\partial^2 F} + \frac{1}{2}\cdot \frac{D_x^2 \bar D_x^2}{16\partial^2 F} \big(\Delta G_\phi\big)_j{}^k \frac{D_x^2 \bar D_x^2}{16\partial^2 F} \big(\Delta G_\phi\big)_k{}^i \frac{D_x^2 \bar D_x^2}{16\partial^2 F} + \ldots \Big) \delta^8_{xy}\qquad\vphantom{\Bigg(}\nonumber\\
&& =  \frac{D_x^2 \bar D_x^2}{4\partial^2 F}\Big(\frac{\Delta G_\phi}{F} - \frac{(\Delta G_\phi)^2}{2 F^2} + \frac{(\Delta G_\phi)^3}{3 F^3} + \ldots\Big)_j{}^i \delta^8_{xy} = \frac{D_x^2 \bar D_x^2}{4\partial^2 F} \Big(\ln \frac{G_\phi}{F}\Big)_j{}^i \delta^8_{xy}.\vphantom{\Bigg(}
\end{eqnarray}

\noindent
Therefore, the first step for calculating the matter contribution to the function (\ref{Delta_Beta}) is to make the replacement

\begin{equation}
\Big(\frac{\delta^2\Gamma}{\delta\phi^{*i}_{\,,x}\, \delta\phi_{i,y}}\Big)^{-1}\bigg|_{\mbox{\scriptsize fields}=0}\ \to \  \frac{D_x^2 \bar D_x^2}{4\partial^2 F} \Big(\ln \frac{G_\phi}{F}\Big)_i{}^i \delta^8_{xy}
\end{equation}

\noindent
in Eq. (\ref{Matter_Part}). We see that no bad terms proportional to $1/q^{2k}$ with $k\ge 2$ appear. Next, following the algorithm described in section \ref{Section_Integrals}, we should insert $\theta^4 (v^B)^2$ to an arbitrary point of the supergraph. Evidently, in this case it is expedient to insert this expression to the point $x$. Moreover, we need to make a replacement

\begin{equation}
\Big(\ln \frac{G_\phi}{F}\Big)_i{}^i =  \Big(\ln \frac{G_\phi}{F}\Big)_j{}^i \delta_i^j \ \to \   \Big(\ln \frac{G_\phi}{F}\Big)_j{}^i C(R)_i{}^j \frac{\partial^2}{\partial q_\mu \partial q^\mu},
\end{equation}

\noindent
where $q_\mu$ is the Minkowski momentum of the matter line which is cut. Note that (as we discussed in section \ref{Subsection_Graphs}) the operator $\partial^2/\partial q_\mu \partial q^\mu$ should act only on $1/q^2$. (After the Wick rotation it gives $-\partial^2/\partial Q_\mu^2$, where $Q_\mu$ is the corresponding Euclidean momentum.) The matter line is cut in the point $x$, so that $q_\mu$ is the momentum of the propagator coming out of this point. Finally the result should be multiplied by $-2\pi/(r{\cal V}_4)$ and differentiated with respect to $\ln\Lambda$ at fixed values of renormalized couplings. This implies that the matter contribution to the function (\ref{Delta_Beta}) can be written in the form

\begin{eqnarray}\label{Delta_Matter_Original}
&& \Delta_{\mbox{\scriptsize matter}}\Big(\frac{\beta}{\alpha_0^2}\Big) = - \frac{i\pi}{8 r{\cal V}_4} C(R)_i{}^j \frac{d}{d\ln\Lambda} \int d^8x\,d^8y\, \big(\theta^4\big)_x \big(v^B\big)^2_x\nonumber\\
&& \qquad\qquad\qquad\qquad\quad \times \int \frac{d^4q}{(2\pi)^4}\, \delta^8_{xy}(q)\, \Big(\ln \frac{G_\phi}{F}\Big)_j{}^i \frac{\partial^2}{\partial q_\mu \partial q^\mu} \Big(\frac{1}{q^2}\Big) D_x^2 \bar D_x^2 \delta^8_{xy},\qquad
\end{eqnarray}

\noindent
where the function $(\ln G_\phi)_i{}^j$ depends on the momentum $q_\mu$ and

\begin{equation}
\delta^8_{xy}(q) \equiv \delta^4(\theta_x-\theta_y) e^{iq_\mu (x^\mu - y^\mu)}.
\end{equation}

\noindent
The momentum integral is calculated in the Euclidean space after the Wick rotation,

\begin{eqnarray}
&& \frac{d}{d\ln\Lambda} \int \frac{d^4q}{(2\pi)^4}\, \delta^8_{xy}(q)\, \Big(\ln \frac{G_\phi}{F}\Big)_j{}^i \frac{\partial^2}{\partial q_\mu \partial q^\mu} \Big(\frac{1}{q^2}\Big) \nonumber\\
&&\ \to\ - 4i\pi^2 \delta^4(\theta_x-\theta_y) \frac{d}{d\ln\Lambda} \int \frac{d^4Q}{(2\pi)^4}\, \Big(\ln \frac{G_\phi}{F}\Big)_j{}^i\, \delta^4(Q)\nonumber\\
&& = - \frac{i}{4\pi^2} \delta^4(\theta_x-\theta_y) \frac{d}{d\ln\Lambda}\big(\ln G_\phi\big)_j{}^i\Big|_{Q=0} = - \frac{i}{4\pi^2} \delta^4(\theta_x-\theta_y) \big(\gamma_\phi\big)_j{}^i(\alpha_0,\lambda_0,Y_0),\qquad
\end{eqnarray}

\noindent
where we took into account that $F(0)=1$ and wrote the result in terms of the anomalous dimension $(\gamma_\phi)_j{}^i$ using Eq. (\ref{Gamma_Phi_Bare_Definition}). Therefore, the expression (\ref{Delta_Matter_Original}) takes the form

\begin{equation}
\Delta_{\mbox{\scriptsize matter}}\Big(\frac{\beta}{\alpha_0^2}\Big) = - \frac{1}{32\pi r{\cal V}_4} C(R)_i{}^j \big(\gamma_\phi\big)_j{}^i(\alpha_0,\lambda_0,Y_0) \int d^8x\,d^8y\,\big(\theta^4\big)_x \big(v^B\big)^2_x\,  \delta^4(\theta_x-\theta_y) D_x^2 \bar D_x^2 \delta^8_{xy}.
\end{equation}

\noindent
Next, it is necessary to use the identity

\begin{equation}\label{Theta_Identity}
\delta^4(\theta_x-\theta_y) D_x^2 \bar D_x^2 \delta^8_{xy} = 4\, \delta^8_{xy}
\end{equation}

\noindent
and calculate the integral over $d^8y$ with the help of the $\delta$-function,

\begin{equation}
\Delta_{\mbox{\scriptsize matter}}\Big(\frac{\beta}{\alpha_0^2}\Big) = - \frac{1}{8\pi r{\cal V}_4} C(R)_i{}^j \big(\gamma_\phi\big)_j{}^i(\alpha_0,\lambda_0,Y_0) \int d^8x\,\big(\theta^4\big)_x \big(v^B\big)^2_x.
\end{equation}

\noindent
Taking into account that

\begin{equation}\label{Coordinate integral}
\int d^8x\, \big(\theta^4\big)_x \big(v^B\big)^2_x = 4{\cal V}_4,
\end{equation}

\noindent
we obtain the final expression for the sum of matter singularities

\begin{equation}\label{Delta_Matter_Final}
\Delta_{\mbox{\scriptsize matter}}\Big(\frac{\beta}{\alpha_0^2}\Big) = - \frac{1}{2\pi r} C(R)_i{}^j \big(\gamma_\phi\big)_j{}^i(\alpha_0,\lambda_0,Y_0).
\end{equation}

\noindent
It exactly coincides with the term containing the anomalous dimension of the matter superfields in the NSVZ equation written in the form (\ref{NSVZ_Equivalent_Form_Bare}) and agrees with another all-loop calculation made in \cite{Stepanyantz:2019lfm} by a different method. This can be considered as a correctness test of the method proposed in this paper.

\subsection{The all-loop sum of ghost singularities}
\hspace*{\parindent}\label{Subsection_Ghost_Singularities}

A contribution of ghost singularities to the function (\ref{Delta_Beta}) is calculated similarly to the matter contribution, but it is necessary to take into account that the ghost superfields are anticommuting. As a starting point we consider a part of the expression (\ref{Gamma_Derivative_In_Terms_Of_Inverse_Functions}) containing the ghost Green functions,

\begin{equation}\label{Ghost_Part}
-\frac{i}{4}\int d^8x\,\bigg\{
\Big(\frac{\delta^2\Gamma}{\delta c^A_{x}\,\delta \bar c^{+A}_{y}}\Big)^{-1}
+ \Big(\frac{\delta^2\Gamma}{\delta \bar c^A_{x}\,\delta c^{+A}_{y}}\Big)^{-1}
\bigg\}\bigg|_{y=x;\ \mbox{\scriptsize fields}=0}.\qquad
\end{equation}

\noindent
The two-point Green functions of the Faddeev--Popov ghosts are obtained by differentiating Eq. (\ref{Gamma2_C}),

\begin{equation}\label{Ghost_Green_Functions}
\frac{\delta^2\Gamma}{\delta \bar c^{A}_{x}\, \delta c^{+B}_{y}}\bigg|_{\mbox{\scriptsize fields}=0} = \frac{\delta^2\Gamma}{\delta c^{A}_{x}\, \delta \bar c^{+B}_{y}}\bigg|_{\mbox{\scriptsize fields}=0} = \delta_{AB} G_c \frac{\bar D_x^2 D^2_y}{16} \delta^8_{xy}.
\end{equation}

\noindent
In the tree approximation the function $G_c$ is equal to 1, so that it is convenient to present it in the form $G_c = 1 + \Delta G_c$, where $\Delta G_c$ is a sum of quantum corrections coming from 1PI supergraphs with two external ghost legs.

By definition, the functions inverse to (\ref{Ghost_Green_Functions}) satisfy the equations

\begin{eqnarray}
&& \int d^6z\, \Big(\frac{\delta^2\Gamma}{\delta c^{+A}_{x}\, \delta \bar c^{D}_{z}}\Big)^{-1} \frac{\delta^2\Gamma}{\delta \bar c^{D}_{z}\, \delta c^{+B}_{y}}\bigg|_{\mbox{\scriptsize fields}=0} = -\frac{1}{2}\delta_{AB} D^2_x \delta^8_{xy};\nonumber\\
&& \int d^6z\, \Big(\frac{\delta^2\Gamma}{\delta \bar c^{+A}_{x}\, \delta c^{D}_{z}}\Big)^{-1} \frac{\delta^2\Gamma}{\delta c^{D}_{z}\, \delta \bar c^{+B}_{y}}\bigg|_{\mbox{\scriptsize fields}=0} = -\frac{1}{2}\delta_{AB} D^2_x \delta^8_{xy}
\end{eqnarray}

\noindent
and are explicitly written as

\begin{equation}
\Big(\frac{\delta^2\Gamma}{\delta \bar c^{A}_{x}\, \delta c^{+B}_{y}}\Big)^{-1}\bigg|_{\mbox{\scriptsize fields}=0} = \Big(\frac{\delta^2\Gamma}{\delta c^{A}_{x}\, \delta \bar c^{+B}_{y}}\Big)^{-1}\bigg|_{\mbox{\scriptsize fields}=0} = \delta_{AB}\,\frac{\bar D^2_x D^2_y}{4\partial^2 G_c} \delta^8_{xy}.
\end{equation}

\noindent
Repeating the transformations similar to the ones made in the previous section for the matter singularities, we conclude that for constructing a ghost contribution to the function (\ref{Delta_Beta}), first, it is necessary to perform a substitution

\begin{eqnarray}
&& \Big(\frac{\delta^2\Gamma}{\delta c^A_{x}\, \delta \bar c^{+A}_{y}}\Big)^{-1}\bigg|_{\mbox{\scriptsize fields}=0}\ \to \  - \delta_{AA} \frac{\bar D_x^2 D_y^2}{4\partial^2} \ln G_c\, \delta^8_{xy};\qquad\nonumber\\
&& \Big(\frac{\delta^2\Gamma}{\delta \bar c^A_{x}\,\delta c^{+A}_{y}}\Big)^{-1}\bigg|_{\mbox{\scriptsize fields}=0}\ \to \  - \delta_{AA} \frac{\bar D_x^2 D_y^2}{4\partial^2} \ln G_c\, \delta^8_{xy}
\end{eqnarray}

\noindent
in the expression (\ref{Ghost_Part}). Next, according to the algorithm described in section \ref{Section_Integrals}, $\theta^4 (v^B)^2$ should be inserted to the point $x$. After this, it is necessary to make the replacement\footnote{The derivatives should act only to the inverse squared  momentum.}

\begin{equation}
\ln G_c\cdot \delta_{AA}\ \to\ \ln G_c\cdot r C_2\, \frac{\partial^2}{\partial q_\mu \partial q^\mu}
\end{equation}

\noindent
and apply the operator (\ref{Operator}) to the resulting expression. Note that the differentiation with respect to $\ln\Lambda$ should be made at fixed values of the renormalized couplings. As a result of this algorithm, we obtain the ghost contribution to the function (\ref{Delta_Beta})

\begin{eqnarray}\label{Delta_Ghost_Original}
&& \Delta_{\mbox{\scriptsize ghost}}\Big(\frac{\beta}{\alpha_0^2}\Big) = \frac{i\pi C_2}{4{\cal V}_4} \frac{d}{d\ln\Lambda} \int d^8x\,d^8y\, \big(\theta^4\big)_x \big(v^B\big)^2_x\nonumber\\
&& \qquad\qquad\qquad\qquad\qquad \times \int \frac{d^4q}{(2\pi)^4}\, \delta^8_{xy}(q) \ln G_c\, \frac{\partial^2}{\partial q_\mu \partial q^\mu} \Big(\frac{1}{q^2}\Big) \bar D_x^2 D_x^2 \delta^8_{xy}.\qquad
\end{eqnarray}

\noindent
(Note that again no bad terms appear in this expression.) Similarly to the case of matter singularities, the momentum integral is calculated after the Wick rotation in the Euclidean space,

\begin{equation}
\frac{d}{d\ln\Lambda} \int \frac{d^4q}{(2\pi)^4}\, \delta^8_{xy}(q) \ln G_c\, \frac{\partial^2}{\partial q_\mu \partial q^\mu} \Big(\frac{1}{q^2}\Big)\ \to\  - \frac{i}{4\pi^2} \delta^4(\theta_x-\theta_y) \gamma_c(\alpha_0,\lambda_0,Y_0).
\end{equation}

\noindent
With the help of this equation the considered contribution to the function (\ref{Delta_Beta}) can be brought to the form

\begin{equation}
\Delta_{\mbox{\scriptsize ghost}}\Big(\frac{\beta}{\alpha_0^2}\Big) = \frac{C_2}{16\pi {\cal V}_4} \gamma_c(\alpha_0,\lambda_0,Y_0) \int d^8x\,d^8y\,\big(\theta^4\big)_x \big(v^B\big)^2_x\,  \delta^4(\theta_x-\theta_y) \bar D_x^2 D_x^2 \delta^8_{xy}.
\end{equation}

\noindent
Then we use the identity (\ref{Theta_Identity}) and calculate the integral over $d^8y$ with the help of resulting $\delta^8_{xy}$,

\begin{equation}
\Delta_{\mbox{\scriptsize ghost}}\Big(\frac{\beta}{\alpha_0^2}\Big) =  \frac{C_2}{4\pi {\cal V}_4} \gamma_c(\alpha_0,\lambda_0,Y_0) \int d^8x\,\big(\theta^4\big)_x \big(v^B\big)^2_x.
\end{equation}

\noindent
According to Eq. (\ref{Coordinate integral}), the remaining integral present in this equation is equal to $4{\cal V}_4$, so that the sum of singular contributions produced by cuts of the Faddeev--Popov ghost propagators takes the form

\begin{equation}\label{Delta_Ghost_Final}
\Delta_{\mbox{\scriptsize ghost}}\Big(\frac{\beta}{\alpha_0^2}\Big) = \frac{C_2}{\pi} \gamma_c(\alpha_0,\lambda_0,Y_0).
\end{equation}

\noindent
This result agrees with the one found in Ref. \cite{Stepanyantz:2019lfm} by a different method and coincides with the term containing the anomalous dimension $\gamma_c$ in Eq. (\ref{NSVZ_Equivalent_Form_Bare}). Note that the sign of the expression (\ref{Delta_Ghost_Final}) is different from the sign of the matter contribution (\ref{Delta_Matter_Final}) due to the anticommutation of the ghost superfields. The factor 2 in the ghost contribution appears because there are two sets of the chiral ghost superfields (the ghost $c$ and the antighost $\bar c$) in the adjoint representation of the gauge group.

\subsection{Effective propagator of the quantum gauge superfield}
\hspace*{\parindent}\label{Subsection_Effective_Propagator}

Taking into account that (for $g=1$) due to the Slavnov--Taylor identity quantum corrections to the two-point Green function of the {\it quantum} gauge superfield are transversal, it is possible to write the corresponding part of the effective action in the form (\ref{Gamma2_Quantum_V}). Substituting the explicit expression for the gauge fixing action, we equivalently present this equation as

\begin{equation}
\Gamma^{(2)}_V = -\frac{1}{4} \int d^8x\, V^A G_V \partial^2\Pi_{1/2} V^A - \frac{1}{32 \xi_0} \int d^8x\, D^2 V^A K \bar D^2 V^A,
\end{equation}

\noindent
where the regulator function $K$ depends on $\partial^2/\Lambda^2$, and the argument $q^2$ of function $G_V$ should be replaced by $-\partial^2$. Differentiating this expression with respect to the quantum gauge superfield, we obtain the corresponding two-point Green function

\begin{equation}
\frac{\delta^2\Gamma}{\delta V_x^A \delta V_y^B}\bigg|_{\mbox{\scriptsize fields}=0} = \delta_{AB} \Big[ - \frac{1}{2} G_V \partial^2\Pi_{1/2} - \frac{1}{32 \xi_0} K \Big(D^2\bar D^2 + \bar D^2 D^2\Big)\Big] \delta^8_{xy}.
\end{equation}

\noindent
By definition, the inverse function satisfies the equation

\begin{equation}
\int d^8z\,\Big(\frac{\delta^2\Gamma}{\delta V_x^A \delta V_z^C}\Big)^{-1} \frac{\delta^2\Gamma}{\delta V_z^C \delta V_y^B}\bigg|_{\mbox{\scriptsize fields}=0} = \delta_{AB} \delta^8_{xy}.
\end{equation}

\noindent
Using the identity (see, e.g., \cite{West:1990tg})

\begin{equation}
0 = \partial^2 + \partial^2\Pi_{1/2} + \frac{1}{16}\Big(D^2 \bar D^2 + \bar D^2 D^2\Big),
\end{equation}

\noindent
which can be proved by (anti)commuting the covariant derivatives, it is possible to demonstrate that the explicit expression for the inverse two-point Green function of the quantum gauge superfield is written as

\begin{equation}\label{Gauge_Propagator}
\Big(\frac{\delta^2\Gamma}{\delta V_x^A \delta V_y^B}\Big)^{-1}\bigg|_{\mbox{\scriptsize fields}=0} = - \delta_{AB}\, \Big[\,\frac{2}{G_V \partial^4} \partial^2\Pi_{1/2} + \frac{\xi_0}{8 \partial^4 K} \Big(D^2\bar D^2 + \bar D^2 D^2\Big)\Big] \delta^8_{xy}.
\end{equation}

\noindent
Thus, we see that the effective propagator is proportional to $Q^{-4}$, where $Q_\mu$ is the corresponding Euclidean momentum.

\subsection{The all-loop sum of singularities produced by the quantum gauge superfield}
\hspace*{\parindent}\label{Subsection_Gauge_Singularities}

A part of the expression (\ref{Gamma_Derivative_In_Terms_Of_Inverse_Functions}) contributed by supergraphs in which one gauge propagator is marked is written as

\begin{equation}\label{Gauge_Part}
\frac{i}{4}\int d^8x\, \Big[ \partial^2\Pi_{1/2} R(\partial^2/\Lambda^2) + \frac{1}{16\xi_0}\Big(\bar D^2 D^2 + D^2 \bar D^2\Big) K(\partial^2/\Lambda^2) \Big]_x
\Big(\frac{\delta^2\Gamma}{\delta V^A_{x}\delta V^A_y}\Big)^{-1} \bigg|_{y=x;\ \mbox{\scriptsize fields}=0},
\end{equation}

\noindent
where the exact propagator of the quantum gauge superfield is given by Eq. (\ref{Gauge_Propagator}). All quantum corrections inside this exact propagator are encoded in the function $G_V$, which is equal to $R(\partial^2/\Lambda^2)$ in the tree approximation. That is why it is convenient to present it in the form

\begin{equation}
G_V = R + \Delta G_V,
\end{equation}

\noindent
where $\Delta G_V$ is a sum of relevant quantum corrections. Exactly as for the matter and ghost contributions, we rewrite the inverse two-point Green function in Eq. (\ref{Gauge_Part}) in the form of a series using the identities

\begin{equation}\label{Transversal_Identity}
\big(\Pi_{1/2}\big)^2 = - \Pi_{1/2};\qquad \Pi_{1/2} D^2 = 0;\qquad \Pi_{1/2} \bar D^2 = 0,
\end{equation}

\noindent
and (\ref{Chiral_Identities}). To simplify the resulting expression, we introduce the notation

\begin{equation}
P \equiv \frac{1}{\partial^4 R} \partial^2\Pi_{1/2} + \frac{\xi_0}{16 \partial^4 K} \Big(D^2\bar D^2 + \bar D^2 D^2\Big).
\end{equation}

\noindent
This expression is proportional to the usual (tree) propagator of the quantum gauge superfield. Then the inverse Green function present in Eq. (\ref{Gauge_Part}) can equivalently be rewritten as

\begin{eqnarray}\label{Gauge_Inverse_Function_Expansion}
&& \Big(\frac{\delta^2\Gamma}{\delta V_x^A \delta V_y^A}\Big)^{-1}\bigg|_{\mbox{\scriptsize fields}=0} = - r\, \Big[\,\frac{2}{G_V \partial^4} \partial^2\Pi_{1/2} + \frac{\xi_0}{8 \partial^4 K} \Big(D^2\bar D^2 + \bar D^2 D^2\Big)\Big] \delta^8_{xy}\qquad\nonumber\\
&& = - 2 r\,P\Big(1 - \partial^2\Pi_{1/2} \Delta G_V P + \partial^2\Pi_{1/2} \Delta G_V P\cdot \partial^2\Pi_{1/2} \Delta G_V P - \ldots \Big)\delta^8_{xy}.\vphantom{\frac{1}{2}}
\end{eqnarray}

\noindent
The factors $\partial^2\Pi_{1/2} G_V$ in this expression correspond to the 1PI subdiagrams (denoted in Fig. \ref{Figure_Graph_With_Coinciding_Momenta} by the gray circles), which are evidently transversal. As we discussed in section \ref{Subsection_Main_Idea}, for constructing a contribution to the function (\ref{Delta_Beta}), one should first divide a term with the $n$-th power of $\Delta G_V$ by $n$. (Certainly, the first term corresponding to the one-loop approximation should be omitted.) Then the function (\ref{Gauge_Inverse_Function_Expansion}) will be replaced by the expression

\begin{eqnarray}
&& - 2 r\, P\Big( - \partial^2\Pi_{1/2} \Delta G_V P + \frac{1}{2}\partial^2\Pi_{1/2} \Delta G_V P \cdot \partial^2\Pi_{1/2} \Delta G_V P - \ldots \Big)\delta^8_{xy}\nonumber\\
&& = - 2 r\, P \sum\limits_{n=1}^\infty \frac{1}{n} \big(-\partial^2\Pi_{1/2} \Delta G_V P\big)^n \delta^8_{xy} = - 2 r\, \frac{1}{\partial^4 R} \partial^2\Pi_{1/2}\sum\limits_{n=1}^\infty \frac{1}{n} \Big(
-\frac{\Delta G_V}{R}\Big)^n \delta^8_{xy}.\qquad
\end{eqnarray}

\noindent
Calculating the sum of this series, we obtain that, at the first step, it is necessary to make in Eq. (\ref{Gauge_Part}) the formal substitution

\begin{equation}
\Big(\frac{\delta^2\Gamma}{\delta V_x^A \delta V_y^A}\Big)^{-1}\bigg|_{\mbox{\scriptsize fields}=0}\ \to \  \frac{2 r}{\partial^4 R}\, \partial^2\Pi_{1/2} \ln \frac{G_V}{R} \delta^8_{xy}.
\end{equation}

\noindent
After this, with the help of Eq. (\ref{Transversal_Identity}) we obtain

\begin{eqnarray}\label{Gauge_Part_Modified}
&& (\ref{Gauge_Part})\ \to\ -\frac{i r}{2}\int d^8x\,  \partial^2\Pi_{1/2} \frac{1}{\partial^2}\, \ln \frac{G_V}{R} \delta^8_{xy}\bigg|_{y=x}
\nonumber\\
&&\qquad\qquad\qquad\qquad
= \frac{i r}{2}\int d^8x\, d^8y\, \int \frac{d^4q}{(2\pi)^4} \delta^8_{xy}(q)\, \partial^2\Pi_{1/2} \frac{1}{q^2}\, \ln \frac{G_V}{R} \delta^8_{xy}.\qquad
\end{eqnarray}

\noindent
It should be noted that all possible bad terms proportional to $1/q^{2k}$ with $k\ge 2$ vanish, and the above expression really contains only the $1/q^2$ singularity. To construct a contribution to the function (\ref{Delta_Beta}) coming from the gauge singularities, we need to modify the expression (\ref{Gauge_Part_Modified}) in a special way. Namely, it is necessary to insert $\theta^4 (v^B)^2$ to the point $x$, apply the operator $C_2 \partial^2/\partial q_\mu \partial q^\mu$ to $1/q^2$ coming from the gauge propagator which is cut in the point $x$, multiply the result by $-2\pi/(r{\cal V}_4)$, and differentiate it with respect to $\ln\Lambda$. The sum of the gauge singularities constructed according to this procedure is given by

\begin{eqnarray}\label{Delta_Gauge_Original}
&& \Delta_{\mbox{\scriptsize gauge}}\Big(\frac{\beta}{\alpha_0^2}\Big) = -\frac{i\pi C_2}{{\cal V}_4} \frac{d}{d\ln\Lambda} \int d^8x\,d^8y\, \big(\theta^4\big)_x \big(v^B\big)^2_x\nonumber\\
&& \qquad\qquad\qquad\qquad\qquad \times \int \frac{d^4q}{(2\pi)^4}\, \delta^8_{xy}(q)\, \ln \frac{G_V}{R}\, \frac{\partial^2}{\partial q_\mu \partial q^\mu} \Big(\frac{1}{q^2}\Big) \big(\partial^2\Pi_{1/2}\big)_x \delta^8_{xy}.\qquad
\end{eqnarray}

\noindent
As earlier, we calculate the momentum integral in the Euclidean space after the Wick rotation taking into account that $R(0)=1$,

\begin{eqnarray}
&& \frac{d}{d\ln\Lambda} \int \frac{d^4q}{(2\pi)^4}\,  \delta^8_{xy}(q)\, \ln \frac{G_V}{R}\, \frac{\partial^2}{\partial q_\mu \partial q^\mu} \Big(\frac{1}{q^2}\Big)\nonumber\\
&&\qquad\quad\ \to\ -\frac{i}{4\pi^2}\, \delta^4(\theta_x-\theta_y)\, \frac{d}{d\ln\Lambda} \ln G_V\Big|_{Q=0} = -\frac{i}{2\pi^2}\, \delta^4(\theta_x-\theta_y)\, \gamma_V(\alpha_0,\lambda_0,Y_0).\qquad
\end{eqnarray}

\noindent
This implies that

\begin{equation}\label{Delta_Gauge_Intermediate}
\Delta_{\mbox{\scriptsize gauge}}\Big(\frac{\beta}{\alpha_0^2}\Big) = - \frac{C_2}{2\pi {\cal V}_4} \gamma_V(\alpha_0,\lambda_0,Y_0) \int d^8x\,d^8y\, \big(\theta^4\big)_x \big(v^B\big)^2_x\, \delta^4(\theta_x-\theta_y) \big(\partial^2\Pi_{1/2}\big)_x \delta^8_{xy}.
\end{equation}

\noindent
For calculating the remaining integral we use the identity

\begin{equation}
\delta^4(\theta_x-\theta_y) \big(\partial^2\Pi_{1/2}\big)_x \delta^8_{xy} = -\frac{1}{8}\, \delta^4(\theta_x-\theta_y) \big(D^a \bar D^2 D_a\big)_x \delta^8_{xy} = - \frac{1}{2}\,\delta^8_{xy}.
\end{equation}

\noindent
Therefore,

\begin{equation}
\int d^8x\,d^8y\, \big(\theta^4\big)_x \big(v^B\big)^2_x\, \delta^4(\theta_x-\theta_y) \big(\partial^2\Pi_{1/2}\big)_x \delta^8_{xy}
= - \frac{1}{2} \int d^8x\, d^8y\, \big(\theta^4\big)_x \big(v^B\big)^2_x\, \delta^8_{xy} = - 2 {\cal V}_4.
\end{equation}

\noindent
Substituting this expression into Eq. (\ref{Delta_Gauge_Intermediate}) we obtain the final result for the sum of singular contributions to the function (\ref{Delta_Beta}) produced by the quantum gauge superfield,

\begin{equation}\label{Delta_Gauge_Final}
\Delta_{\mbox{\scriptsize gauge}}\Big(\frac{\beta}{\alpha_0^2}\Big) = \frac{C_2}{\pi} \gamma_V(\alpha_0,\lambda_0,Y_0).
\end{equation}

\noindent
We see that it coincides with the corresponding term in Eq. (\ref{NSVZ_Equivalent_Form_Bare}) in exact agreement with the guess made in Ref. \cite{Stepanyantz:2016gtk}.

\subsection{The $\beta$-function defined in terms of the bare couplings}
\hspace*{\parindent}\label{Subsection_Result}

The overall result for the function (\ref{Delta_Beta}) is obtained by summing the contributions produced by cuts of matter, ghost, and gauge propagators, which are given by Eqs. (\ref{Delta_Matter_Final}), (\ref{Delta_Ghost_Final}), and (\ref{Delta_Gauge_Final}), respectively,

\begin{eqnarray}
&& \frac{\beta(\alpha_0,\lambda_0,Y_0)}{\alpha_0^2} - \frac{\beta_{\mbox{\scriptsize 1-loop}}(\alpha_0)}{\alpha_0^2} = \Delta_{\mbox{\scriptsize matter}}\Big(\frac{\beta}{\alpha_0^2}\Big) + \Delta_{\mbox{\scriptsize ghost}}\Big(\frac{\beta}{\alpha_0^2}\Big) + \Delta_{\mbox{\scriptsize gauge}}\Big(\frac{\beta}{\alpha_0^2}\Big)\qquad\nonumber\\
&& = - \frac{1}{2\pi r} C(R)_i{}^j \big(\gamma_\phi\big)_j{}^i(\alpha_0,\lambda_0,Y_0) + \frac{1}{\pi} C_2 \gamma_c(\alpha_0,\lambda_0,Y_0) + \frac{1}{\pi} C_2 \gamma_V(\alpha_0,\lambda_0,Y_0).
\end{eqnarray}

\noindent
Taking into account that the one-loop contribution to the $\beta$-function is given by Eq. (\ref{Beta_1Loop}), we see that the resulting expression for the $\beta$-function defined in terms of the bare couplings coincides with the NSVZ equation written in the form (\ref{NSVZ_Equivalent_Form_Bare}). Certainly, it is highly important that the theory is regularized by higher covariant derivatives. As we have already mentioned, for RGFs defined in terms of the bare couplings the NSVZ equation does not hold in the case of using the regularization by dimensional reduction \cite{Aleshin:2016rrr} due to a different structure of loop integrals \cite{Aleshin:2015qqc}.

Note that, in fact, in the previous sections we have also proved the equation (\ref{NSVZ_For_Green_Functions_With_1Loop}) relating the two-point Green function of the considered theory in the limit of the vanishing external momenta.

To obtain the NSVZ relation in the usual form, one needs to involve the non-renormalization theorem for the triple gauge-ghost vertices. Namely, following Ref. \cite{Stepanyantz:2016gtk}, we differentiate Eq. (\ref{ZZZ_Constraint}) (which is a consequence of this theorem) with respect to $\ln\Lambda$ at fixed values of renormalized couplings and obtain the equation

\begin{equation}\label{Beta_Triple_Relation}
\beta(\alpha_0,\lambda_0,Y_0) = 2\alpha_0\Big(\gamma_c(\alpha_0,\lambda_0,Y_0) + \gamma_V(\alpha_0,\lambda_0,Y_0)\Big).
\end{equation}

\noindent
Excluding the sum $\gamma_c+\gamma_V$ from Eqs. (\ref{NSVZ_Equivalent_Form_Bare}) and (\ref{Beta_Triple_Relation}) we obtain the NSVZ relation in the original form of the relation between the $\beta$-function and the anomalous dimension of the matter superfields,

\begin{equation}\label{NSVZ_Exact_Beta_Function_Bare}
\frac{\beta(\alpha_0,\lambda_0,Y_0)}{\alpha_0^2} = - \frac{3 C_2 - T(R) + C(R)_i{}^j \big(\gamma_\phi\big)_j{}^i(\alpha_0,\lambda_0,Y_0)/r}{2\pi(1- C_2\alpha_0/2\pi)}.
\end{equation}

\noindent
Thus, we have proved that it is valid in all loops for RGFs defined in terms of the bare couplings in the case of using the regularization by higher covariant derivatives. Note that this regularization is not uniquely defined, because for introducing it we use three arbitrary higher-derivative regulator functions $R(x)$, $F(x)$, and $K(x)$, see Eqs. (\ref{Action_Regularized_Without_G}) and (\ref{Term_For_Fixing_Gauge}), and two numerical parameters $a=M/\Lambda$ and $a_\varphi = M_\varphi/\Lambda$, which are the ratios of the Pauli--Villars masses to the dimensionful parameter in the higher derivative terms. The NSVZ equations (\ref{NSVZ_Equivalent_Form_Bare}) and (\ref{NSVZ_Exact_Beta_Function_Bare}) for RGFs defined in terms of the bare couplings hold for any choice of all these functions (provided they are equal to 1 at $x=0$ and rapidly tend to infinity at $x\to\infty$) and parameters. However, all RGFs entering these equations in general can depend on the functions $R(x)$ and $F(x)$. Also they can depend on $a$ and $a_\varphi$. The explicit form of this dependence for the two-loop anomalous dimension and for the three-loop $\beta$-function\footnote{In Ref. \cite{Kazantsev:2020kfl} the three-loop $\beta$-function was constructed with the help of the NSVZ equation without a direct calculation of superdiagrams. However, this paper confirms the correctness of the result.} can be found in Ref. \cite{Kazantsev:2020kfl}. The function $K(x)$ appears in the gauge fixing term and is also present in the action for the Nielsen--Kallosh ghosts. The calculations in the lowest orders (see, e.g., \cite{Stepanyantz:2019lyo,Aleshin:2020gec}) demonstrate that the $\beta$-function and the anomalous dimension of the matter superfields do not depend on it. Possibly, this is a consequence of general theorems about the gauge dependence of the effective action (see, e.g., \cite{Batalin:2019wkb} and references therein), although the detailed superfield analyses in the supersymmetric case has not yet been done. Note that the gauge independence of the matter superfield anomalous dimension presumably follows from the fact that in supersymmetric theories it is related to the mass anomalous dimension due to the non-renormalization of the superpotential \cite{Grisaru:1979wc}. However, the anomalous dimensions $\gamma_c$ and $\gamma_V$ seem to depend on the function $K$.\footnote{Although the explicit calculation of the two-loop contribution to $\gamma_c$ was made in \cite{Kazantsev:2018kjx} for $K(x)=R(x)$, it reveals a non-trivial dependence of $\gamma_c$ on the gauge parameter $\xi_0$. Therefore, for $K(x)\ne R(x)$ one can expect the dependence of $\gamma_c$ on the function $K(x)$.} Nevertheless, both sides of the NSVZ equations have the same dependence on all regulator functions and parameters, so that the results discussed above are valid independently of their particular form.

The subtraction scheme in which the NSVZ equations hold for RGFs defined in terms of the renormalized couplings will be constructed in the next section.

\section{NSVZ scheme for RGFs defined in terms of the renormalized couplings}
\hspace*{\parindent}\label{Section_NSVZ_Scheme}

We have proved that the NSVZ relations (\ref{NSVZ_Exact_Beta_Function}) and (\ref{NSVZ_Equivalent_Form}) are satisfied by RGFs defined in terms of the bare couplings in the case of using the higher covariant derivative regularization described in section \ref{Section_N=1_Gauge_Theories}. Note that these RGFs are scheme-independent \cite{Kataev:2013eta}, so that the NSVZ relations for them are valid independently of a renormalization prescription. However, the standard RGFs are defined in terms of the renormalized couplings and depend on both a regularization and a subtraction scheme. It is known that the NSVZ relation for RGFs defined in terms of the renormalized couplings is satisfied only in certain subtraction schemes, which are called the NSVZ schemes. In particular, the $\overline{\mbox{DR}}$-scheme is not the NSVZ scheme \cite{Jack:1996vg,Jack:1996cn,Jack:1998uj}. However, using the results described above it is possible to demonstrate that an NSVZ scheme can be obtained in all orders with the help of the HD+MSL prescription \cite{Stepanyantz:2016gtk}. For completeness, here we briefly explain, how this statement can be obtained.

In terms of the renormalized couplings RGFs are defined by the equations

\begin{eqnarray}\label{RGFs_Renormalized}
&& \widetilde\beta(\alpha,\lambda,Y) \equiv \left.\frac{d \alpha}{d\ln\mu}\right|_{\alpha_0,\lambda_0,Y_0 = \mbox{\scriptsize const}};\qquad\quad\ \ \widetilde\gamma_V(\alpha,\lambda,Y) \equiv \left.  \frac{d\ln Z_V}{d\ln\mu}\right|_{\alpha_0,\lambda_0,Y_0 = \mbox{\scriptsize const}};\nonumber\\
&& \widetilde\gamma_c(\alpha,\lambda,Y) \equiv \left. \frac{d\ln Z_c}{d\ln\mu}\right|_{\alpha_0,\lambda_0,Y_0 = \mbox{\scriptsize const}};\qquad\ \  (\widetilde\gamma_\phi)_i{}^j(\alpha,\lambda,Y) \equiv \left. \frac{d(\ln Z_\phi)_i{}^j}{d\ln\mu}\right|_{\alpha_0,\lambda_0,Y_0 = \mbox{\scriptsize const}}.\qquad
\end{eqnarray}

\noindent
Unlike Eqs. (\ref{Beta_Bare_Definition}) --- (\ref{Gamma_Phi_Bare_Definition}), the derivatives in Eq. (\ref{RGFs_Renormalized}) are taken with respect to $\ln \mu$ (instead of $\ln\Lambda$) at fixed values of the bare couplings (instead of the renormalized ones).

The key observation made in \cite{Kataev:2013eta} is that both definitions of RGFs give the same functions (of different arguments) if certain boundary conditions are imposed on the renormalization constants. In the non-Abelian case considered here these boundary conditions (in the point $x_0$ which is a fixed value of $\ln\Lambda/\mu$) take the form

\begin{eqnarray}\label{Boundary_Conditions}
&& \alpha(\alpha_0,\lambda_0,Y_0,\ln\Lambda/\mu \to x_0) = \alpha_0;\qquad\ Z_V(\alpha_0,\lambda_0,Y_0,\ln\Lambda/\mu \to x_0) = 1;\qquad\vphantom{\Big(}\nonumber\\
&& Z_c(\alpha_0,\lambda_0,Y_0,\ln\Lambda/\mu\to x_0) = 1;\qquad\ (Z_\phi)_i{}^j(\alpha_0,\lambda_0,Y_0,\ln\Lambda/\mu \to x_0) = \delta_i^j;\vphantom{\Big(}\nonumber\\
&&\qquad\qquad\qquad\qquad Y(\alpha_0,\lambda_0,Y_0,\ln\Lambda/\mu \to x_0) = Y_0.\vphantom{\Big(}
\end{eqnarray}

\noindent
We also assume that the renormalization of the Yukawa couplings is related to the renormalization of the matter superfields by Eq. (\ref{Lambda_Renormalization}). Note that, due to the boundary conditions (\ref{Boundary_Conditions}) and the nonrenormalization of the triple ghost-gauge vertices, Eq. (\ref{ZZZ_Constraint}) is satisfied automatically, because in this case the equation

\begin{equation}
\frac{d}{d\ln\Lambda} (Z_\alpha^{-1/2} Z_c Z_V) = 0
\end{equation}

\noindent
has the only solution $Z_\alpha^{-1/2} Z_c Z_V=1$. Similarly, it is easy to see that the condition $Z_\xi Z_V^2=1$ is also satisfied automatically.

The conditions (\ref{Boundary_Conditions}) define a class of the HD+MSL-like schemes. Here HD means that the higher covariant derivative method is used for regularizing the theory under consideration. This is always assumed in this paper. The HD+MSL scheme is obtained for $x_0 = 0$. In this case all renormalization constants include only powers of $\ln\Lambda/\mu$ and all finite constants which define a subtraction scheme vanish. The schemes corresponding to $x_0\ne 0$ are related to the HD+MSL scheme by the redefinition of the renormalization point $\mu = \exp(-x_0)\mu_{\mbox{\scriptsize HD+MSL}}$ exactly as in the Abelian case considered in \cite{Goriachuk:2018cac}.

Under the boundary conditions (\ref{Boundary_Conditions}) RGFs (\ref{RGFs_Renormalized}) can be related to RGFs (\ref{Beta_Bare_Definition}) --- (\ref{Gamma_Phi_Bare_Definition}) by the equations

\begin{eqnarray}\label{RGFs_Relations}
&& \widetilde\beta(\alpha,\lambda,Y) = \beta(\alpha_0\to \alpha, \lambda_0\to\lambda,Y_0\to Y);\vphantom{\Big(}\nonumber\\
&& \widetilde\gamma_V(\alpha,\lambda,Y) = \gamma_V(\alpha_0\to \alpha, \lambda_0\to\lambda,Y_0\to Y);\vphantom{\Big(}\nonumber\\
&& \widetilde\gamma_c(\alpha,\lambda,Y) = \gamma_c(\alpha_0\to \alpha, \lambda_0\to\lambda,Y_0\to Y);\vphantom{\Big(}\nonumber\\
&& (\widetilde\gamma_\phi)_i{}^j(\alpha,\lambda,Y) = (\gamma_\phi)_i{}^j(\alpha_0\to \alpha, \lambda_0\to\lambda,Y_0\to Y),\qquad\vphantom{\Big(}
\end{eqnarray}

\noindent
where the arrows point that it is necessary to make a formal change of the argument, say, to write $\alpha$ instead of $\alpha_0$, etc.

All these equations can be proved repeating the argumentation of \cite{Kataev:2013eta} (in which similar equations were derived in the Abelian case). For example, taking into account that $\alpha = \alpha_0 Z_\alpha$, we obtain

\begin{eqnarray}\label{Beta_Relation}
&&\hspace*{-4mm} \widetilde\beta(\alpha,\lambda,Y) = \frac{d\alpha}{d\ln\mu} \bigg|_{\alpha_0,\lambda_0,Y_0 = \mbox{\scriptsize const}} = \frac{d}{d\ln\mu} \Big(\alpha_0 Z_\alpha(\alpha,\lambda,Y,\ln\Lambda/\mu)\Big)\bigg|_{\alpha_0,\lambda_0,Y_0 = \mbox{\scriptsize const}}\quad\nonumber\\
&&\hspace*{-4mm} = \alpha_0 Z_\alpha \bigg(\frac{\partial \ln Z_\alpha}{\partial\ln\mu} + \frac{\partial \ln Z_\alpha}{\partial \alpha}\, \frac{d\alpha}{d\ln\mu} + \frac{\partial \ln Z_\alpha}{\partial \lambda^{ijk}}\, \frac{d\lambda^{ijk}}{d\ln\mu} + \frac{\partial \ln Z_\alpha}{\partial \lambda^*_{ijk}}\, \frac{d\lambda^*_{ijk}}{d\ln\mu} + \frac{\partial \ln Z_\alpha}{\partial Y}\, \frac{dY}{d\ln\mu}\bigg).\qquad
\end{eqnarray}

\noindent
Here the derivative $d/d\ln\mu$ acts on both $\ln\Lambda/\mu$ inside $\alpha$, $\lambda$, $Y$ and the explicitly written $\ln\Lambda/\mu$. The partial derivative $\partial/\partial\ln\mu$, by contrast, acts only on the explicitly written $\ln\Lambda/\mu$ and does not act on $\ln\Lambda/\mu$ inside $\alpha$, $\lambda$, and $Y$. Therefore,

\begin{equation}\label{Z_Derivatives}
\frac{\partial \ln Z_\alpha}{\partial\ln\mu} = -\frac{\partial \ln Z_\alpha}{\partial\ln\Lambda} \equiv -\frac{d\ln Z_\alpha}{d\ln\Lambda}\bigg|_{\alpha,\lambda,Y=\mbox{\scriptsize const}} = \frac{d\ln (Z_\alpha)^{-1}}{d\ln\Lambda}\bigg|_{\alpha,\lambda,Y=\mbox{\scriptsize const}}.
\end{equation}

\noindent
Next, we consider Eq. (\ref{Beta_Relation}) in the point $\ln\Lambda/\mu = x_0$. Due to the boundary conditions (\ref{Boundary_Conditions}) $\ln Z_\alpha(\alpha,\lambda,Y,\ln\Lambda/\mu\to x_0) = 0$ for all values of $\alpha$, $\lambda$, and $Y$. This implies that

\begin{eqnarray}\label{Z_Derivatives_In_X0}
&& \frac{\partial \ln Z_\alpha}{\partial \alpha}\bigg|_{\ln\Lambda/\mu=x_0} = 0;\qquad\qquad \frac{\partial \ln Z_\alpha}{\partial \lambda^{ijk}}\bigg|_{\ln\Lambda/\mu=x_0} = 0;\qquad\nonumber\\
&& \frac{\partial \ln Z_\alpha}{\partial \lambda^*_{ijk}}\bigg|_{\ln\Lambda/\mu=x_0} = 0;\qquad\qquad \frac{\partial \ln Z_\alpha}{\partial Y}\bigg|_{\ln\Lambda/\mu=x_0} = 0.
\end{eqnarray}

\noindent
Taking into account that the left hand side of Eq. (\ref{Beta_Relation}) does not depend on the value of $\ln\Lambda/\mu$ at fixed values of the renormalized couplings due to the renormalization group equation, from Eqs. (\ref{Beta_Relation}), (\ref{Z_Derivatives}), and (\ref{Z_Derivatives_In_X0}) we obtain

\begin{equation}
\widetilde\beta(\alpha,\lambda,Y) = \alpha \left.\frac{d\ln (Z_\alpha)^{-1}}{d\ln\Lambda}\right|_{\alpha,\lambda,Y=\mbox{\scriptsize const}} = \left.\frac{d\alpha_0}{d\ln\Lambda}\right|_{\alpha,\lambda,Y=\mbox{\scriptsize const}} = \beta(\alpha_0,\lambda_0,Y_0)\Big|_{\ln\Lambda/\mu = x_0}.
\end{equation}

\noindent
However, according to Eq. (\ref{Boundary_Conditions}), in the point $\ln\Lambda/\mu= x_0$ the values of the bare and renormalized couplings coincide,

\begin{equation}\label{Couplings_Relation}
\alpha_0\Big|_{\ln\Lambda/\mu = x_0} = \alpha;\qquad \lambda_0^{ijk}\Big|_{\ln\Lambda/\mu = x_0} = \lambda^{ijk};\qquad Y_0\Big|_{\ln\Lambda/\mu = x_0} = Y.
\end{equation}

\noindent
(For the Yukawa couplings it is also necessary to use Eq. (\ref{Lambda_Renormalization}), which relates the renormalization of the Yukawa couplings to the renormalization of the chiral matter superfields.)

Taking into account that the left hand side of Eq. (\ref{Beta_Relation}) is considered as a  function of the renormalized couplings, the right hand side should also be expressed in terms of them using Eq. (\ref{Couplings_Relation}). This implies that we should formally replace the bare couplings by the renormalized ones, so that

\begin{equation}
\widetilde\beta(\alpha,\lambda,Y) = \beta(\alpha_0\to\alpha,\lambda_0\to\lambda,Y_0\to Y).
\end{equation}

The other equations in (\ref{RGFs_Relations}) can be proved in a similar way. For example, with the help of the chain rule for the derivative with respect to $\ln\mu$ we can present the anomalous dimension of the quantum gauge superfield in the form

\begin{equation}
\widetilde\gamma_V(\alpha,\lambda,Y) = \frac{\partial \ln Z_V}{\partial\ln\mu} + \frac{\partial \ln Z_V}{\partial \alpha}\, \frac{d\alpha}{d\ln\mu} + \frac{\partial \ln Z_V}{\partial \lambda^{ijk}}\, \frac{d\lambda^{ijk}}{d\ln\mu} + \frac{\partial \ln Z_V}{\partial \lambda^*_{ijk}}\, \frac{d\lambda^*_{ijk}}{d\ln\mu} + \frac{\partial \ln Z_V}{\partial Y}\, \frac{dY}{d\ln\mu}.
\end{equation}

\noindent
Again we consider this equation in the point $\ln\Lambda/\mu = x_0$ and take into account that due to Eq. (\ref{Boundary_Conditions}) $\ln Z_V(\alpha,\lambda,Y,\ln\Lambda/\mu\to x_0) = 0$. Then, repeating the above argumentation, we obtain

\begin{equation}
\widetilde\gamma_V(\alpha,\lambda,Y) = \left. - \frac{d\ln Z_V}{d\ln\Lambda}\right|_{\alpha,\lambda,Y=\mbox{\scriptsize const}} = \gamma_V(\alpha_0\to\alpha,\lambda_0\to\lambda,Y_0\to Y).
\end{equation}

Earlier we have demonstrated that RGFs defined in terms of the bare couplings satisfy the NSVZ relations (\ref{NSVZ_Equivalent_Form_Bare}) and (\ref{NSVZ_Exact_Beta_Function_Bare}) for theories regularized by higher covariant derivatives independently of a renormalization prescription. (Let us recall that these RGFs are scheme-independent for a fixed regularization.) After the formal change of arguments $\alpha_0\to \alpha$, $\lambda_0 \to \lambda$, $Y_0\to Y$ these equalities certainly remain valid. Therefore, from Eq. (\ref{RGFs_Relations}) we conclude that in the case of using the higher covariant derivative regularization supplemented by the renormalization prescription (\ref{Boundary_Conditions}) RGFs defined in terms of the renormalized couplings also satisfy the equations

\begin{eqnarray}
&&\hspace*{-8mm} \frac{\widetilde\beta(\alpha,\lambda,Y)}{\alpha^2} = - \frac{1}{2\pi}\Big(3 C_2 - T(R) - 2C_2 \widetilde\gamma_c(\alpha,\lambda,Y) - 2C_2 \widetilde\gamma_V(\alpha,\lambda,Y) + \frac{1}{r} C(R)_i{}^j \big(\widetilde\gamma_\phi\big)_j{}^i(\alpha,\lambda,Y)\Big);\nonumber\\
\label{NSVZ_Renormalized}
&&\hspace*{-8mm} \widetilde\beta(\alpha,\lambda,Y) = - \frac{\alpha^2\Big(3 C_2 - T(R) + C(R)_i{}^j
\big(\widetilde\gamma_\phi\big)_j{}^i(\alpha,\lambda,Y)/r\Big)}{2\pi(1- C_2\alpha/2\pi)}.
\end{eqnarray}

\noindent
This implies that in the non-Abelian case the prescription (\ref{Boundary_Conditions}) also provides the NSVZ scheme in all loops. The HD+MSL prescription is obtained by imposing the boundary conditions (\ref{Boundary_Conditions}) with $x_0=0$. For other values of $x_0$ we obtain a family of schemes which differ from HD+MSL by redefinitions of the normalization point $\mu$. Certainly, in these schemes the NSVZ relation is also valid in all loops.\footnote{For ${\cal N}=1$ SYM theory without matter superfields any NSVZ scheme belongs to this family \cite{Goriachuk:2020wyn}.} These statements agree with the explicit calculations (of some scheme dependent terms in RGFs) made in Refs. \cite{Shakhmanov:2017soc,Kazantsev:2018nbl}.

Let us also note that the HD+MSL scheme can supplement various versions of the higher covariant derivative regularization, which differ in a particular form of the regulator functions $R$, $F$, and $K$ and values of the parameters $a$ and $a_\varphi$. This implies that we obtain a set of in general different NSVZ renormalization prescription, which can certainly be related by finite renormalizations.

\section{Verifications in the lowest orders}
\hspace*{\parindent}\label{Section_Explicit_Calculation}

In this section we verify the general argumentation discussed above by explicit calculations in the lowest orders of the perturbation theory made with the help of the higher covariant derivative regularization. For this purpose we will use the results for various groups of supergraphs obtained earlier. Namely, using the method described in section \ref{Subsection_Idea} the two-loop $\beta$-function for a general ${\cal N}=1$ supersymmetric gauge theory with a simple gauge group has been calculated in Ref. \cite{Stepanyantz:2019lyo}. Also using this method the parts of the three-loop $\beta$-function containing the Yukawa couplings and ghost loops have been found in Refs. \cite{Stepanyantz:2019ihw} and \cite{Kuzmichev:2019ywn}, respectively. Note that before this the part of the three-loop $\beta$-function containing the Yukawa couplings has also been calculated with the help of the standard technique in Refs. \cite{Shakhmanov:2017soc,Kazantsev:2018nbl}. For ${\cal N}=1$ SQED with $N_f$ flavors the complete three-loop $\beta$-function in a general $\xi$-gauge has been calculated by the algorithm of section \ref{Subsection_Idea} in Ref. \cite{Aleshin:2020gec}. Here all these calculations are used for checking the exact results described in the previous sections. Note that we will not verify that the equations (\ref{NSVZ_Equivalent_Form_Bare}), (\ref{NSVZ_Exact_Beta_Function_Bare}), and (\ref{NSVZ_Renormalized}) really hold, because this has already been done in Refs. \cite{Shakhmanov:2017soc,Kazantsev:2018nbl,Kuzmichev:2019ywn,Stepanyantz:2019lyo,Aleshin:2020gec}. The main purpose of this section is to test the argumentation which was used for the all-loop derivation of the NSVZ equations at intermediate steps.

\subsection{The two-loop approximation}
\hspace*{\parindent}\label{Subsection_Two_Loop}

\begin{figure}[h]
\begin{picture}(0,4)
\put(1.5,2.2){\includegraphics[scale=0.21,clip]{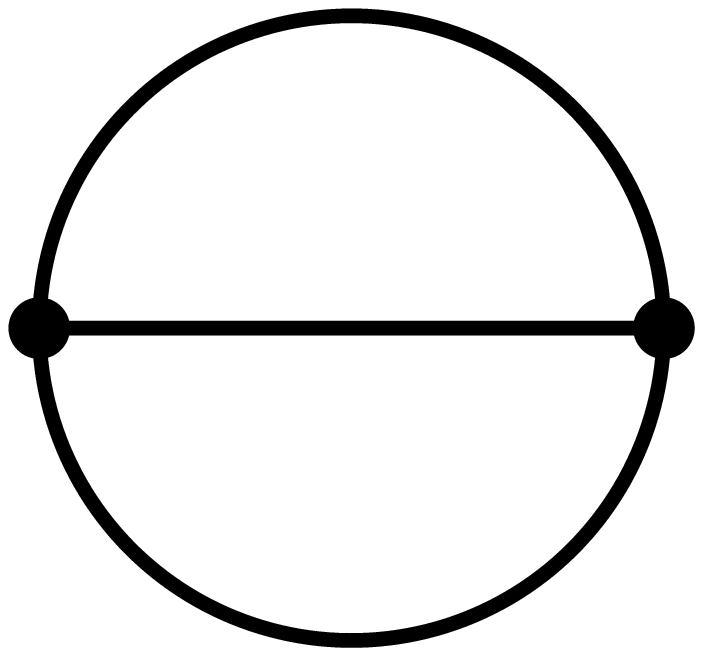}}
\put(1.3,3.6){B1}
\put(4.5,2.2){\includegraphics[scale=0.21,clip]{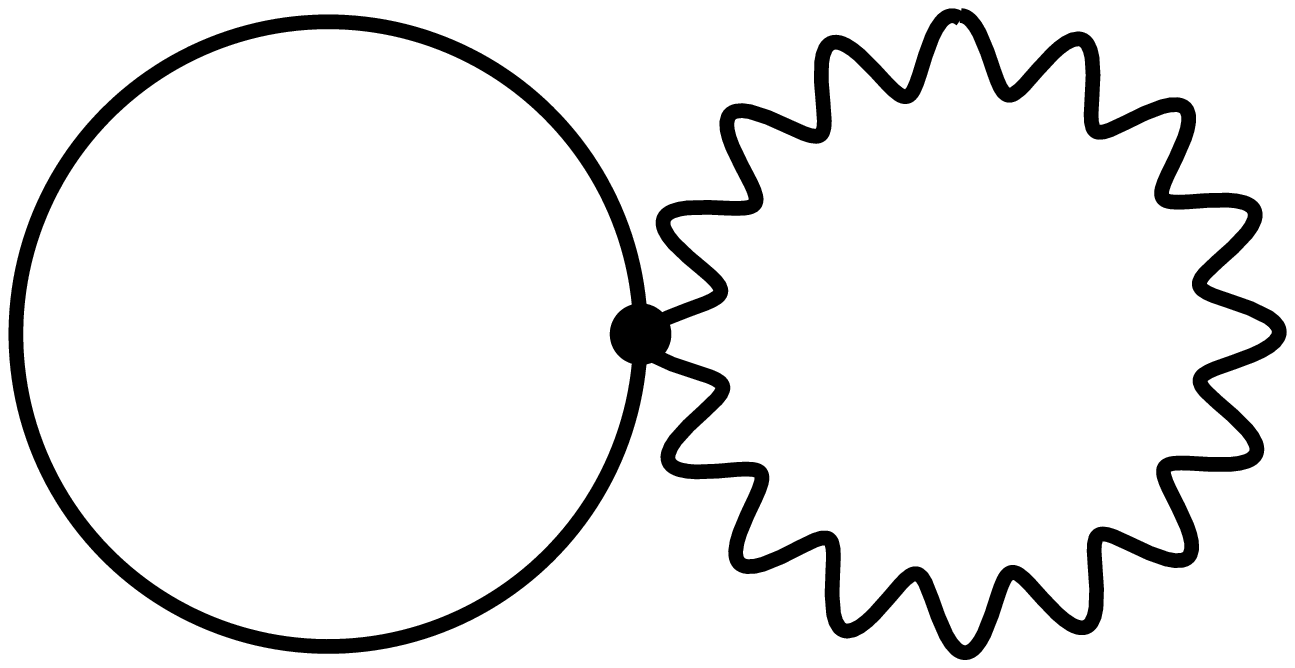}}
\put(4.1,3.6){B2}
\put(8.8,2.2){\includegraphics[scale=0.185,clip]{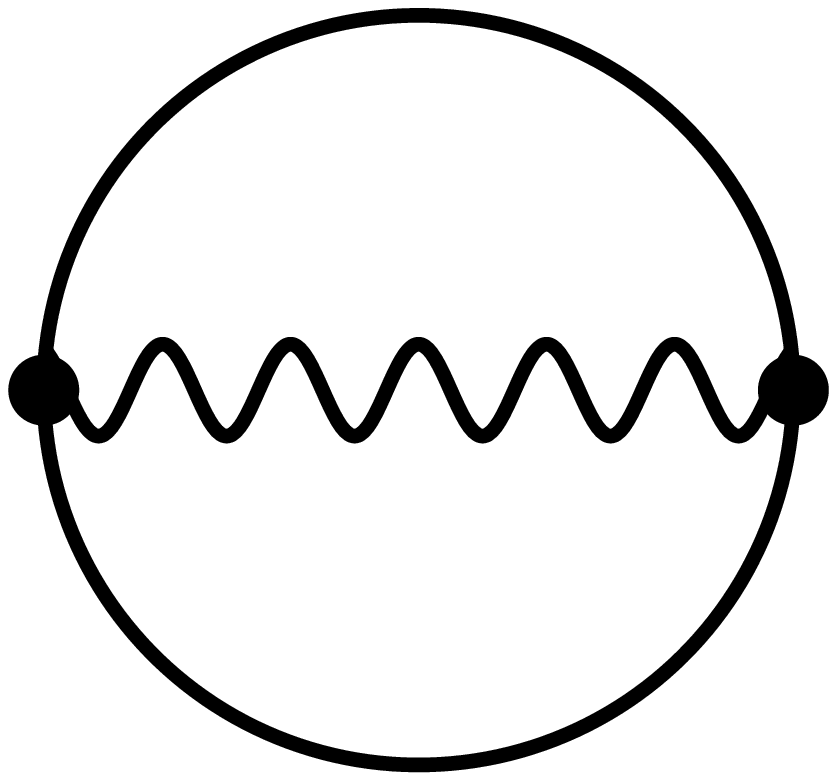}}
\put(8.4,3.6){B3}
\put(11.9,2.2){\includegraphics[scale=0.21,clip]{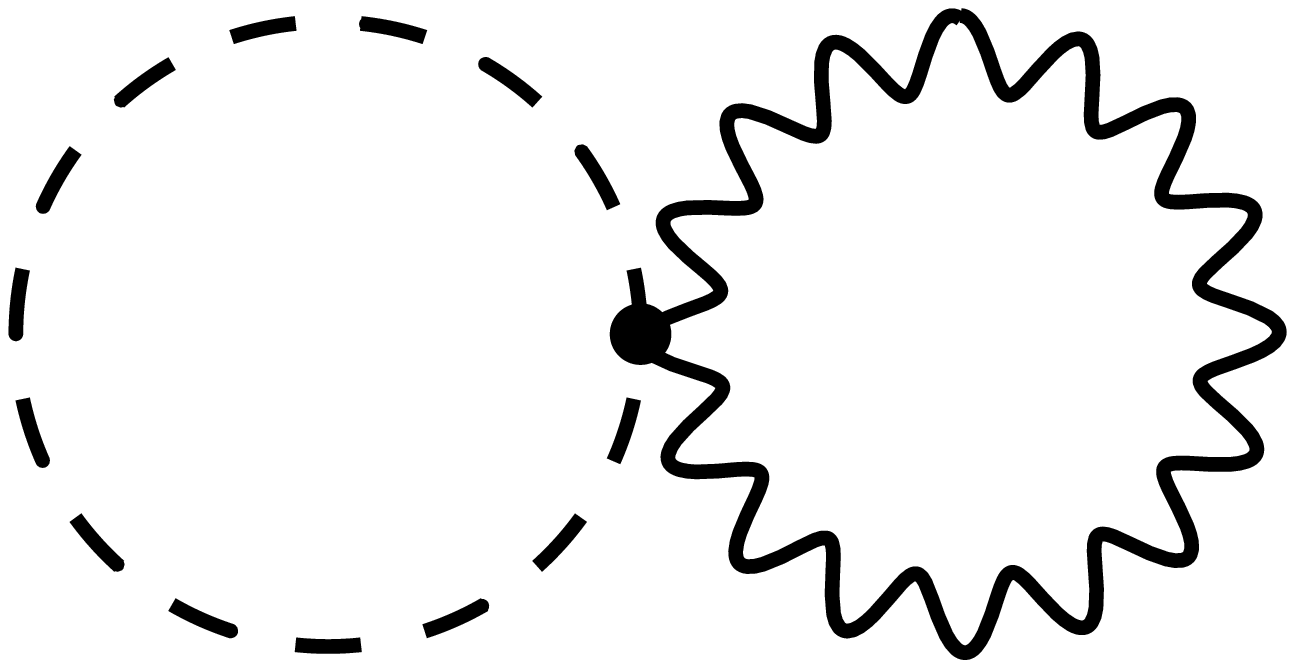}}
\put(11.5,3.6){B4}
\put(3.5,0){\includegraphics[scale=0.21,clip]{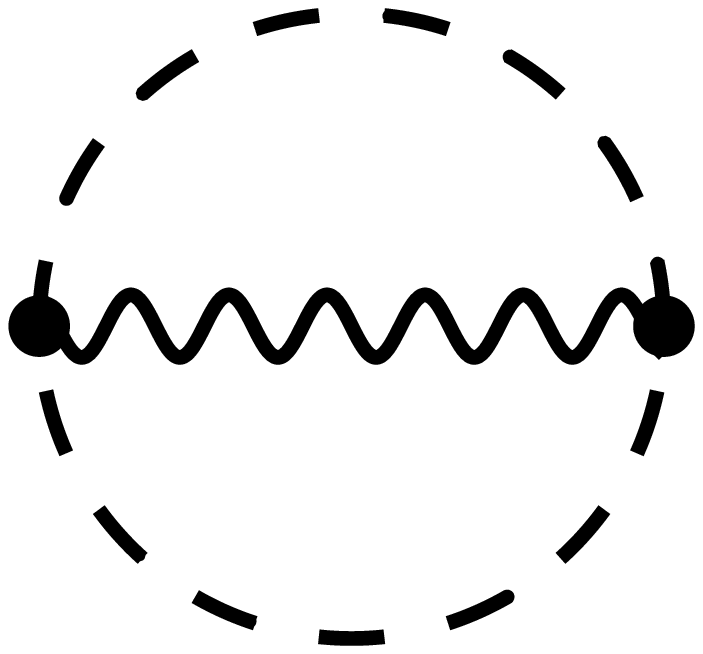}}
\put(3.1,1.4){B5}
\put(6.5,0){\includegraphics[scale=0.21,clip]{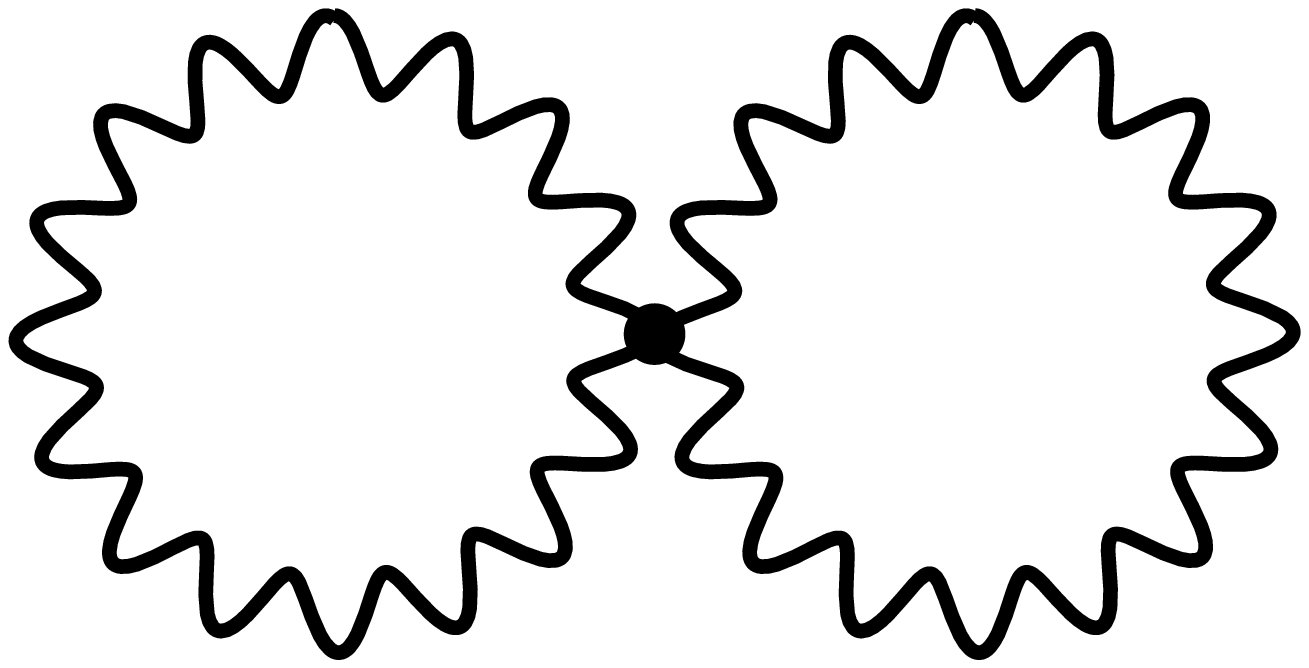}}
\put(6.1,1.4){B6}
\put(11.0,0){\includegraphics[scale=0.185,clip]{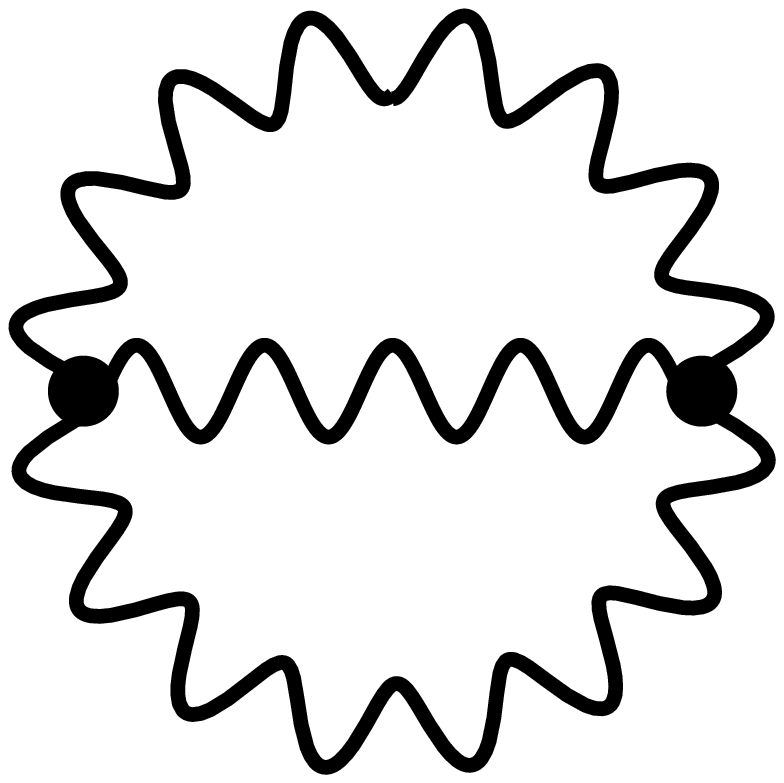}}
\put(10.6,1.4){B7}
\end{picture}
\caption{Vacuum supergraphs generating the two-loop contribution to the $\beta$-function.}\label{Figure_Two_Loop}
\end{figure}

First, we reanalyse the two-loop calculation made in Ref. \cite{Stepanyantz:2019lyo}. The corresponding contribution to the $\beta$-function is generated by the vacuum supergraphs presented in Fig. \ref{Figure_Two_Loop}. The standard technique requires calculating all superdiagrams contributing to the two-point Green function of the background gauge superfield. They are obtained from the ones presented in Fig. \ref{Figure_Two_Loop} by attaching two external $\bm{V}$-legs in all possible ways. The method of Ref. \cite{Stepanyantz:2019ihw} considerably simplifies the calculation, because it deals only with vacuum supergraphs. In this paper we have made some modifications, which will be verified here. This allows checking the general argumentation used for deriving the NSVZ equation and illustrating it by explicit calculations.

Let us start with the supergraph B1. The expression for it has been calculated in Ref. \cite{Stepanyantz:2019ihw}. If we include the factor $-2\pi/(r{\cal V}_4)\cdot d/d\ln\Lambda$, the result can be written as

\begin{equation}
\mbox{B1}\ \to\ -\frac{2\pi}{r{\cal V}_4} \frac{d}{d\ln\Lambda} \cdot \frac{2}{3} {\cal V}_4 \int \frac{d^4Q}{(2\pi)^4} \frac{d^4K}{(2\pi)^4}  \frac{\lambda_0^{ijk} \lambda^*_{0ijk}}{Q^2 F_Q K^2 F_K (Q+K)^2 F_{Q+K}},
\end{equation}

\noindent
where $F_K\equiv F(K^2/\Lambda^2)$. The integrand here contains three inverse squared momenta. Therefore, we should sum up three (equal) expressions obtained by replacing one of these inverse squared momenta $1/P^2$ (together with the corresponding $\delta$-symbol $\delta_m^n$) by $4\pi^2 C(R)_m{}^n \delta^4(P)$. This gives the contribution to the function (\ref{Delta_Beta}) of the form

\begin{eqnarray}
&& \Delta_{\mbox{\scriptsize Yukawa}}\Big(\frac{\beta}{\alpha_0^2}\Big) = -\frac{4\pi}{r} C(R)_i{}^m \frac{d}{d\ln\Lambda} \int \frac{d^4Q}{(2\pi)^4} \frac{d^4K}{(2\pi)^4}\,  4\pi^2 \delta^4(Q)\frac{\lambda_0^{ijk} \lambda^*_{0mjk}}{F_Q K^2 F_K (Q+K)^2 F_{Q+K}}\qquad\nonumber\\
&& = -\frac{1}{\pi r} C(R)_i{}^j \frac{d}{d\ln\Lambda} \int \frac{d^4K}{(2\pi)^4}  \frac{\lambda_0^{imn} \lambda^*_{0jmn}}{K^4 F_K^2}.
\end{eqnarray}

\noindent
This result exactly agrees with the one obtained in Ref. \cite{Shakhmanov:2017soc} by the direct calculation of two-point superdiagrams with two external $\bm{V}$-legs.

The supergraphs B2 and B3 containing a matter loop produce two different contributions. If a matter loop corresponds to the superfields $\phi_i$ and to the Pauli--Villars superfields $\Phi_i$, then (for an arbitrary value of $\xi_0$) the result for the (properly modified) vacuum supergraphs multiplied by the operator (\ref{Operator}) is given by

\begin{eqnarray}\label{Matter1}
&&\hspace*{-5mm} \mbox{B2} + \mbox{B3}\ \to\ \frac{4\pi}{r}\, \mbox{tr}\, C(R)\, \frac{d}{d\ln\Lambda} \int \frac{d^4Q}{(2\pi)^4} \frac{d^4K}{(2\pi)^4} \frac{e_0^2}{K^2 R_K} \bigg\{\frac{1}{2Q^2(Q+K)^2} + \frac{1}{((K+Q)^2-Q^2)}\nonumber\\
&&\hspace*{-5mm} \times \bigg[\frac{F_{K+Q}^2}{2((K+Q)^2 F_{K+Q}^2 + M^2)} - \frac{F_Q^2}{2(Q^2 F_Q^2 + M^2)} - \frac{M^2 F_{K+Q}'}{\Lambda^2 F_{K+Q} ((K+Q)^2 F_{K+Q}^2 + M^2)}\qquad\nonumber\\
&&\hspace*{-5mm} + \frac{M^2 F_Q'}{\Lambda^2 F_Q (Q^2 F_Q^2 + M^2)}\bigg] \bigg\},
\end{eqnarray}

\noindent
where $R_K\equiv R(K^2/\Lambda^2)$. Unlike the separate supergraphs B2 and B3, it does not contain bad singularities proportional to the inverse momenta to the fourth power and terms which depend on $\xi_0$. Next, we should find a sum of the expressions obtained from (\ref{Matter1}) either by replacing $1/K^2$ (which comes from the gauge propagator) by $4\pi^2 C_2 \delta^4(K)$ or by replacing $\delta_i^j/Q^2$ or $\delta_i^j/(Q+K)^2$ (coming from the propagators of the superfields $\phi_i$) by $4\pi^2 C(R)_i{}^j \delta^4(Q)$ or $4\pi^2 C(R)_i{}^j \delta^4(Q+K)$, respectively. Note that no replacement should be made for the non-singular Pauli--Villars propagators. The above procedure gives the result

\begin{eqnarray}
&& \Delta_{\mbox{\scriptsize matter}}\Big(\frac{\beta}{\alpha_0^2}\Big) = \frac{64\pi^4}{r}\,\mbox{tr}\left(C(R)^2\right) \frac{d}{d\ln\Lambda} \int \frac{d^4Q}{(2\pi)^4} \frac{d^4K}{(2\pi)^4} \frac{\alpha_0}{K^2 R_K} \bigg\{\frac{1}{2(Q+K)^2}\,\delta^4(Q)  \nonumber\\
&& + \frac{1}{2Q^2}\, \delta^4(Q+K) \bigg\}
+ \frac{64\pi^4}{r}\,C_2\, \mbox{tr}\,C(R)\, \frac{d}{d\ln\Lambda} \int \frac{d^4Q}{(2\pi)^4} \frac{d^4K}{(2\pi)^4}\, \delta^4(K) \frac{\alpha_0}{R_K}  \nonumber\\
&& \times \bigg\{\frac{1}{2Q^2(Q+K)^2} + \frac{1}{((K+Q)^2-Q^2)} \bigg[\frac{F_{K+Q}^2}{2((K+Q)^2 F_{K+Q}^2 + M^2)} - \frac{F_Q^2}{2(Q^2 F_Q^2 + M^2)} \qquad\nonumber\\
&& - \frac{M^2 F_{K+Q}'}{\Lambda^2 F_{K+Q} ((K+Q)^2 F_{K+Q}^2 + M^2)} + \frac{M^2 F_Q'}{\Lambda^2 F_Q (Q^2 F_Q^2 + M^2)}\bigg] \bigg\}.
\end{eqnarray}

\noindent
After calculating the integrals of the $\delta$-functions and some transformations we obtain the expression

\begin{eqnarray}
&& \Delta_{\mbox{\scriptsize matter}}\Big(\frac{\beta}{\alpha_0^2}\Big) = \frac{4}{r}\,\mbox{tr}\left(C(R)^2\right) \frac{d}{d\ln\Lambda} \int \frac{d^4K}{(2\pi)^4} \frac{\alpha_0}{K^4 R_K} \qquad\nonumber\\
&&\qquad\qquad\qquad + \frac{1}{2r} \,C_2\, \mbox{tr}\,C(R)\,\frac{d}{d\ln\Lambda} \int \frac{d^4Q}{(2\pi)^4} \frac{\partial^2}{\partial Q_\mu^2}\bigg[\frac{\alpha_0 }{Q^2} \ln\Big(1+\frac{M^2}{Q^2 F_Q^2}\Big)\bigg],\qquad
\end{eqnarray}

\noindent
which exactly coincides with the one found in Ref. \cite{Stepanyantz:2019lyo} with the help of a different prescription.

The solid lines in the supergraphs B2 and B3 can also stand for the propagators of the Pauli--Villars superfields $\varphi_a$. In this case the calculation of the vacuum superdiagrams with an insertion of $\theta^4 (v^B)^2$ multiplied by the operator (\ref{Operator}) for an arbitrary value of $\xi_0$ gives

\begin{eqnarray}
&&\hspace*{-5mm} \mbox{B2} + \mbox{B3}\ \to\ 4\pi C_2 \frac{d}{d\ln\Lambda} \int \frac{d^4Q}{(2\pi)^4} \frac{d^4K}{(2\pi)^4} \frac{e_0^2}{K^2 R_K} \bigg\{\frac{1}{(Q^2+M_\varphi^2)((Q+K)^2+M_\varphi^2)}\nonumber\\
&&\hspace*{-5mm}  - \frac{1}{((K+Q)^2-Q^2)} \bigg[\frac{R_{K+Q}^2}{2((K+Q)^2 R_{K+Q}^2 + M_\varphi^2)} - \frac{R_Q^2}{2(Q^2 R_Q^2 + M^2)} -\frac{1}{\Lambda^2 R_{K+Q}}\qquad\nonumber\\
&&\hspace*{-5mm} \times \frac{M_\varphi^2 R_{K+Q}'}{((K+Q)^2 R_{K+Q}^2 + M_\varphi^2)} + \frac{M_\varphi^2 R_Q'}{\Lambda^2 R_Q (Q^2 R_Q^2 + M_\varphi^2)} + \frac{R_{K+Q}'}{\Lambda^2 R_{K+Q}} - \frac{R_Q'}{\Lambda^2 R_Q} \bigg] \bigg\}.
\end{eqnarray}

\noindent
In this case all matter propagators are massive and do not produce singularities. Therefore, it is only necessary to make the replacement $1/K^2\ \to\ 4\pi^2 C_2 \delta^4(K)$, after which the contribution of the Pauli--Villars superfields $\varphi_a$ to the function (\ref{Delta_Beta}) takes the form

\begin{eqnarray}
&& \Delta_\varphi\Big(\frac{\beta}{\alpha_0^2}\Big) = C_2^2 \frac{d}{d\ln\Lambda} \int \frac{d^4Q}{(2\pi)^4}\bigg\{\frac{6\alpha_0}{Q^4} - \frac{\partial^2}{\partial Q_\mu^2}\bigg[\frac{\alpha_0}{Q^2}\ln\Big(1+ \frac{M_\varphi^2}{Q^2}\Big)\nonumber\\
&&\qquad\qquad\qquad\qquad\qquad\qquad\qquad\qquad\quad - \frac{\alpha_0}{2Q^2} \ln \Big(1+\frac{M_\varphi^2}{Q^2 R_Q^2}\Big) - \frac{\alpha_0}{Q^2} \ln R_Q \bigg]\bigg\}.\qquad
\end{eqnarray}

\noindent
This expression again coincides with the one found in Ref. \cite{Stepanyantz:2019lyo} with the help of a different technique.\footnote{The first term is not essential in the two-loop approximation, because $d\alpha_0/d\ln\Lambda = O(\alpha_0^2)$.}

The contribution of the vacuum supergraphs B4, B5, B6, and B7 is written in the form

\begin{eqnarray}
&&\hspace*{-5mm} \mbox{B4} + \mbox{B5} + \mbox{B6} + \mbox{B7}\ \to\ 4\pi C_2\, \frac{d}{d\ln\Lambda} \int \frac{d^4Q}{(2\pi)^4} \frac{d^4K}{(2\pi)^4} \frac{e_0^2}{R_K R_Q} \bigg\{-\frac{R_K}{2 Q^2 K^2 (K+Q)^2}\nonumber\\
&&\hspace*{-5mm} - \frac{1}{2Q^2 K^2} \bigg(\frac{R_Q-R_K}{Q^2-K^2}\bigg)
- \frac{1}{R_{K+Q} K^2} \Big(1-\frac{Q^2}{2(K+Q)^2}\Big) \bigg(\frac{R_Q-R_K}{Q^2-K^2}\bigg)\bigg(\frac{R_{K+Q}-R_Q}{(K+Q)^2-Q^2}\bigg)\nonumber\\
&&\hspace*{-5mm} - \frac{1}{R_{K+Q} (K+Q)^2}\bigg(\frac{R_Q-R_K}{Q^2-K^2}\bigg)^2 +\frac{2}{K^2 \big((K+Q)^2-Q^2\big)^2} \bigg[ R_{K+Q} - R_Q - R_Q' \Big(\frac{(K+Q)^2}{\Lambda^2} \nonumber\\
&&\hspace*{-5mm} - \frac{Q^2}{\Lambda^2}\Big)\bigg] - \frac{Q_\mu K^\mu}{Q^2 K^2} \bigg[\frac{R_{K+Q}}{\big((K+Q)^2-K^2\big) \big((K+Q)^2-Q^2\big)} + \frac{R_{K}}{\big(K^2-(K+Q)^2\big) \big(K^2-Q^2\big)} \nonumber\\
&&\hspace*{-5mm} + \frac{R_{Q}}{\big(Q^2-(K+Q)^2\big)\big(Q^2-K^2\big)}\bigg]
\bigg\}.
\end{eqnarray}

\noindent
We see that all bad terms containing inverse momenta to the fourth power really cancel each other, although they are present in expressions for the separate supergraphs. This agrees with the general argumentation of section \ref{Section_Singularities}. Also all terms dependent on the gauge parameter $\xi_0$ cancel each other. To find a contribution to the function (\ref{Delta_Beta}) we should replace one of the gauge or ghost inverse squared momenta by the corresponding $\delta$-function multiplied by $4\pi^2 C_2$ and sum up all expression thus obtained. The result is

\begin{eqnarray}
&&\hspace*{-5mm} \Delta_{\mbox{\scriptsize gauge+ghost}}\Big(\frac{\beta}{\alpha_0^2}\Big) = 64\pi^4 C_2^2\, \frac{d}{d\ln\Lambda} \int \frac{d^4Q}{(2\pi)^4} \frac{d^4K}{(2\pi)^4} \frac{\alpha_0}{R_K R_Q} \bigg\{-\frac{R_K}{2 K^2 (K+Q)^2}\, \delta^4(Q) \nonumber\\
&&\hspace*{-5mm}  -\frac{R_K}{2 Q^2 (K+Q)^2}\, \delta^4(K) -\frac{R_K}{2 Q^2 K^2}\, \delta^4(K+Q) - \bigg(\frac{1}{2 K^2}\,\delta^4(Q) + \frac{1}{2Q^2} \delta^4(K) \bigg)\bigg(\frac{R_Q-R_K}{Q^2-K^2}\bigg)\nonumber\\
&& \hspace*{-5mm} - \frac{1}{R_{K+Q}} \bigg[\,\delta^4(K) \Big(1-\frac{Q^2}{2(K+Q)^2}\Big) - \frac{Q^2}{2 K^2}\delta^4(K+Q) \bigg] \bigg(\frac{R_Q-R_K}{Q^2-K^2}\bigg)\bigg(\frac{R_{K+Q}-R_Q}{(K+Q)^2-Q^2}\bigg)\nonumber\\
&&\hspace*{-5mm} - \frac{1}{R_{K+Q}}\, \delta^4(K+Q) \bigg(\frac{R_Q-R_K}{Q^2-K^2}\bigg)^2 +\frac{2}{\big((K+Q)^2-Q^2\big)^2}\,\delta^4(K) \bigg[ - R_Q' \Big(\frac{(K+Q)^2}{\Lambda^2}  - \frac{Q^2}{\Lambda^2}\Big) \nonumber\\
&&\hspace*{-5mm} +R_{K+Q} - R_Q \bigg] - \bigg(\frac{Q_\mu K^\mu}{K^2}\,\delta^4(Q) + \frac{Q_\mu K^\mu}{Q^2}\,\delta^4(K) \bigg)\bigg[\frac{R_{K+Q}}{\big((K+Q)^2-K^2\big) \big((K+Q)^2-Q^2\big)} \nonumber\\
&&\hspace*{-5mm}  + \frac{R_{K}}{\big(K^2-(K+Q)^2\big) \big(K^2-Q^2\big)} + \frac{R_{Q}}{\big(Q^2-(K+Q)^2\big)\big(Q^2-K^2\big)}\bigg]
\bigg\}.
\end{eqnarray}

\noindent
After calculating the integrals of the $\delta$-functions this expression can be rewritten as

\begin{eqnarray}
&& \Delta_{\mbox{\scriptsize gauge+ghost}}\Big(\frac{\beta}{\alpha_0^2}\Big) = 4 C_2^2\, \frac{d}{d\ln\Lambda} \int \frac{d^4Q}{(2\pi)^4}\, \bigg\{-\frac{3\alpha_0}{2 Q^4} - \frac{\alpha_0}{\Lambda^4 R_Q^2} (R_Q')^2 + \frac{\alpha_0}{\Lambda^4 R_Q} R_Q''\bigg\}\qquad\nonumber\\
&& = C_2^2 \frac{d}{d\ln\Lambda}\int \frac{d^4Q}{(2\pi)^4}\bigg\{-\frac{6\alpha_0}{Q^4} + \frac{\partial^2}{\partial Q_\mu^2} \Big[\frac{\alpha_0}{Q^2}\ln R_Q\Big]\bigg\}.
\end{eqnarray}

\noindent
Again we see that the result coincides with the one obtained by the method of Ref. \cite{Stepanyantz:2019ihw} in Ref. \cite{Stepanyantz:2019lyo}. This confirms the correctness of the argumentation used in this paper for deriving the NSVZ equation.

The overall two-loop expression for the $\beta$-function reads as

\begin{eqnarray}
&& \frac{\beta(\alpha_0,\lambda_0,Y_0)}{\alpha_0^2} - \frac{\beta_{\mbox{\scriptsize 1-loop}}(\alpha_0)}{\alpha_0^2} = \int \frac{d^4Q}{(2\pi)^4} \frac{d}{d\ln\Lambda} \bigg\{ - \alpha_0 C_2^2\, \frac{\partial^2}{\partial Q_\mu^2}\bigg[\frac{1}{2Q^2}\ln\Big(1+\frac{M_\varphi^2}{Q^2 R_Q^2}\Big)  \qquad\nonumber\\
&& - \frac{1}{Q^2} \ln\Big(1+\frac{M_\varphi^2}{Q^2}\Big)\bigg] + \frac{\alpha_0}{2r}\, C_2\,\mbox{tr}\, C(R)\, \frac{\partial^2}{\partial Q_\mu^2}\bigg[\frac{1}{Q^2}\ln\Big(1+\frac{M^2}{Q^2 F_Q^2}\Big)\bigg] + \frac{4\alpha_0}{r}\, \mbox{tr}\left(C(R)^2\right) \nonumber\\
&& \times \frac{1}{Q^4 R_Q} - \frac{1}{\pi r} \lambda_0^{imn} \lambda^*_{0jmn} C(R)_i{}^j \frac{1}{Q^2 F_Q^2}\bigg\}
+ O(\alpha_0^2,\alpha_0\lambda_0^2,\lambda_0^4)
\end{eqnarray}

\noindent
and coincides with the one obtained in Ref. \cite{Stepanyantz:2019lyo}. Certainly, this expression satisfies the NSVZ equations (\ref{NSVZ_Equivalent_Form_Bare}) and (\ref{NSVZ_Exact_Beta_Function_Bare}) and agrees with the previous calculations (first made in Ref. \cite{Jones:1974pg}).

\subsection{The three-loop approximation: supergraphs with Yukawa vertices}
\hspace*{\parindent}\label{Subsection_Three_Loop_Yukawa}

\begin{figure}[h]
\begin{picture}(0,3)
\put(1.2,0.3){\includegraphics[scale=0.10,clip]{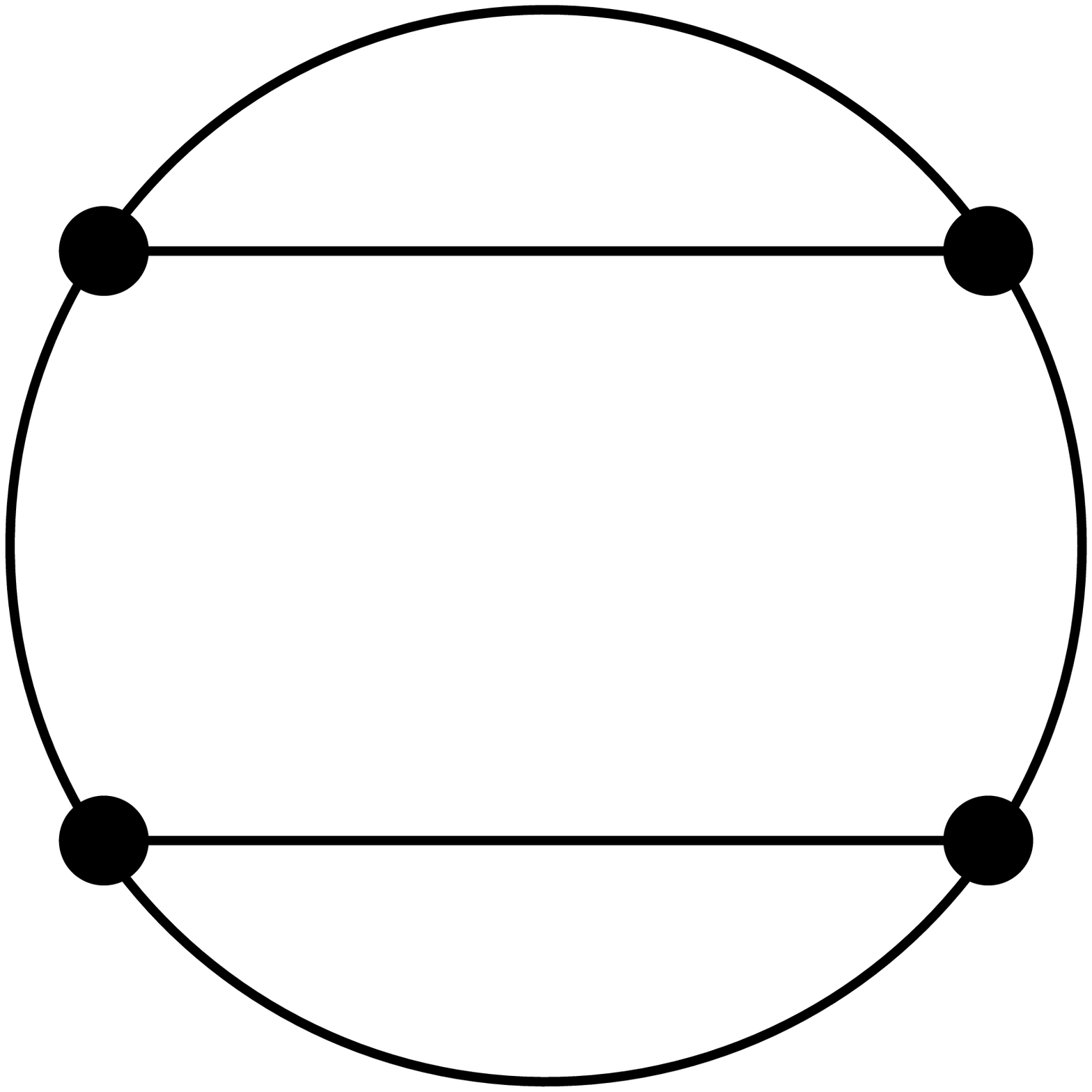}}
\put(0.8,2.0){B8} \put(3,1.05){\small $K_\mu$}\put(1.9,-0.1){\small $L_\mu$} \put(1.9,2.2){\small $Q_\mu$}
\put(5.2,0.3){\includegraphics[scale=0.10,clip]{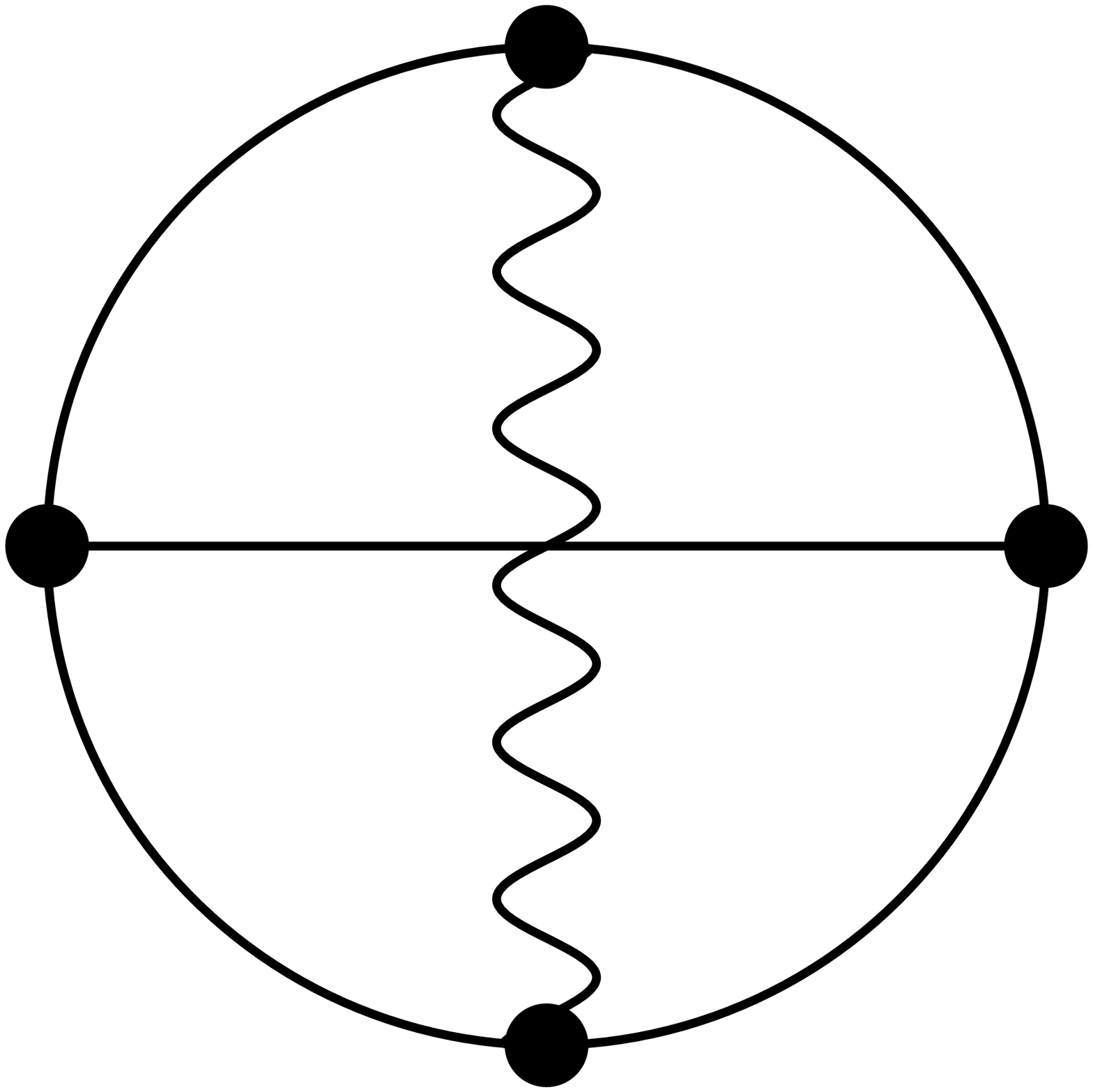}}
\put(4.8,2.0){B9} \put(5.55,0.75){\small $K_\mu$}\put(5.45,1.35){\small $L_\mu$} \put(6.8,1.8){\small $Q_\mu$}
\put(9.2,0.3){\includegraphics[scale=0.10,clip]{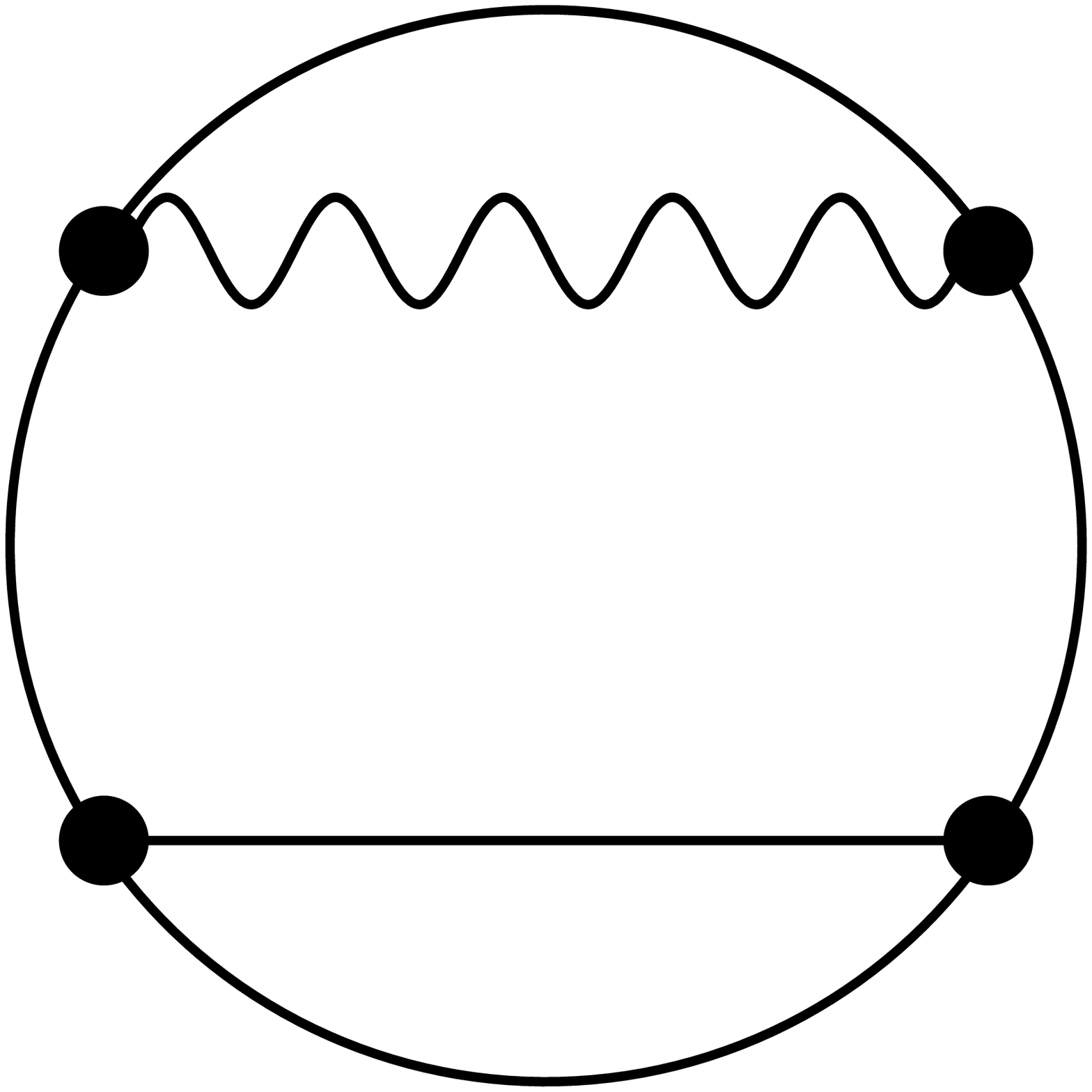}}
\put(8.7,2.0){B10} \put(9.9,1.2){\small $K_\mu$}\put(10,-0.1){\small $L_\mu$} \put(11,1.05){\small $Q_\mu$}
\put(13.2,-0.1){\includegraphics[scale=0.21,clip]{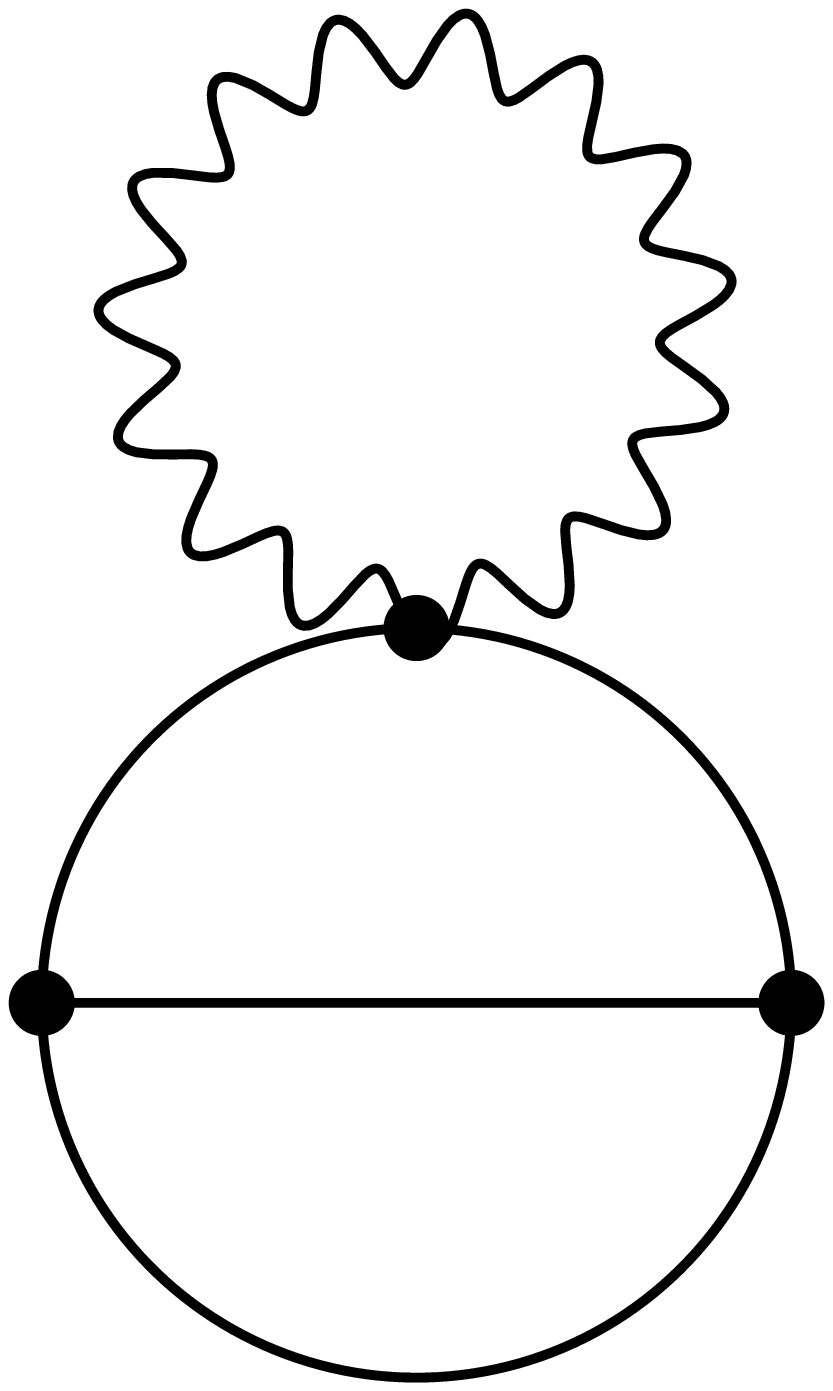}}
\put(12.5,2.0){B11} \put(14.8,2.0){\small $K_\mu$}\put(14.78,0){\small $L_\mu$} \put(14.8,1.2){\small $Q_\mu$}
\end{picture}
\caption{The three-loop vacuum supergraphs containing vertices with the Yukawa couplings.}\label{Figure_3Loop_Yukawa}
\end{figure}

Next, we will verify the argumentation of this paper on the example of the three-loop contribution to $\beta$-function which contains the Yukawa couplings. It is generated by the supergraphs B8 --- B11 presented in Fig. \ref{Figure_3Loop_Yukawa}. The direct calculation of the two-point superdiagrams obtained from them by attaching two external lines of the background gauge superfield $\bm{V}$ has been made in Refs. \cite{Shakhmanov:2017soc,Kazantsev:2018nbl} in the Feynman gauge $\xi_0=1$. Subsequently, the result has been reobtained by the method described in section \ref{Subsection_Idea} in Ref. \cite{Stepanyantz:2019ihw}. Now we will demonstrate how the modification proposed in this paper works in this case.

First, we consider the supergraph B8 which is quartic in the Yukawa couplings. The result for the modified vacuum supergraph is given by the expression

\begin{eqnarray}\label{B8_Vacuum}
&& B8\ \to\ \frac{4\pi}{r} \frac{d}{d\ln\Lambda} \int \frac{d^4Q}{(2\pi)^4} \frac{d^4K}{(2\pi)^4} \frac{d^4L}{(2\pi)^4} \lambda_0^{ijk} \lambda^*_{0ijl} \lambda_0^{mnl} \lambda^*_{0mnk} \nonumber\\
&& \qquad\qquad\qquad\qquad\qquad \times \frac{1}{Q^2 F_Q (K+Q)^2 F_{K+Q} L^2 F_L (K+L)^2 F_{K+L} K^2 F_K^2},\qquad
\end{eqnarray}

\noindent
which does not contain bad terms proportional to the inverse momenta to the fourth power. As earlier, to construct the corresponding contribution to the function (\ref{Delta_Beta}), we should sum all expressions obtained from (\ref{B8_Vacuum}) by replacing squared inverse momenta multiplied by $\delta$-symbols coming from the corresponding propagator by momentum $\delta$-functions multiplied by $4\pi^2 C(R)$. The tensor structure of the resulting factors can easily be viewed from the structure of the graph under consideration. After constructing and calculating the integrals of the $\delta$-functions we obtain the required contribution

\begin{eqnarray}
&& \Delta_{\mbox{\scriptsize B8}}\Big(\frac{\beta}{\alpha_0^2}\Big) = \frac{1}{\pi r} C(R)_p{}^l \frac{d}{d\ln\Lambda} \int \frac{d^4Q}{(2\pi)^4} \frac{d^4L}{(2\pi)^4}\, \lambda_0^{ijk} \lambda^*_{0ijl} \lambda_0^{mnp} \lambda^*_{0mnk} \frac{1}{Q^4 F_Q^2 L^4 F_L^2}\nonumber\\
&& + \frac{4}{\pi r} C(R)_i{}^p \frac{d}{d\ln\Lambda} \int \frac{d^4K}{(2\pi)^4} \frac{d^4L}{(2\pi)^4}\, \lambda_0^{ijk} \lambda^*_{0pjl} \lambda_0^{mnl} \lambda^*_{0mnk} \frac{1}{K^4 F_K^3 L^2 F_L (K+L)^2 F_{K+L}},\qquad
\end{eqnarray}

\noindent
which exactly coincides with the corresponding result of Refs. \cite{Shakhmanov:2017soc,Stepanyantz:2019ihw} found by different methods.

According to \cite{Stepanyantz:2019ihw} the properly modified vacuum supergraph B9 is given by the expression

\begin{eqnarray}\label{B9_Vaccum}
&& B9\ \to\ -\frac{8\pi}{r} \frac{d}{d\ln\Lambda} \int \frac{d^4Q}{(2\pi)^4} \frac{d^4K}{(2\pi)^4} \frac{d^4L}{(2\pi)^4}\, e_0^2 \lambda_0^{ijk} \lambda^*_{0imn} (T^A)_j{}^m (T^A)_k{}^n\nonumber\\
&& \times \frac{N(Q,K,L)}{K^2 R_K L^2 F_L Q^2 F_Q (Q+K)^2 F_{Q+K} (Q-L)^2 F_{Q-L} (Q+K-L)^2 F_{Q+K-L}},\qquad
\end{eqnarray}

\noindent
where $K_\mu$ corresponds to the propagator of the quantum gauge superfield and, following Ref. \cite{Kazantsev:2018nbl}, we use the notation

\begin{eqnarray}
&&\hspace*{-5mm} N(Q,K,L) \equiv L^2 F_{Q+K} F_{Q+K-L} - Q^2 \Big((Q+K)^2 - L^2\Big) F_{Q+K-L} \frac{F_{Q+K}-F_Q}{(Q+K)^2-Q^2}\nonumber\\
&&\hspace*{-5mm} - (Q-L)^2 \Big((Q+K-L)^2-L^2\Big) F_{Q+K} \frac{F_{Q+K-L}-F_{Q-L}}{(Q+K-L)^2-(Q-L)^2} + Q^2 (Q-L)^2  \nonumber\\
&&\hspace*{-5mm} \times \Big(L^2 - (Q+K)^2 - (Q+K-L)^2\Big)\bigg(\frac{F_{Q+K}-F_Q}{(Q+K)^2-Q^2}\bigg)\bigg(\frac{F_{Q+K-L}-F_{Q-L}}{(Q+K-L)^2-(Q-L)^2}\bigg).\qquad
\end{eqnarray}

\noindent
Again, Eq. (\ref{B9_Vaccum}) does not contain bad terms. Next, we proceed according to the algorithm of section \ref{Subsection_Graphs}. In this case the relevant replacements are $1/K^2 \to 4\pi^2 C_2 \delta^4(K)$ and $\delta_i^j/P^2 \to 4\pi^2 C(R)_i{}^j \delta^4(P)$, where $P_\mu$ stands for momenta of the matter superfields, namely, $Q_\mu$, $(Q+K)_\mu$, $(Q-L)_\mu$, and $(Q+K-L)_\mu$. After some transformations involving the identities

\begin{eqnarray}
&&\hspace*{-5mm} \lambda_0^{ijk} \lambda^*_{0imn} (T^A)_j{}^m (T^A)_k{}^n = -\frac{1}{2} \lambda_0^{imn} \lambda^*_{0jmn} C(R)_i{}^j;\\
&&\hspace*{-5mm} \lambda_0^{ijk} \lambda^*_{0imn} C(R)_j{}^l (T^A)_l{}^m (T^A)_k{}^n = -\frac{1}{2} \lambda_0^{imn} \lambda^*_{0jmn} \big(C(R)^2\big)_i{}^j;\\
&&\hspace*{-5mm} \lambda_0^{ijk} \lambda^*_{0lmn} C(R)_i{}^l (T^A)_j{}^m (T^A)_k{}^n = \frac{1}{2} \lambda_0^{imn} \lambda^*_{0jmn} \big(C(R)^2\big)_i{}^j - \lambda_0^{ijm} \lambda^*_{0klm} C(R)_i{}^k C(R)_j{}^l, \qquad
\end{eqnarray}

\noindent
which follow from Eq. (\ref{Yukawa_Constraint}), we get the result

\begin{eqnarray}
&&\hspace*{-5mm} \Delta_{\mbox{\scriptsize B9}}\Big(\frac{\beta}{\alpha_0^2}\Big) = \frac{1}{\pi r}\, \frac{d}{d\ln\Lambda} \int \frac{d^4Q}{(2\pi)^4} \frac{d^4L}{(2\pi)^4}\, e_0^2 \lambda_0^{imn}\lambda^*_{0jmn} C_2 C(R)_i{}^j \frac{N(Q,0,L)}{Q^4 F_Q^2 L^2 F_L (Q-L)^4 F_{Q-L}^2}\nonumber\\
&&\hspace*{-5mm} - \frac{1}{\pi r}\, \frac{d}{d\ln\Lambda} \int \frac{d^4Q}{(2\pi)^4} \frac{d^4K}{(2\pi)^4}\, e_0^2 \Big(\lambda_0^{imn}\lambda^*_{0jmn} \big(C(R)^2\big)_i{}^j - 2 \lambda_0^{ijm} \lambda^*_{0klm} C(R)_i{}^k C(R)_j{}^l\Big) \frac{1}{K^2 R_K}  \nonumber\\
&&\hspace*{-5mm} \times \frac{N(Q,K,0)}{Q^4 F_Q^2 (Q+K)^4 F_{Q+K}^2} + \frac{4}{\pi r}\, \frac{d}{d\ln\Lambda} \int \frac{d^4K}{(2\pi)^4} \frac{d^4L}{(2\pi)^4}\, e_0^2 \lambda_0^{imn}\lambda^*_{0jmn} \big(C(R)^2\big)_i{}^j
\frac{1}{K^4 R_K F_K L^4}
\nonumber\\
&&\hspace*{-5mm} \times \frac{N(0,K,L)}{F_L^2 (K-L)^2 F_{K-L}},
\end{eqnarray}

\noindent
which agrees with the one found in \cite{Kazantsev:2018nbl,Stepanyantz:2019ihw} by different methods.

The supergraph B10 is given by the expression

\begin{eqnarray}
&& B10\ \to\ -\frac{8\pi}{r} \frac{d}{d\ln\Lambda} \int \frac{d^4Q}{(2\pi)^4} \frac{d^4K}{(2\pi)^4} \frac{d^4L}{(2\pi)^4}\, e_0^2 \lambda_0^{ijk} \lambda^*_{0ijl} (T^A)_k{}^m (T^A)_m{}^l\nonumber\\
&&\qquad\qquad\qquad\qquad\qquad\qquad\qquad \times \frac{L(Q,Q+K)}{K^2 R_K Q^2 F_Q^2 (Q+L)^2 F_{Q+L} (Q+K)^2 F_{Q+K} L^2 F_L},\qquad\quad
\end{eqnarray}

\noindent
where $K_\mu$ denotes the momentum of the quantum gauge superfield propagator and, again following Ref. \cite{Kazantsev:2018nbl},

\begin{equation}
L(Q,P) \equiv F_Q F_P + \frac{F_P-F_Q}{P^2-Q^2}\, \Big(F_Q Q^2 + F_P P^2\Big) + 2 Q^2 P^2 \bigg(\frac{F_P-F_Q}{P^2-Q^2}\bigg)^2.
\end{equation}

\noindent
In this case the algorithm of section \ref{Subsection_Graphs} produces the expression

\begin{eqnarray}
&&\hspace*{-5mm} \Delta_{\mbox{\scriptsize B10}}\Big(\frac{\beta}{\alpha_0^2}\Big) = -\frac{2}{\pi r} \frac{d}{d\ln\Lambda} \int \frac{d^4Q}{(2\pi)^4} \frac{d^4L}{(2\pi)^4}\, e_0^2\, \lambda_0^{imn} \lambda^*_{0jmn} C_2 C(R)_i{}^j \frac{L(Q,Q)}{Q^4 F_Q^3 L^2 F_L (Q+L)^2 F_{Q+L}} \nonumber\\
&&\hspace*{-5mm} -\frac{2}{\pi r} \frac{d}{d\ln\Lambda} \int \frac{d^4K}{(2\pi)^4} \frac{d^4L}{(2\pi)^4}\, e_0^2\, \lambda_0^{imn} \lambda^*_{0jmn} \big(C(R)^2\big)_i{}^j \bigg(\frac{L(0,K)}{K^4 R_K F_K L^4 F_L^2} + \frac{L(K,0)}{K^4 R_K F_K^2 L^2 F_L} \nonumber\\
&&\hspace*{-5mm} \times\frac{1}{(K+L)^2 F_{K+L}}\bigg) - \frac{4}{\pi r} \frac{d}{d\ln\Lambda} \int \frac{d^4Q}{(2\pi)^4} \frac{d^4K}{(2\pi)^4}\, e_0^2\, \lambda_0^{ijm} \lambda^*_{0klm} C(R)_i{}^k C(R)_j{}^l \frac{L(Q,Q+K)}{K^2 R_K Q^4 F_Q^3} \qquad\nonumber\\
&&\hspace*{-5mm} \times \frac{1}{(Q+K)^2 F_{Q+K}},
\end{eqnarray}

\noindent
which also coincides with the results of Refs. \cite{Kazantsev:2018nbl,Stepanyantz:2019ihw}.

The expression for the last supergraph B11 reads as

\begin{eqnarray}
&& B11\ \to\ \frac{8\pi}{r} \frac{d}{d\ln\Lambda} \int \frac{d^4Q}{(2\pi)^4} \frac{d^4K}{(2\pi)^4} \frac{d^4L}{(2\pi)^4}\, e_0^2 \lambda_0^{ijk} \lambda^*_{0ijl} (T^A)_k{}^m (T^A)_m{}^l\nonumber\\
&&\qquad\qquad\qquad\qquad\qquad\qquad\qquad\qquad\qquad\quad \times \frac{K(Q,K)}{K^2 R_K Q^2 F_Q^2 L^2 F_L (Q+L)^2 F_{Q+L}},\qquad
\end{eqnarray}

\noindent
where the momentum of the gauge superfield propagator is denoted by $K_\mu$ and

\begin{equation}
K(Q,K) \equiv \frac{F_{Q+K}-F_Q-2Q^2 F_Q'/\Lambda^2}{(Q+K)^2 - Q^2} + \frac{2Q^2 (F_{Q+K}-F_Q)}{((Q+K)^2-Q^2)^2}.
\end{equation}

\noindent
Replacing the squared inverse momenta and the corresponding $\delta$-symbols  according to the prescription given in section \ref{Subsection_Graphs} we obtain the part of the function (\ref{Delta_Beta}),

\begin{eqnarray}
&& \Delta_{\mbox{\scriptsize B11}}\Big(\frac{\beta}{\alpha_0^2}\Big) = \frac{2}{\pi r} \frac{d}{d\ln\Lambda} \int \frac{d^4Q}{(2\pi)^4} \frac{d^4L}{(2\pi)^4}\, e_0^2\, \lambda_0^{imn} \lambda^*_{0jmn} C_2 C(R)_i{}^j \frac{K(Q,0)}{Q^2 F_Q^2 L^2 F_L (Q+L)^2 F_{Q+L}}\nonumber\\
&& + \frac{2}{\pi r} \frac{d}{d\ln\Lambda} \int \frac{d^4K}{(2\pi)^4} \frac{d^4L}{(2\pi)^4}\, e_0^2\, \lambda_0^{imn} \lambda^*_{0jmn} \big(C(R)^2\big)_i{}^j \frac{K(0,K)}{K^2 R_K L^4 F_L^2}\nonumber\\
&& + \frac{4}{\pi r} \frac{d}{d\ln\Lambda} \int \frac{d^4Q}{(2\pi)^4} \frac{d^4K}{(2\pi)^4}\, e_0^2\, \lambda_0^{ijm} \lambda^*_{0klm} C(R)_i{}^k C(R)_j{}^l \frac{K(Q,K)}{K^2 R_K Q^4 F_Q^3}.
\end{eqnarray}

\noindent
Again, it agrees with the calculations made previously.

\subsection{The three-loop approximation: supergraphs with ghost loops}
\hspace*{\parindent}

All three-loop vacuum supergraphs containing loops of the Faddeev--Popov ghosts have been calculated in Ref. \cite{Kuzmichev:2019ywn}. These supergraphs are presented in Fig. \ref{Figure_3Loop_Ghosts}. Note that in this approximation the first nonlinear term in the function ${\cal F}(V)$ (see Eq. (\ref{Nonlinear_Function_F})) is essential. It generates the vertex

\begin{equation}
- \frac{3}{4} e_0^2\, y_0\, G^{ABCD} \int d^8x\, (\bar c^A + \bar c^{+A}) V^C V^D (c^B - c^{+B})
\end{equation}

\noindent
present in the supergraph B21, see Ref. \cite{Kazantsev:2018kjx} for details. This vertex is denoted by a cross.

Expressions for the supergraphs in Fig. \ref{Figure_3Loop_Ghosts} contain $1/K^4$ singularities, where $K_\mu$ is the Euclidean momentum of the quantum gauge superfield. As in the two-loop approximation, these singularities should disappear after adding the purely gauge vacuum supergraphs. However, the sum of the three-loop diagrams containing only the gauge propagators has not yet been calculated with the higher covariant derivative regularization. Nevertheless, the sum of the supergraphs with ghost loops does not contain any singularities proportional to the inverse ghost or matter momenta to the fourth power. This implies that it is possible to compare the sums of singularities coming from the cuts of ghost and matter lines with the corresponding anomalous dimensions $\gamma_c(\alpha_0,\lambda_0,Y_0)$ and $(\gamma_\phi)_i{}^j(\alpha_0,\lambda_0,Y_0)$. With the help of the method described in section \ref{Subsection_Idea} this has been done in Ref. \cite{Kuzmichev:2019ywn}. Also it is possible to use these results for checking the modification of the algorithm discussed in section \ref{Subsection_Graphs}. This is made as follows:

\begin{figure}[p]
\begin{picture}(0,21)
\put(2.4,-0.4){$+$}
\put(3.2,-1.0){\includegraphics[scale=0.165,clip]{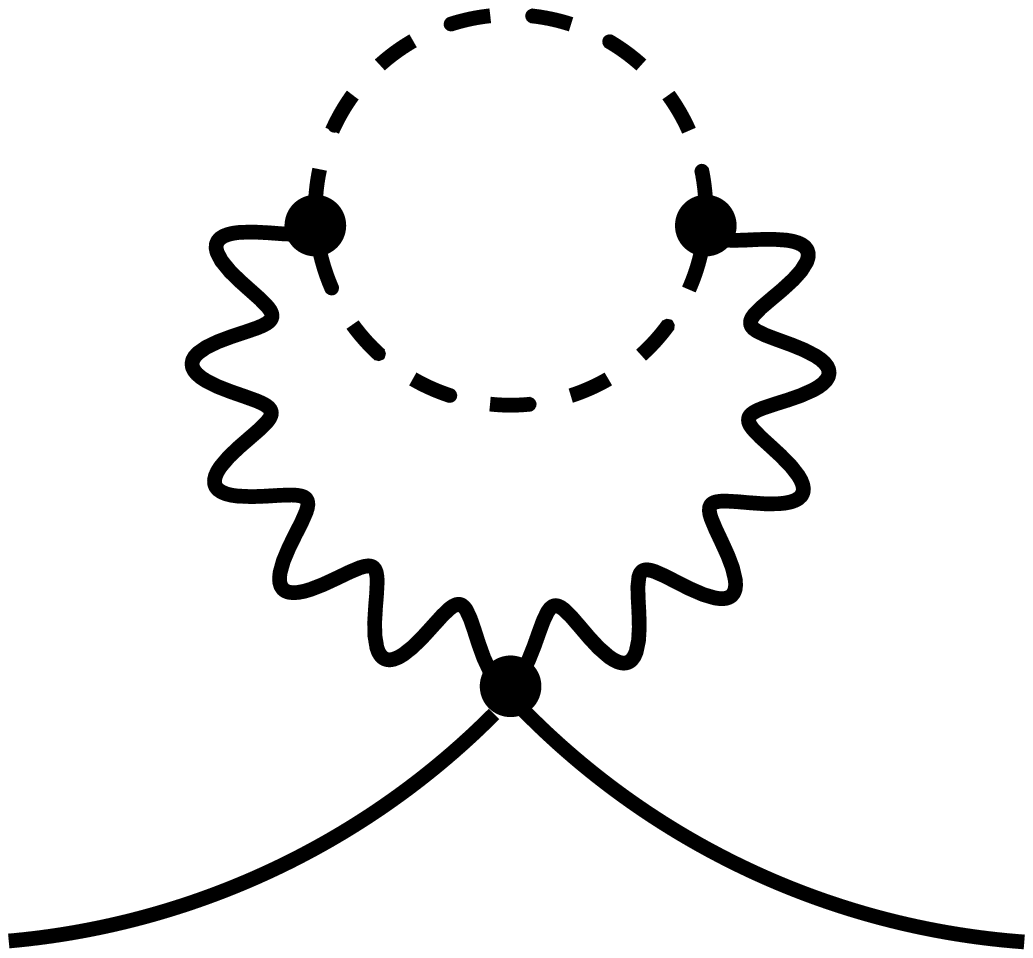}}
\put(5.5,-0.4){$+$}
\put(6.4,-1.1){\includegraphics[scale=0.168,clip]{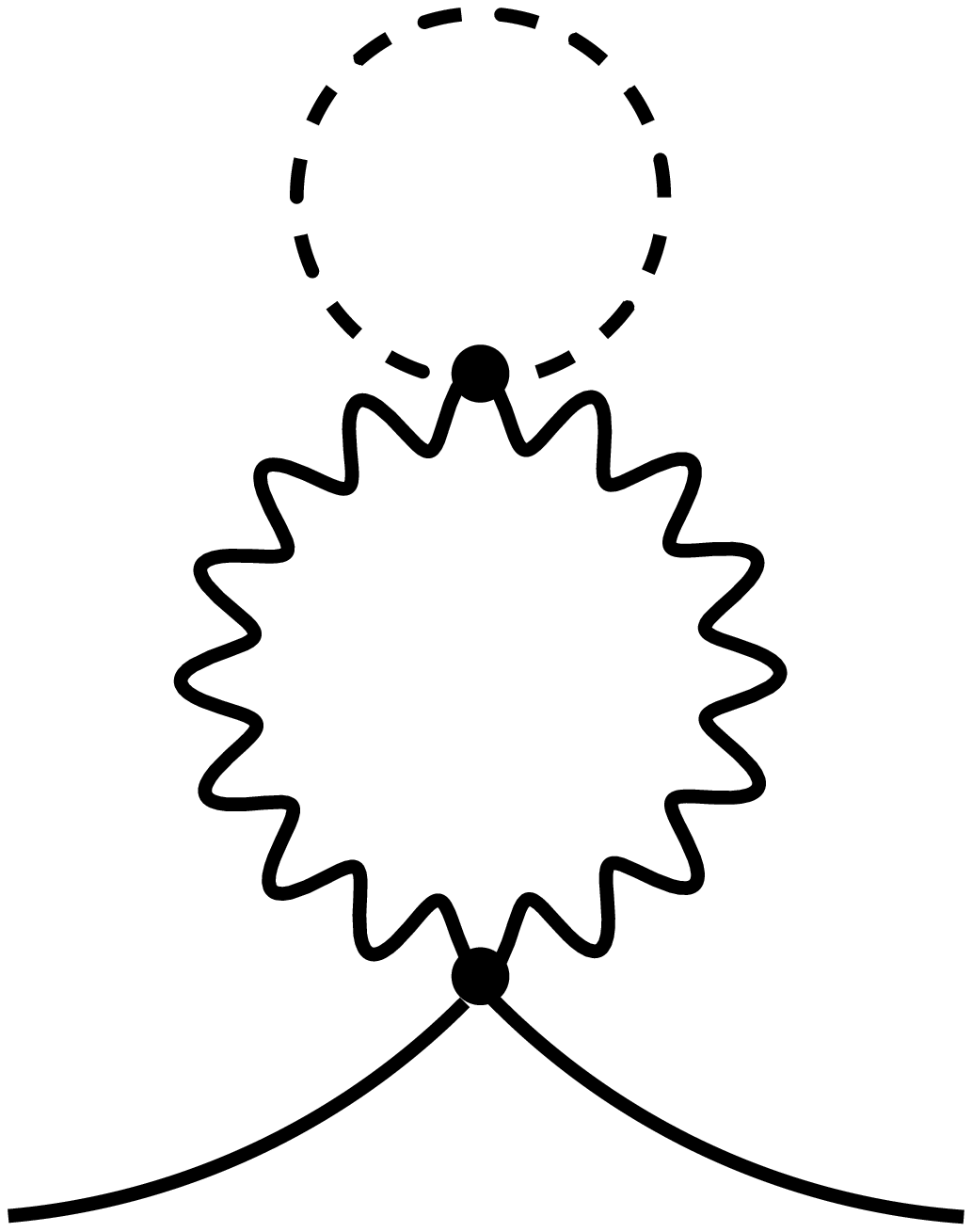}}
\put(8.8,-0.4){$+$}
\put(9.5,-0.9){\includegraphics[scale=0.165,clip]{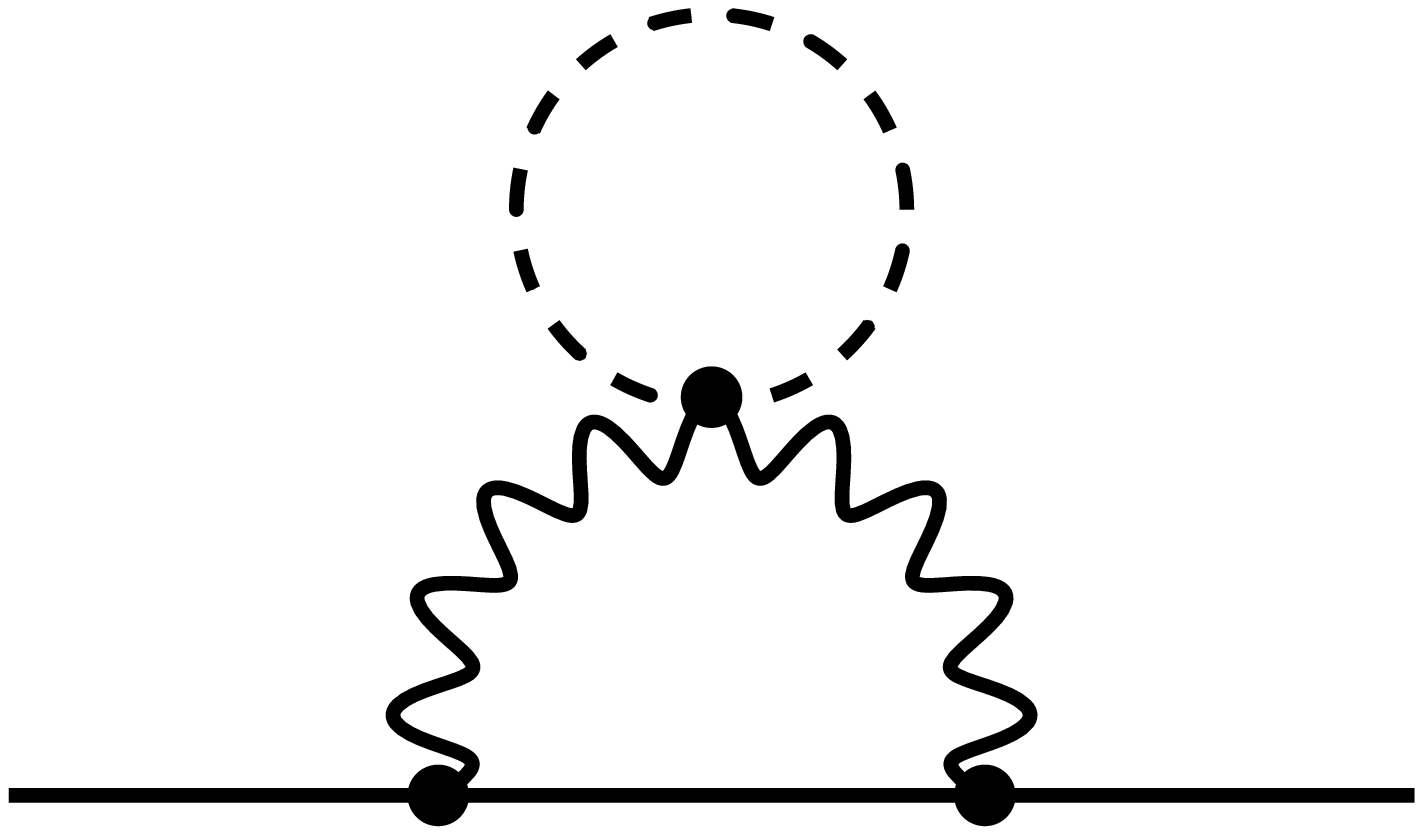}}
\put(12.35,-0.4){$+$}
\put(13.2,-0.9){\includegraphics[scale=0.16,clip]{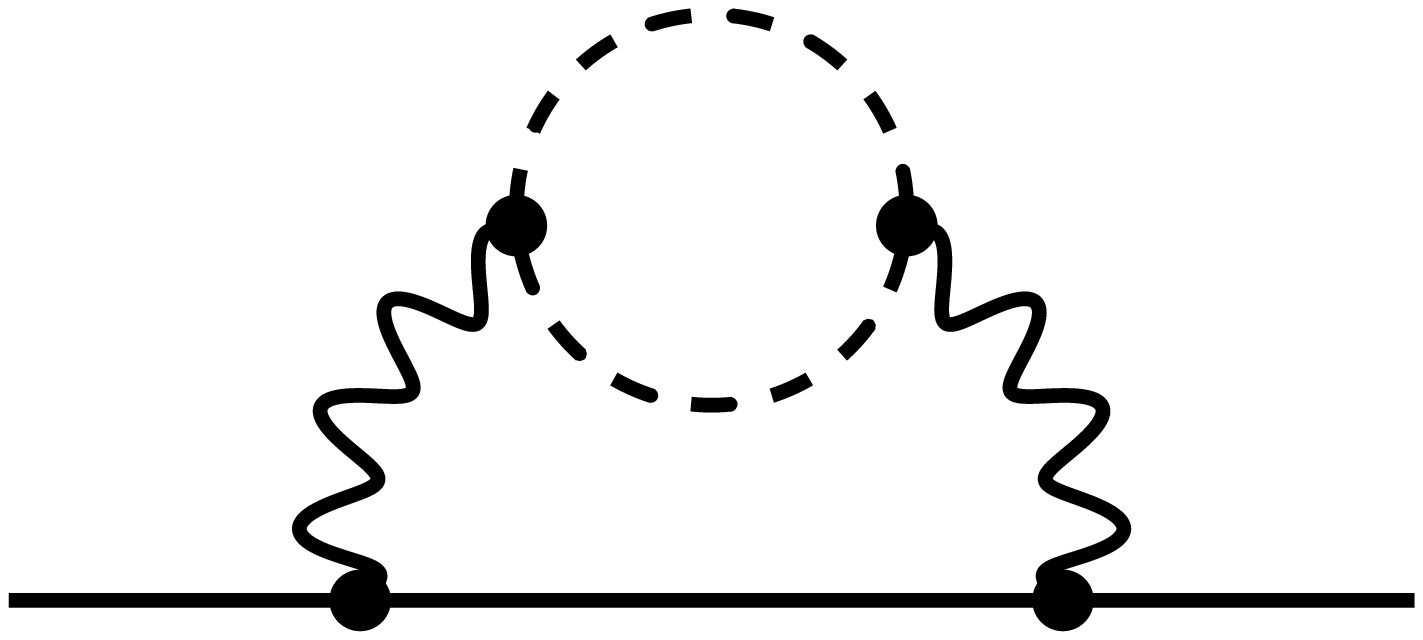}}

\put(0.7,1){\includegraphics[scale=0.2,clip]{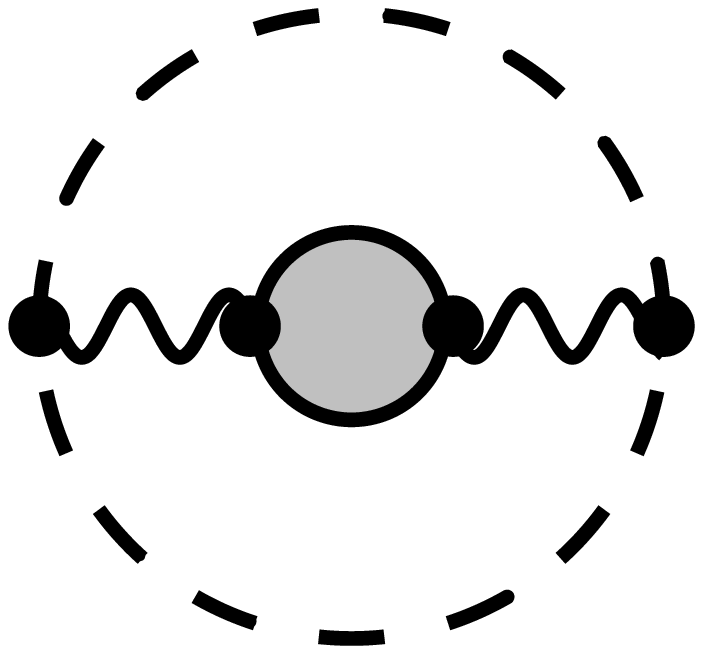}}
\put(0,2.3){B22}
\put(2.7,1.6){$+$}
\put(3.7,1){\includegraphics[scale=0.185,clip]{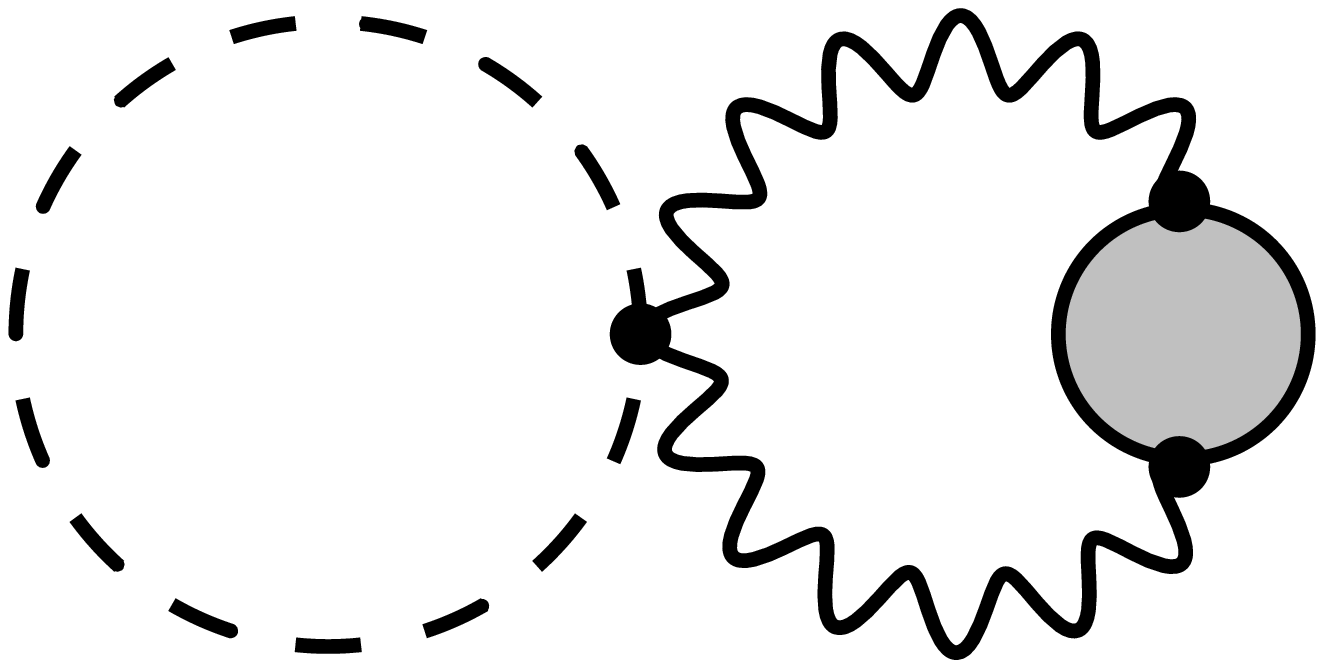}}
\put(3,2.3){B23}
\put(7.0,1.7){\vector(1,0){1.5}}
\put(9.5,1.2){\includegraphics[scale=0.13,clip]{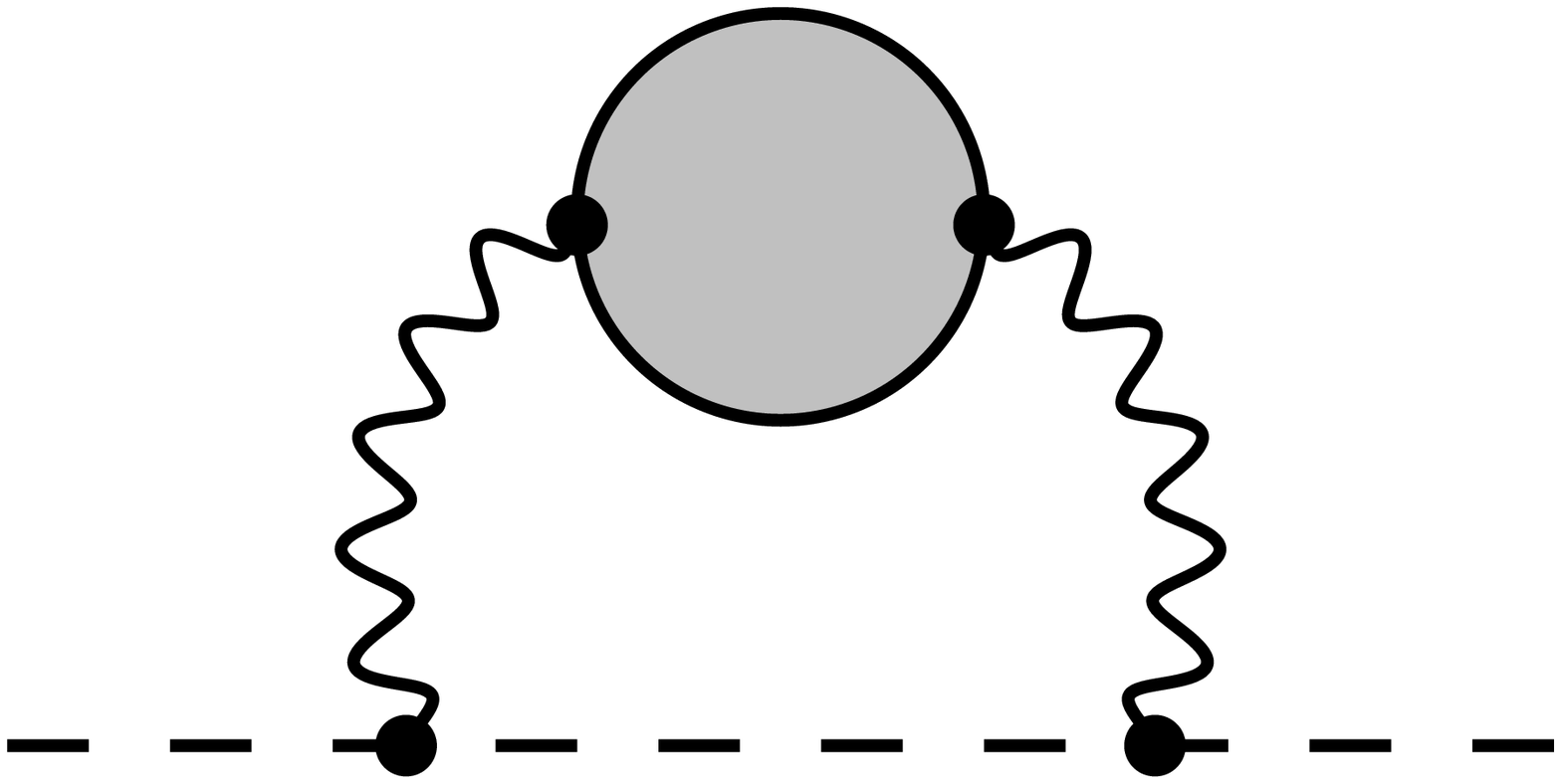}}
\put(12.35,1.6){$+$}
\put(13.2,1.17){\includegraphics[scale=0.12,clip]{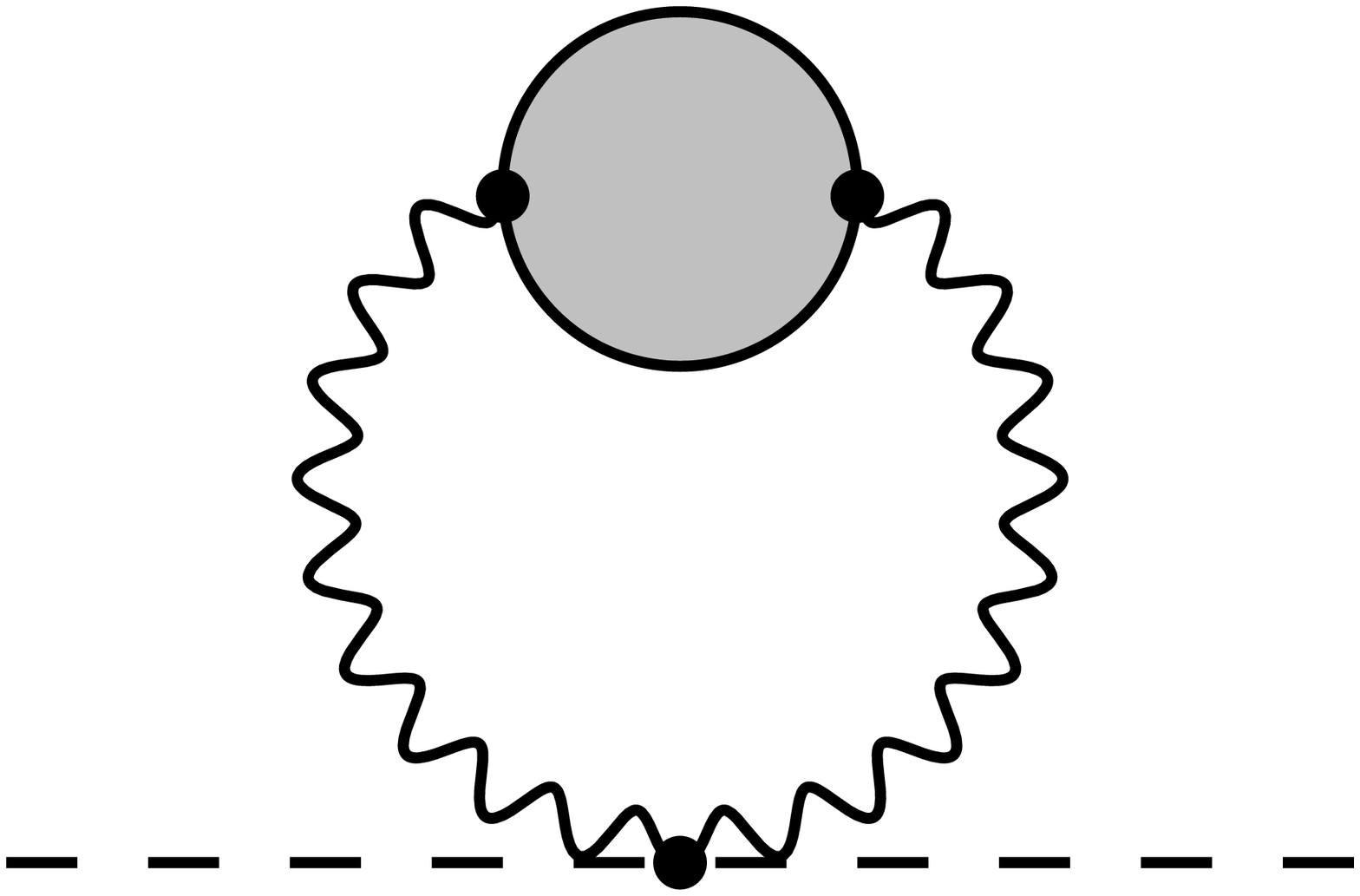}}

\put(0.5,2.9){\includegraphics[scale=0.18,clip]{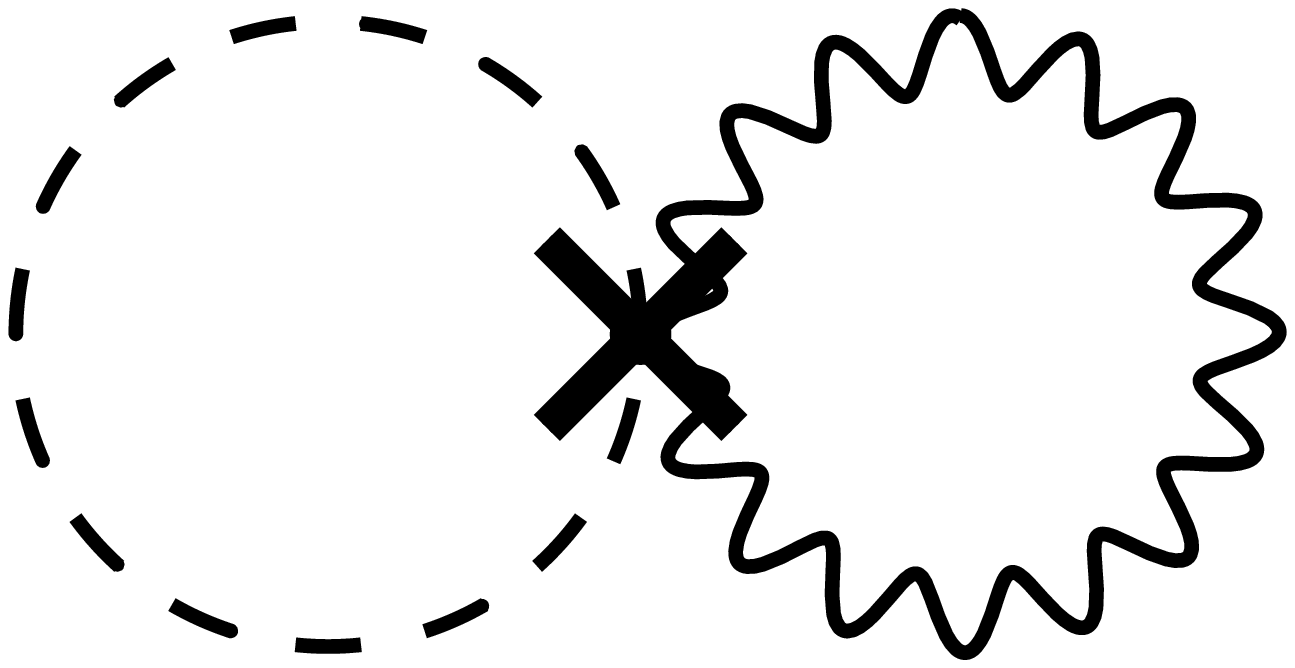}}
\put(0,4.2){B21}
\put(3.9,3.6){\vector(1,0){1.6}}
\put(6.5,3.1){\includegraphics[scale=0.13,clip]{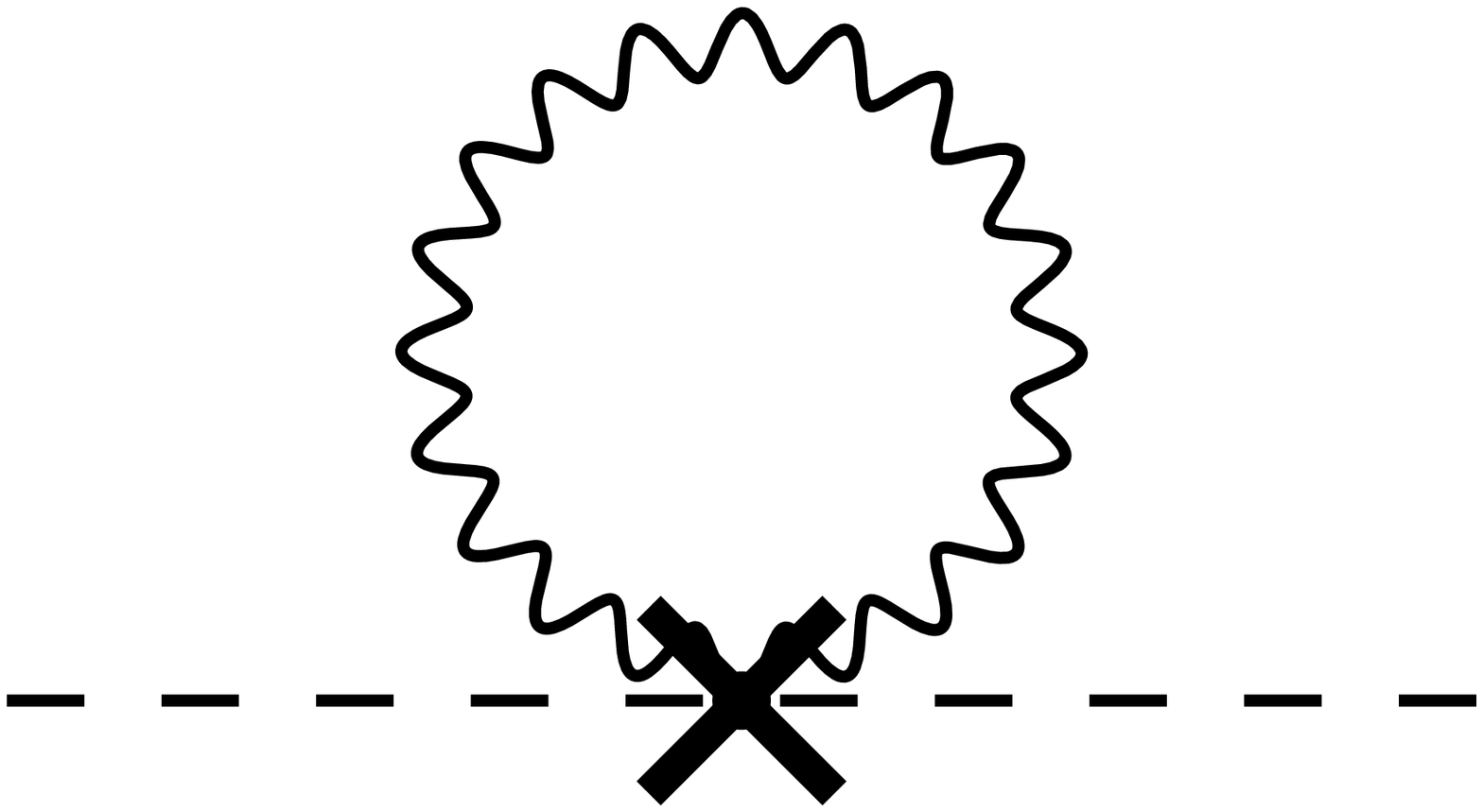}}

\put(0.5,4.8){\includegraphics[scale=0.2,clip]{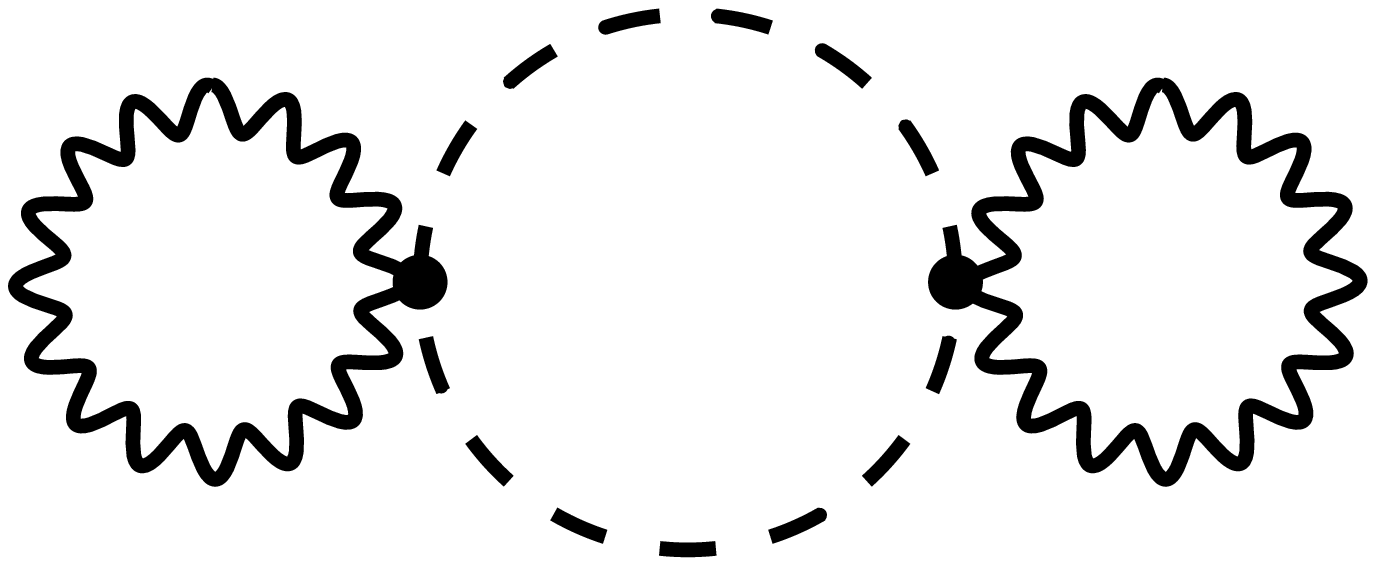}}
\put(0,6.1){B20}
\put(3.9,5.5){\vector(1,0){1.6}}
\put(6.2,5.4){${\displaystyle -\frac{1}{2}\ \Bigg(}$}
\put(7.4,5.1){\includegraphics[scale=0.125,clip]{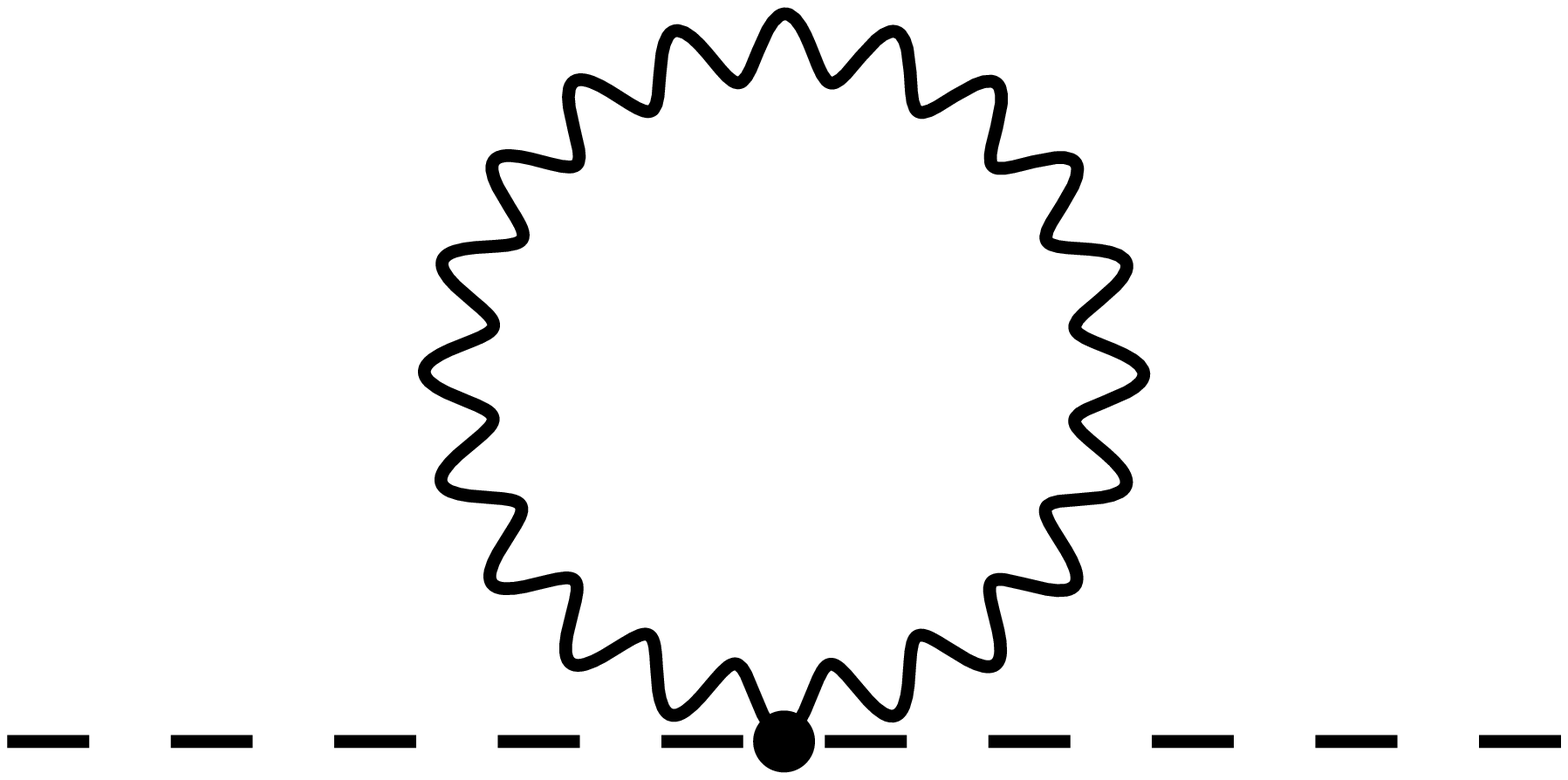}}
\put(9.9,5.4){${\displaystyle\Bigg)^{\mbox{2}}}$}

\put(1,6.7){\includegraphics[scale=0.2,clip]{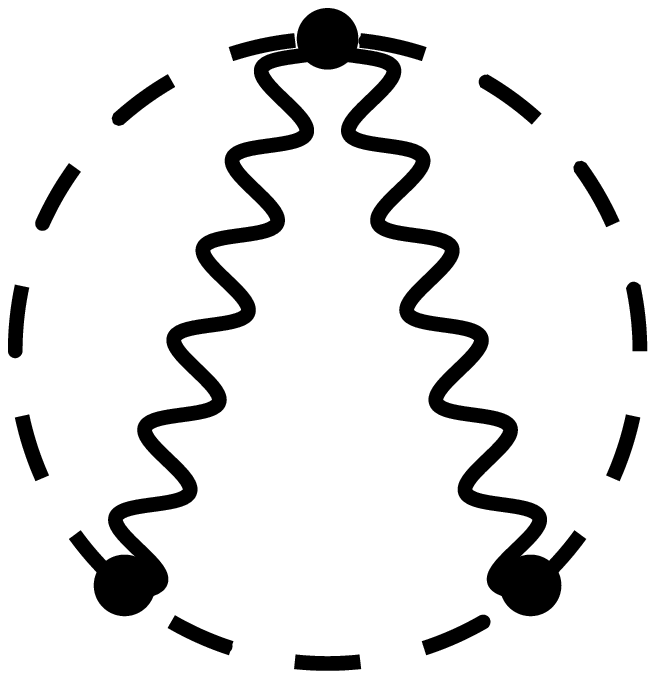}}
\put(0,8.0){B19}
\put(3.5,7.4){\vector(1,0){1.5}}
\put(6,6.9){\includegraphics[scale=0.13,clip]{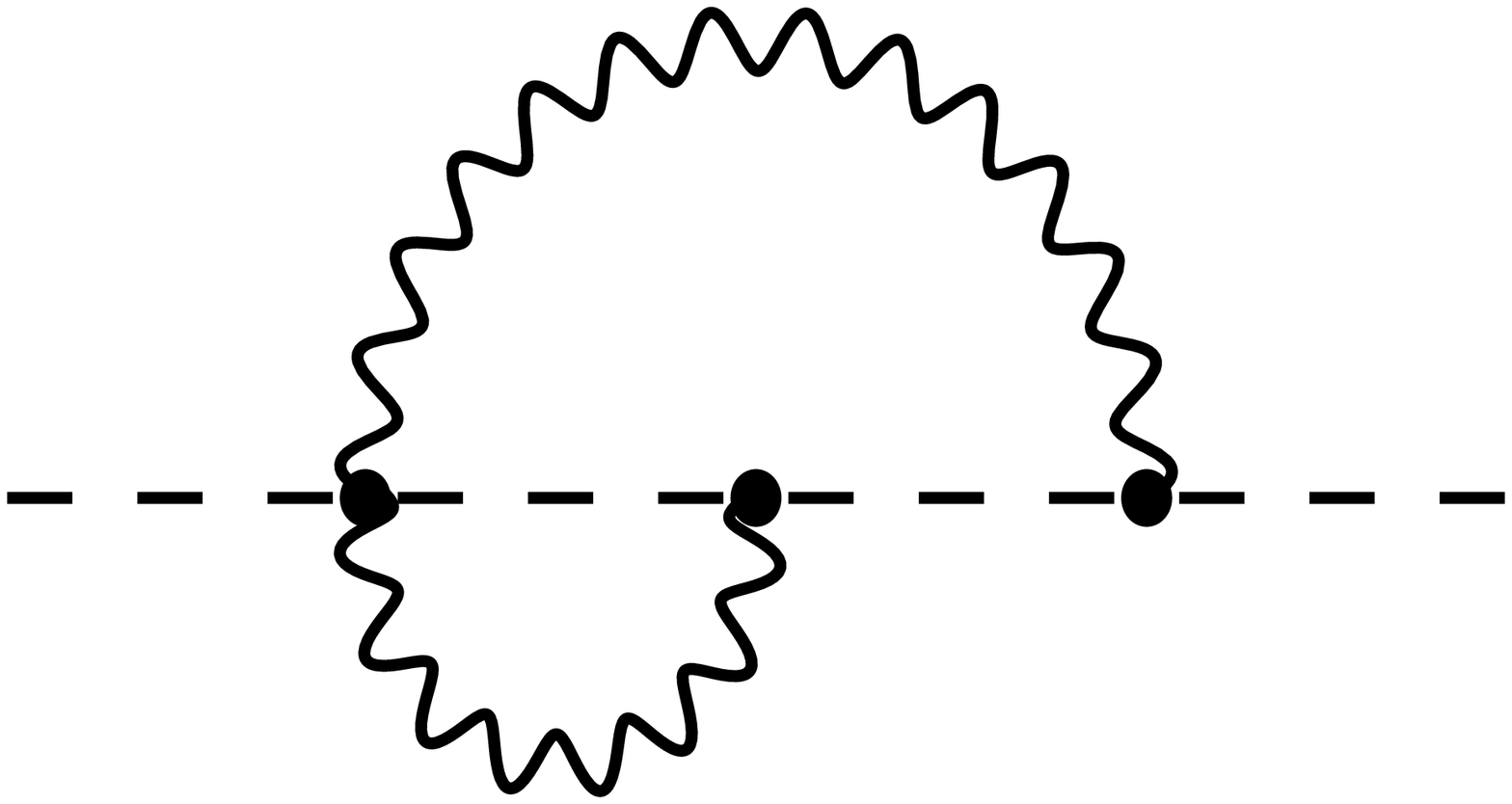}}
\put(8.8,7.3){$+$}
\put(9.4,6.9){\includegraphics[scale=0.13,clip]{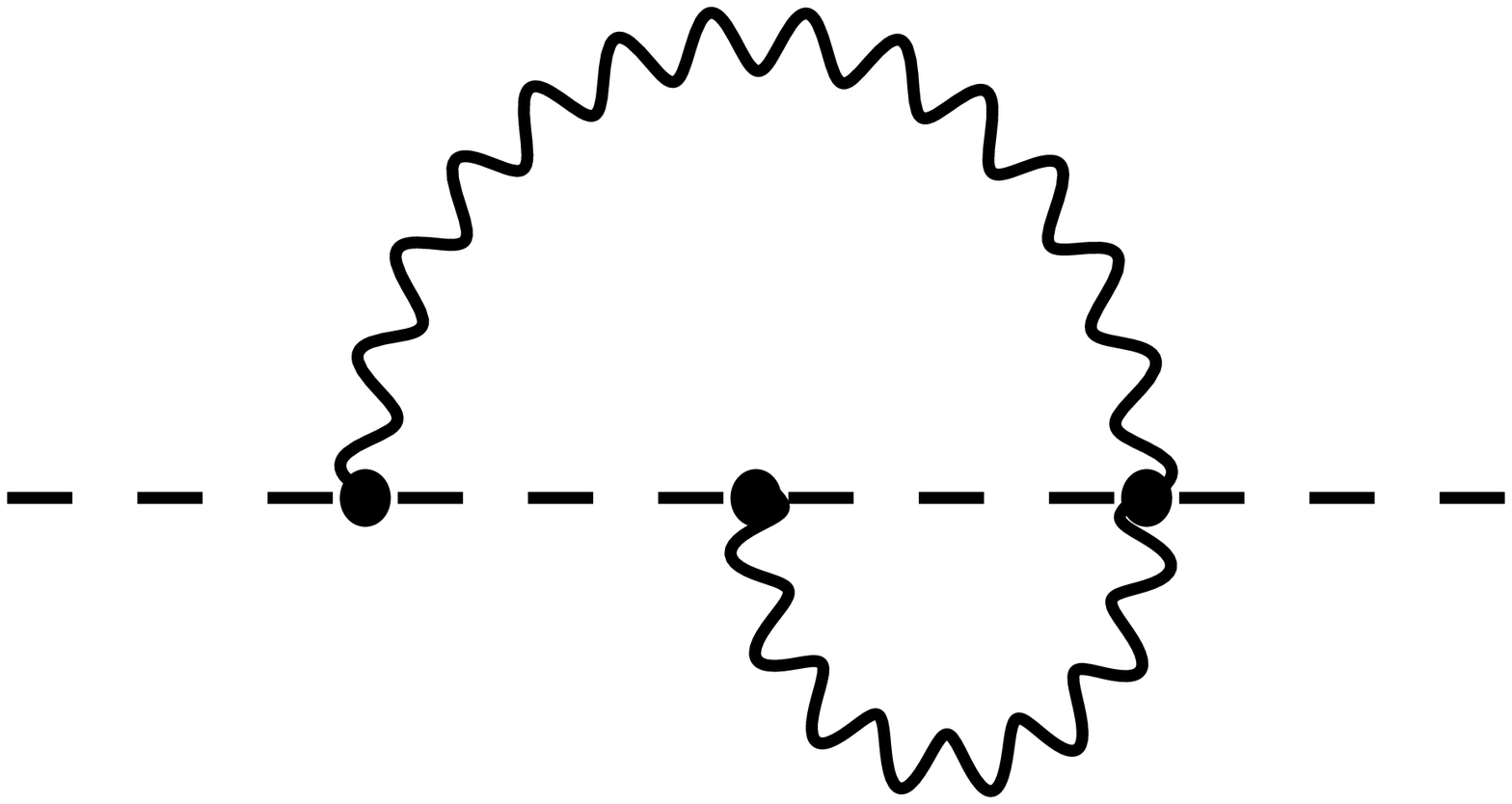}}
\put(12.2,7.3){$+$}
\put(12.85,6.8){\includegraphics[scale=0.14,clip]{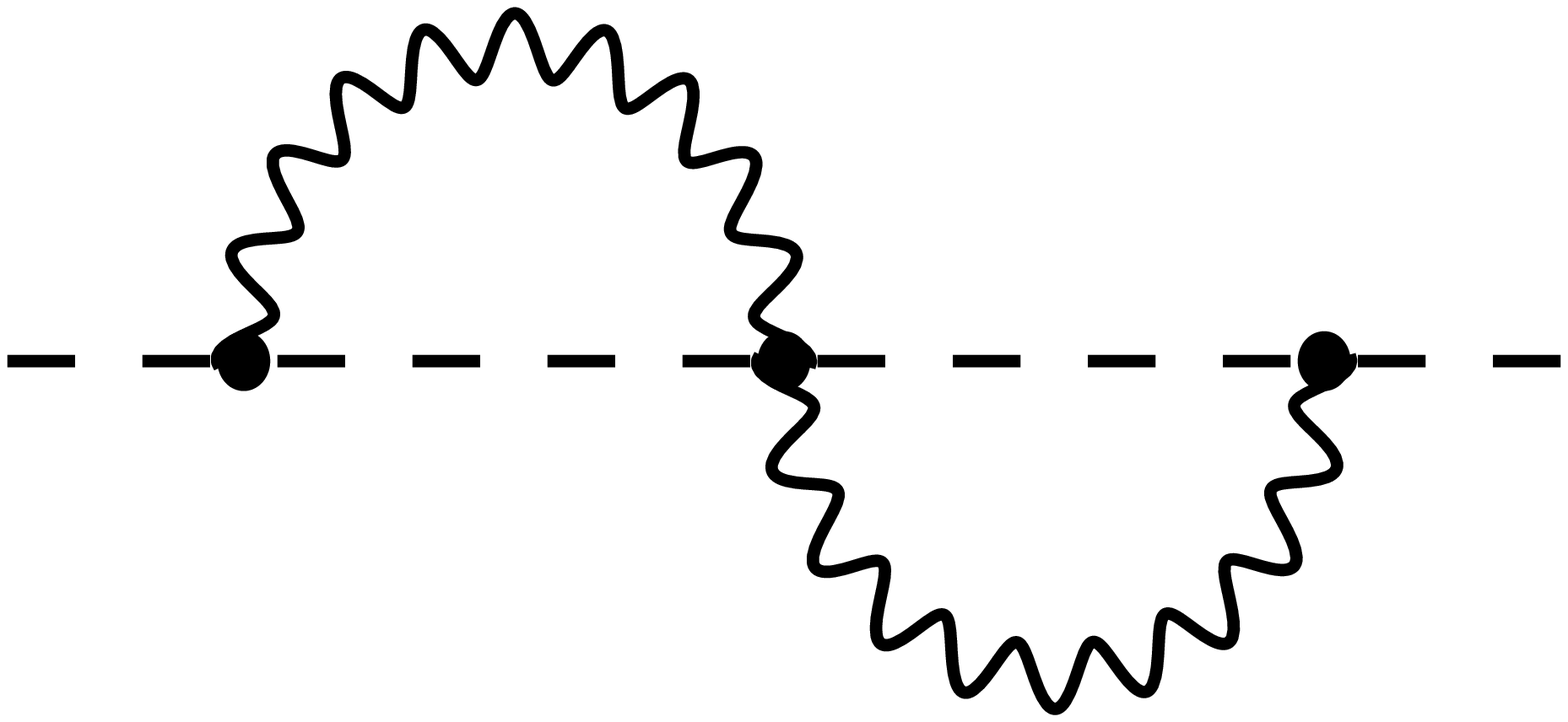}}

\put(1,8.6){\includegraphics[scale=0.2,clip]{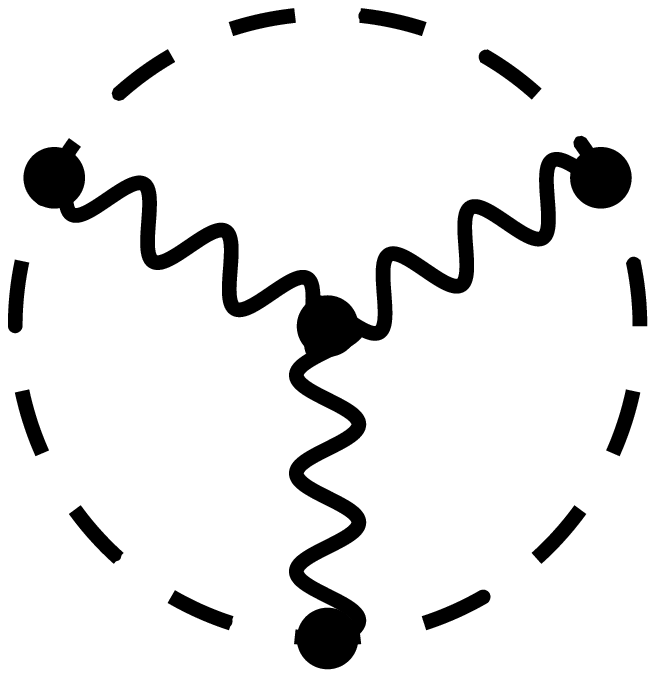}}
\put(0,9.9){B18}
\put(3.5,9.3){\vector(1,0){2}}
\put(6.5,9.0){\includegraphics[scale=0.13,clip]{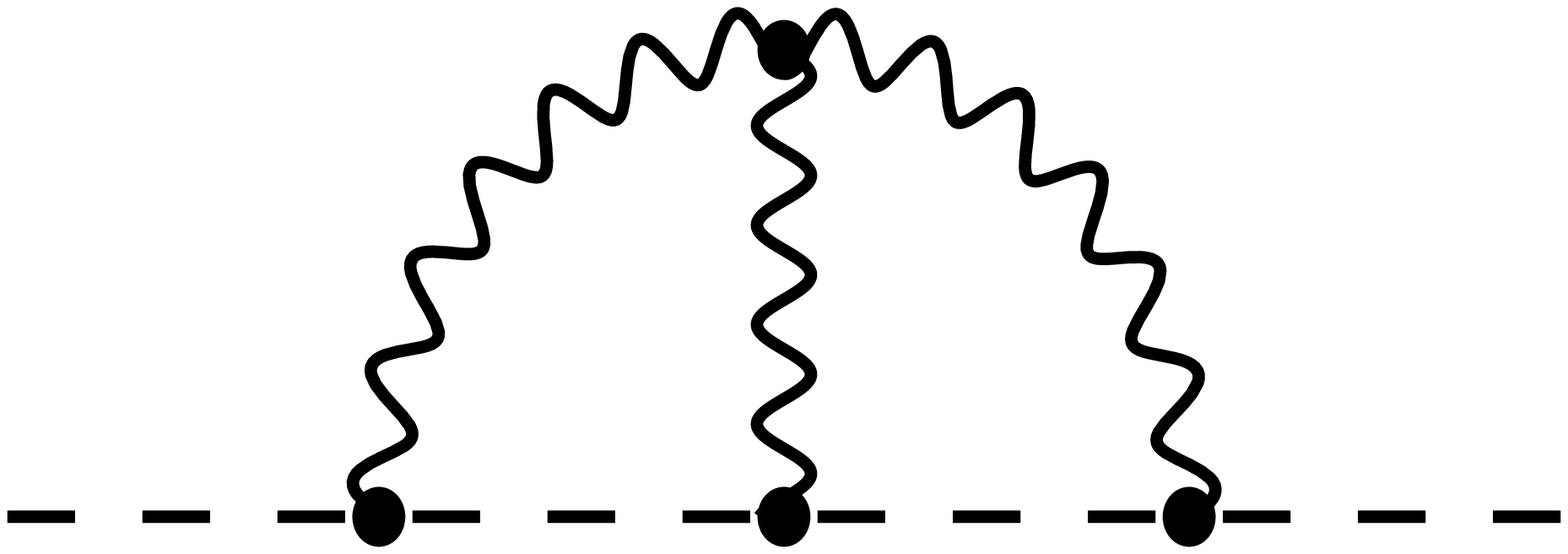}}

\put(1,10.5){\includegraphics[scale=0.2,clip]{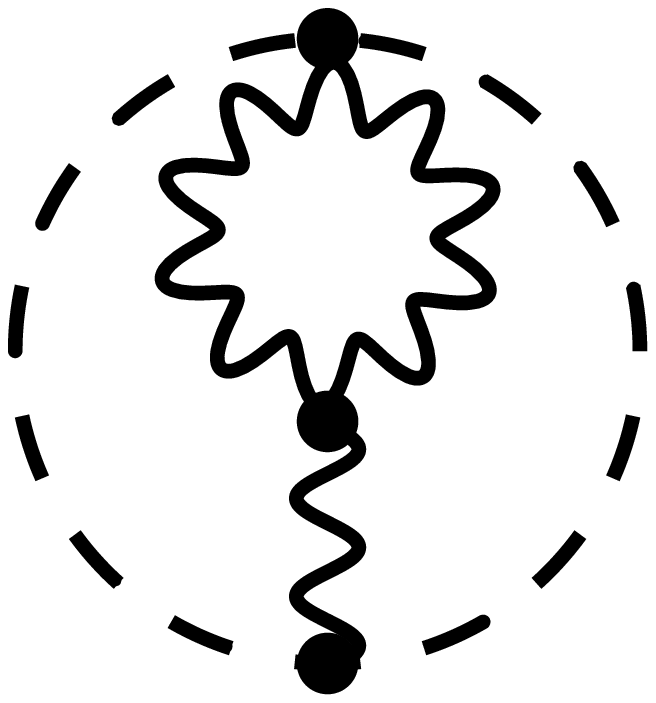}}
\put(0,11.8){B17}
\put(3.5,11.2){\vector(1,0){2}}
\put(6.5,10.9){\includegraphics[scale=0.13,clip]{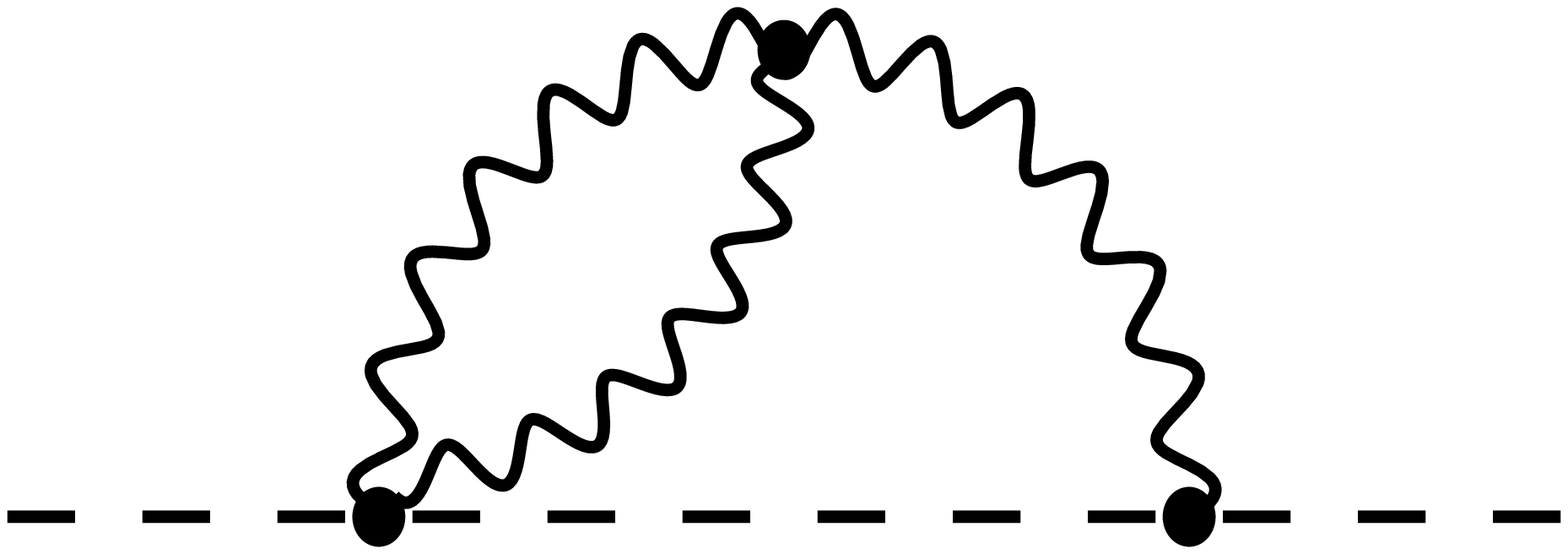}}
\put(9.5,11.1){$+$}
\put(10.3,10.9){\includegraphics[scale=0.13,clip]{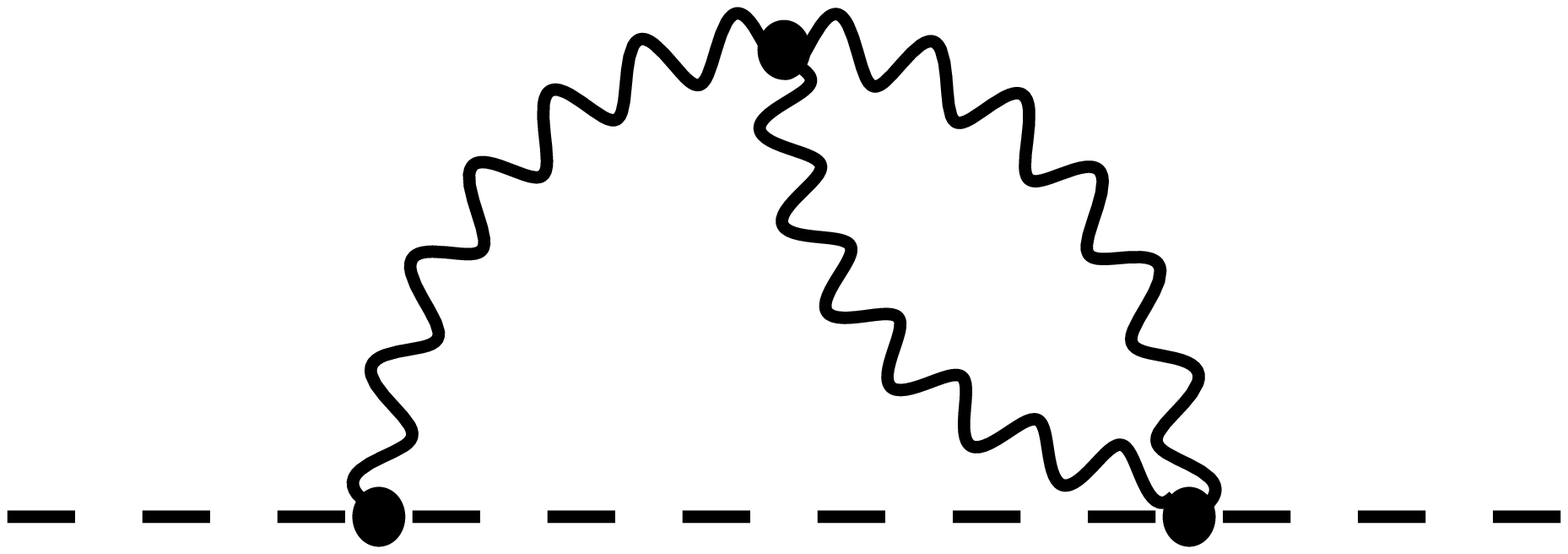}}

\put(1,12.4){\includegraphics[scale=0.2,clip]{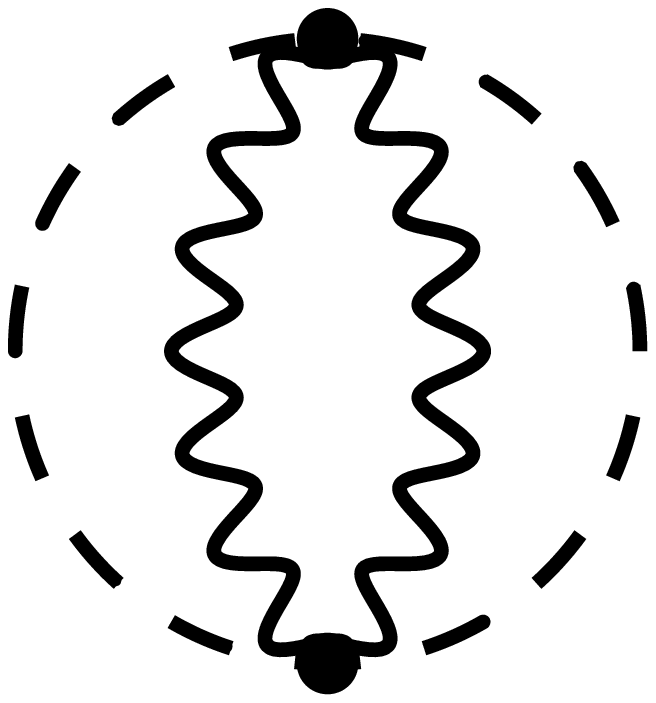}}
\put(0,13.7){B16}
\put(3.5,13.1){\vector(1,0){2}}
\put(6.5,12.4){\includegraphics[scale=0.14,clip]{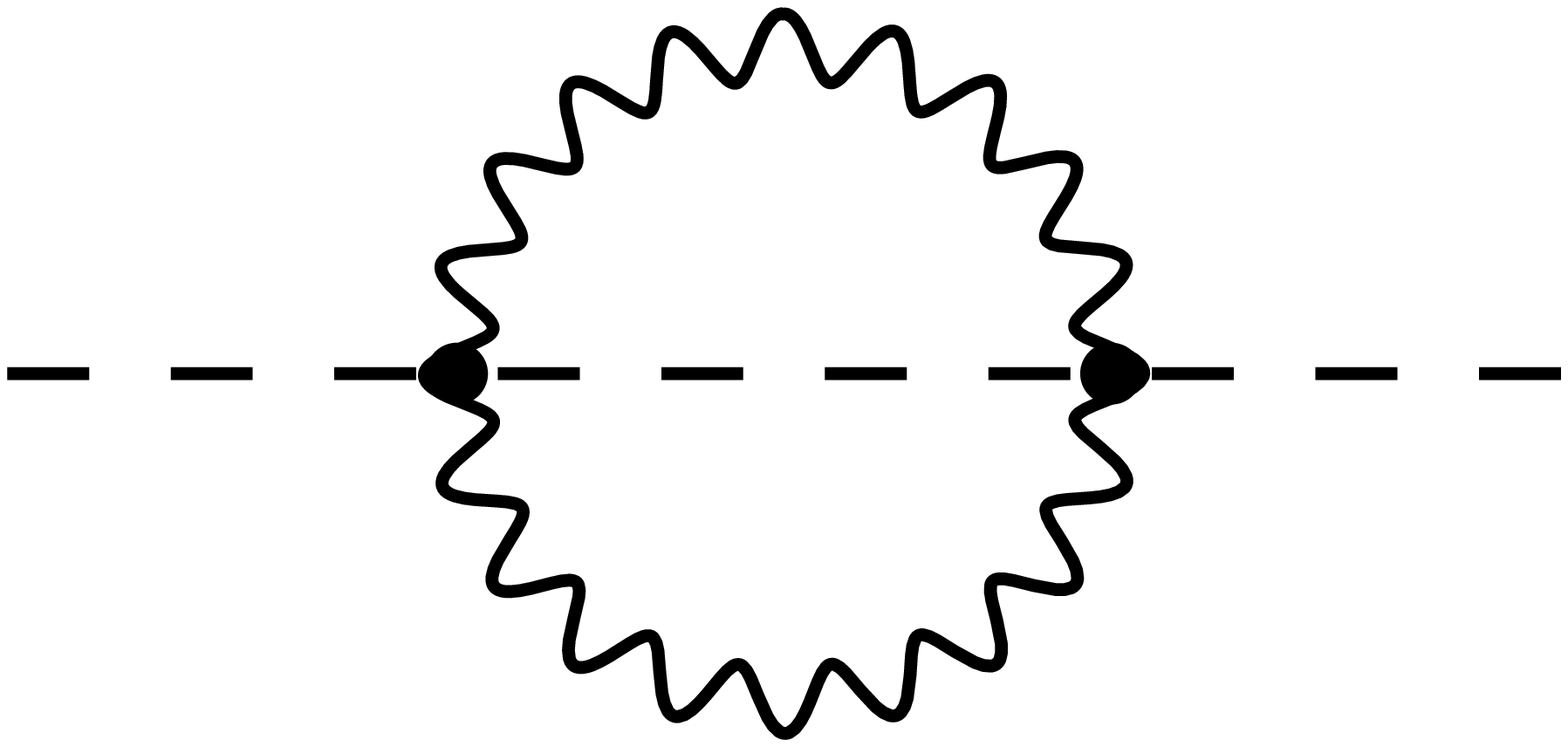}}

\put(1.1,14.3){\includegraphics[scale=0.2,clip]{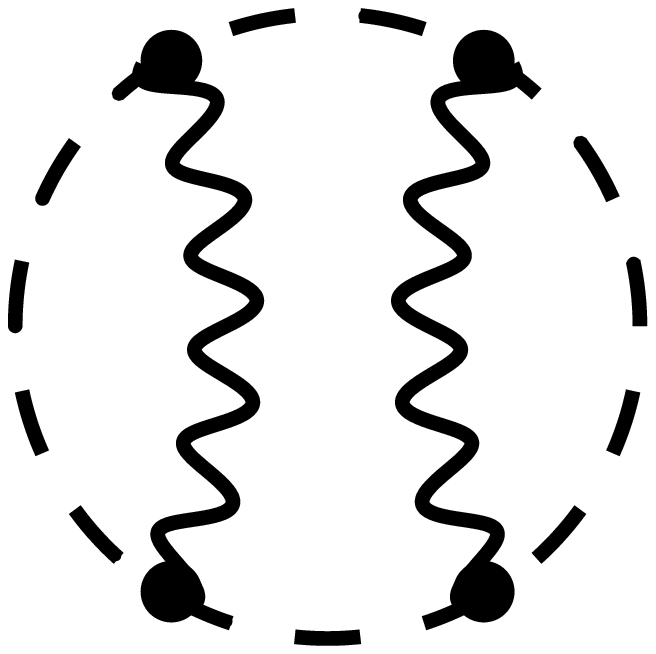}}
\put(0,15.6){B15}
\put(3.5,15.0){\vector(1,0){2}}
\put(6.5,14.5){\includegraphics[scale=0.13,clip]{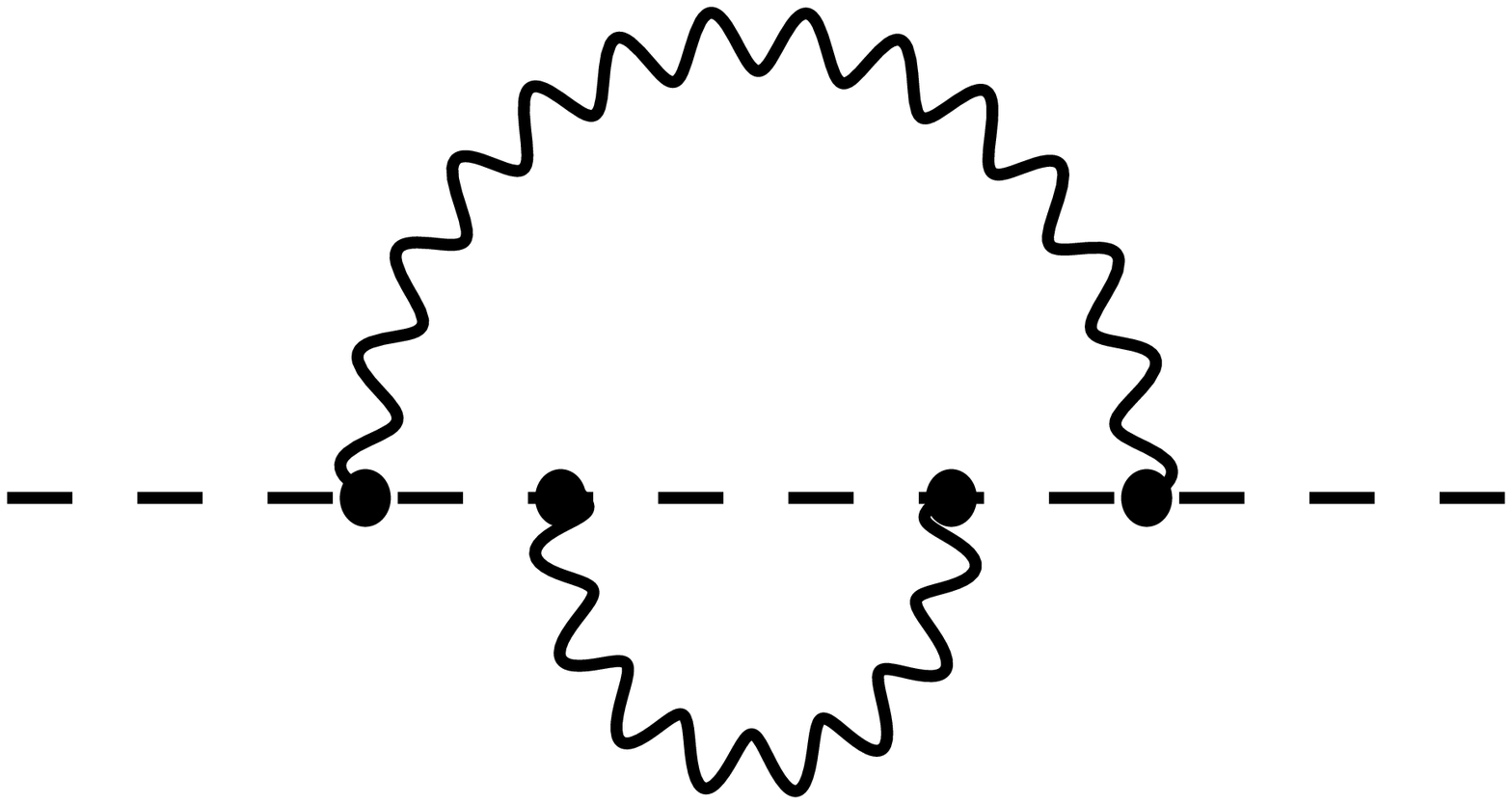}}
\put(9.4,14.9){$-$}
\put(10.0,14.9){${\displaystyle \frac{1}{2}\ \Bigg(}$}
\put(10.9,14.7){\includegraphics[scale=0.13,clip]{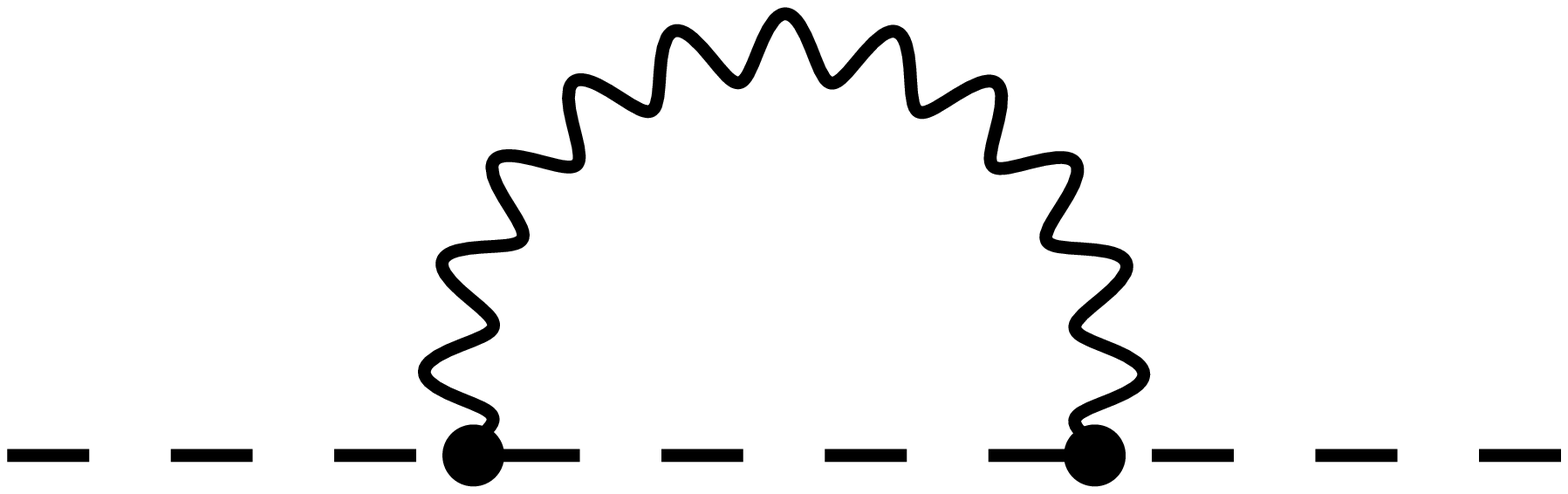}}
\put(13.4,14.9){${\displaystyle\Bigg)^{\mbox{2}}}$}

\put(1.1,16.2){\includegraphics[scale=0.2,clip]{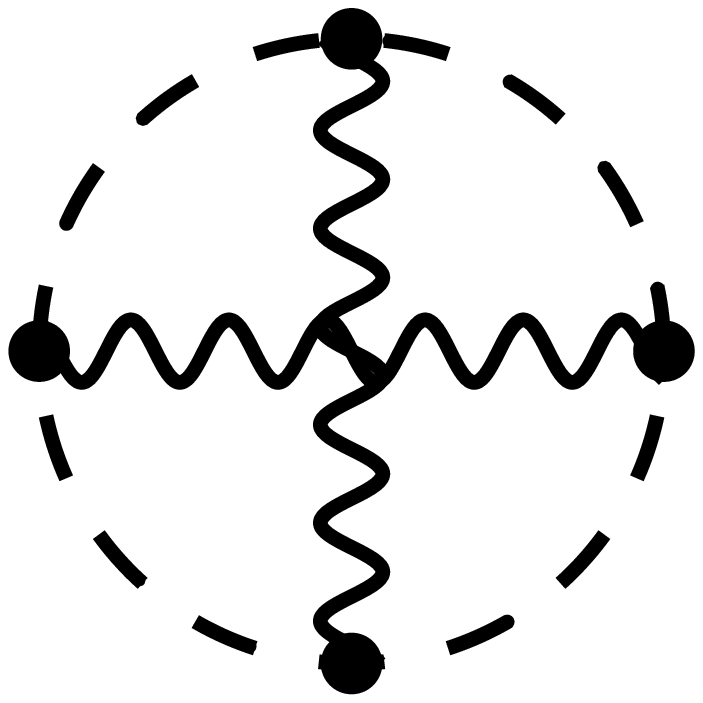}}
\put(0,17.5){B14}
\put(3.5,16.9){\vector(1,0){2}}
\put(6.5,16.2){\includegraphics[scale=0.13,clip]{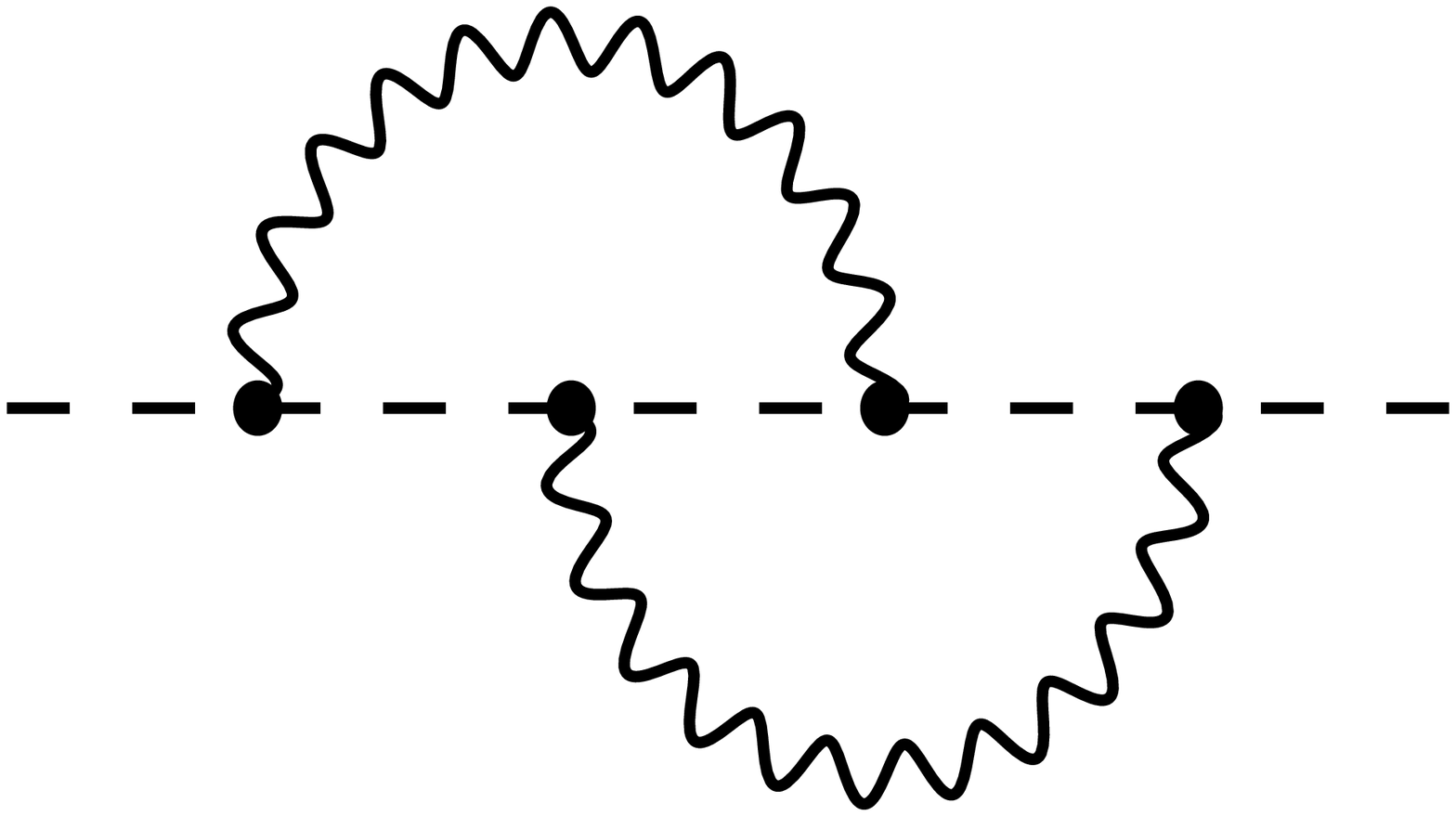}}

\put(0.7,18.1){\includegraphics[scale=0.2,clip]{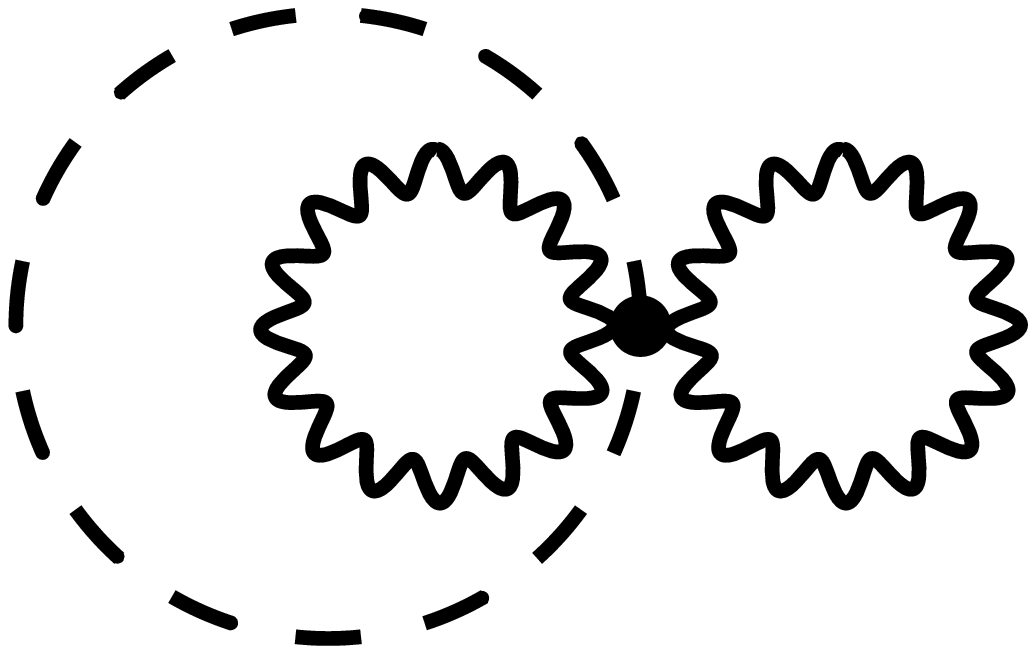}}
\put(0,19.4){B13}
\put(3.5,18.8){\vector(1,0){2}}
\put(6.6,17.8){\includegraphics[scale=0.13,clip]{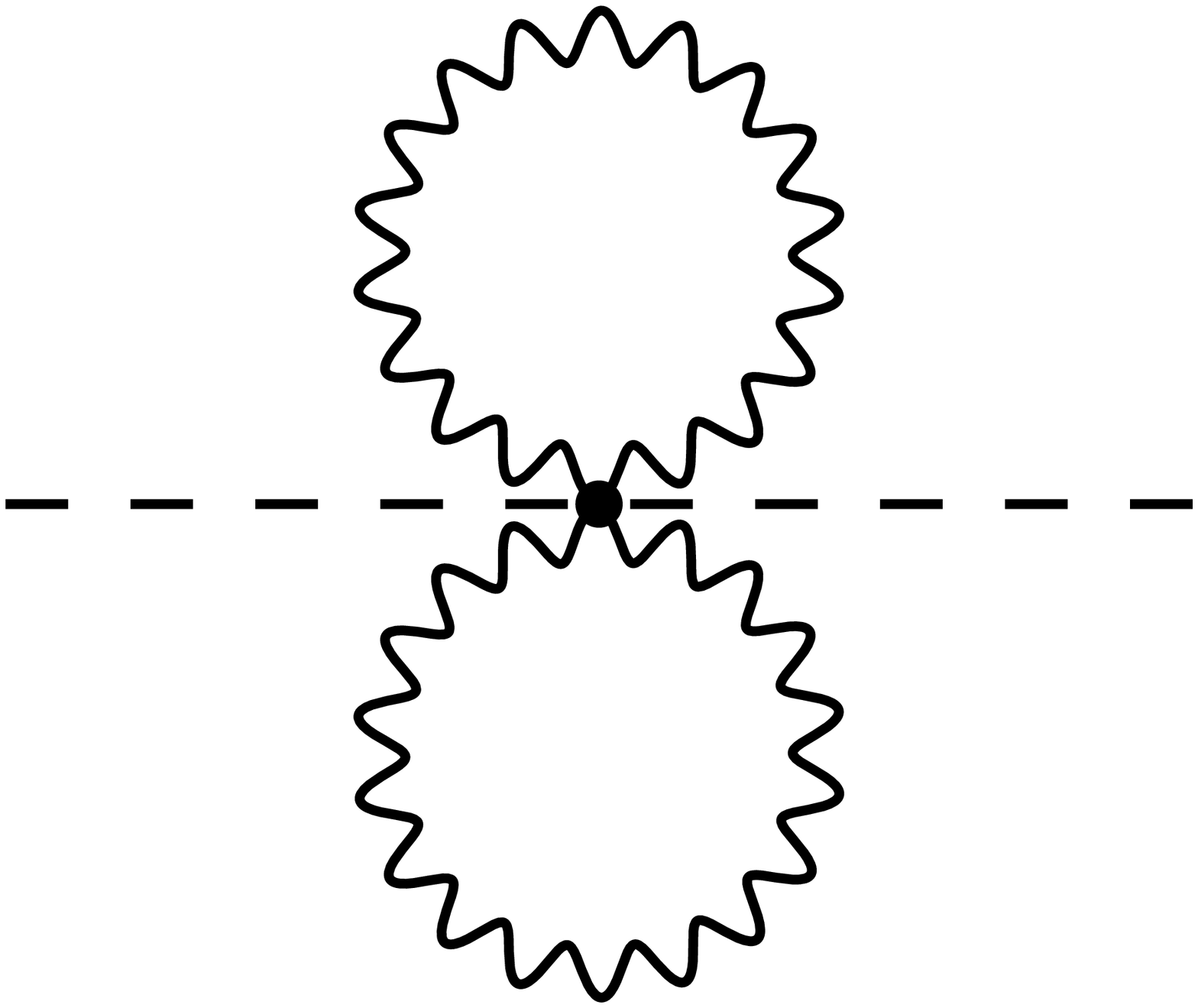}}

\put(0.5,20){\includegraphics[scale=0.19,clip]{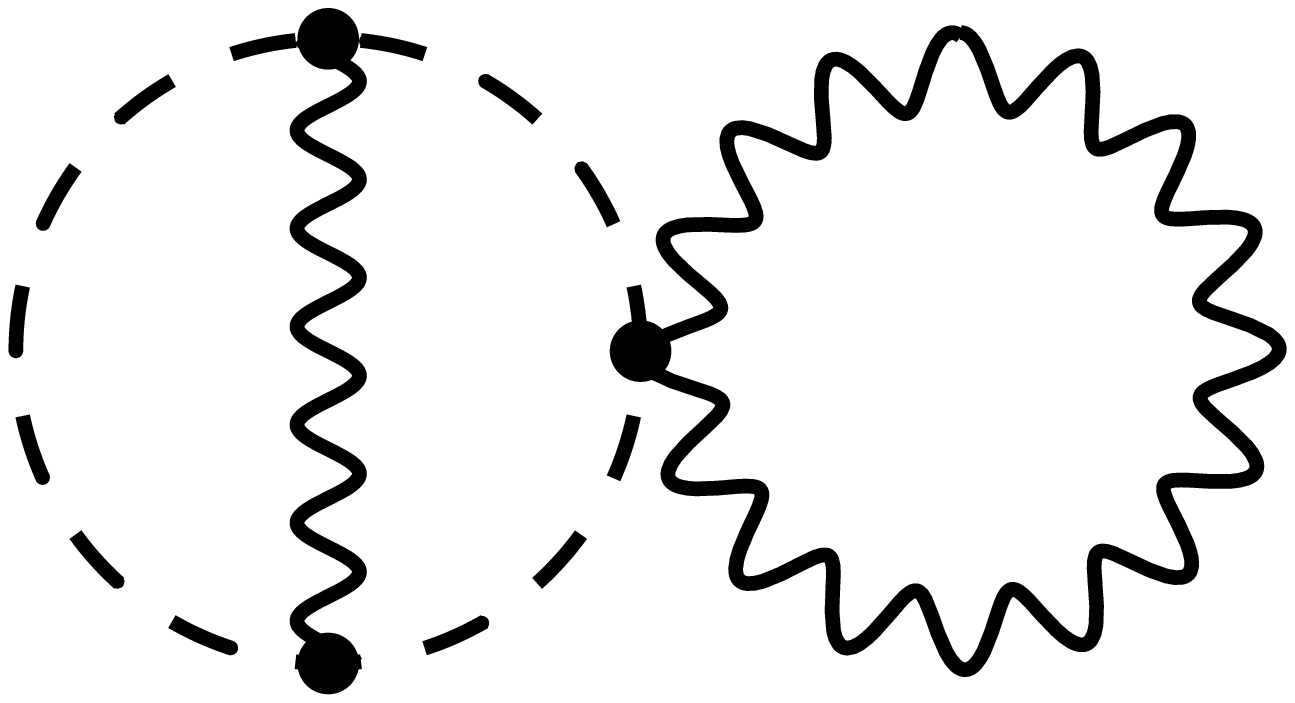}}
\put(0,21.3){B12}
\put(3.5,20.7){\vector(1,0){1}}
\put(5.1,19.9){\includegraphics[scale=0.14,clip]{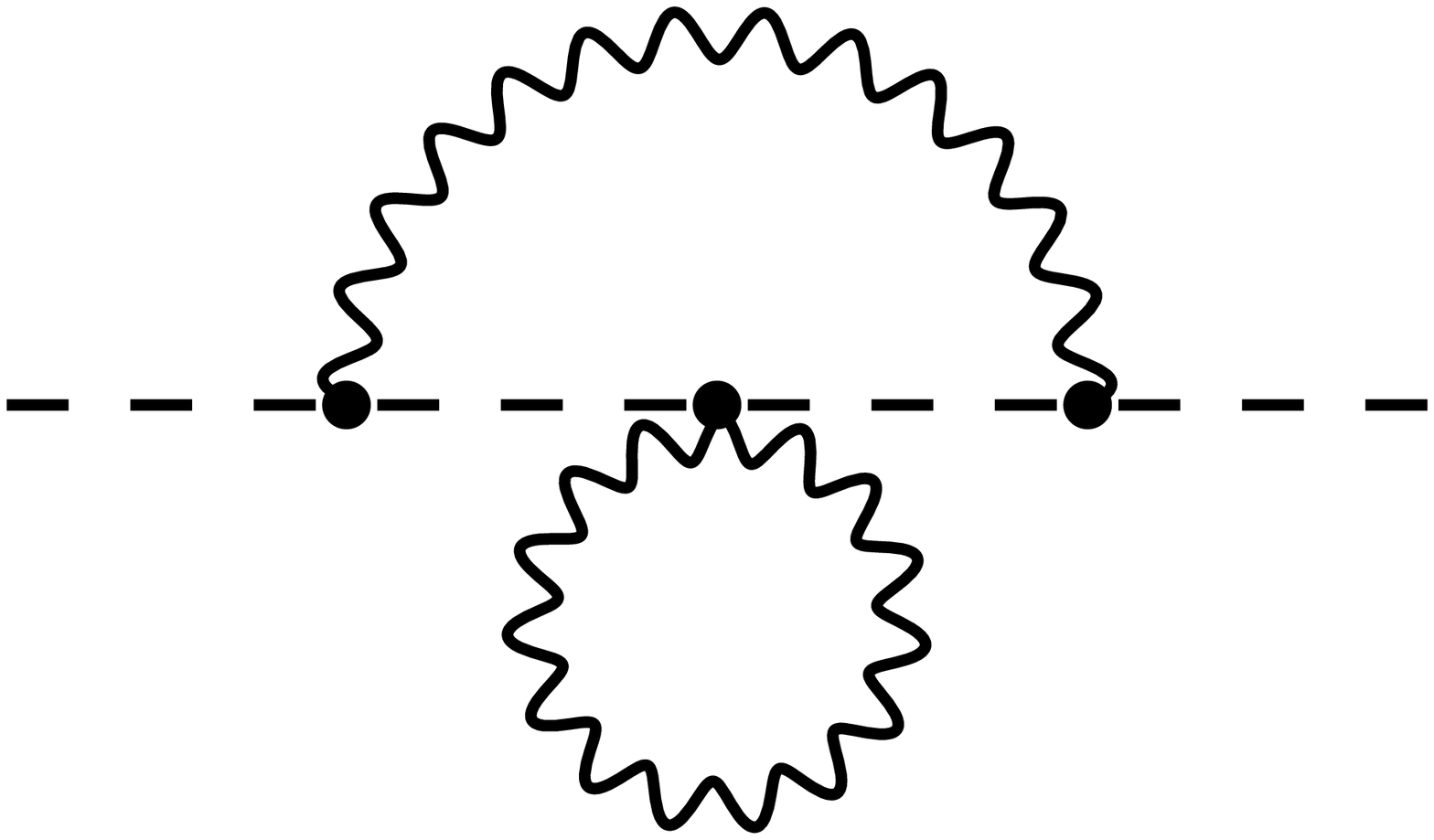}}
\put(8.1,20.6){$-$}
\put(8.6,20.6){${\displaystyle\Bigg(}$}
\put(9.1,20.3){\includegraphics[scale=0.14,clip]{ghost_gamma1.eps}}
\put(12.0,20.6){$\times$}
\put(12.7,20.3){\includegraphics[scale=0.125,clip]{ghost_gamma2.eps}}
\put(15.15,20.6){${\displaystyle\Bigg)}$}
\end{picture}
\vspace*{10mm}
\caption{The three-loop vacuum supergraphs containing a ghost loop (or ghost loops) and the corresponding superdiagrams contributing to the ghost and matter anomalous dimensions. The gray circles denote insertions of the one-loop polarization operator of the quantum gauge superfield, see Ref. \cite{Kazantsev:2017fdc} for details.}\label{Figure_3Loop_Ghosts}
\end{figure}

1. First, we consider a vacuum supergraph containing a ghost loop (or loops) with an insertion of $\theta^4 (v^B)^2$ and calculate it using the $D$-algebra.

2. Next, we replace one of inverse squared momenta (say, $1/Q^2$) of a {\it ghost} propagator by $4\pi^2 C_2 \delta^4(Q)$. If matter propagators are present in the considered supergraph, then $\delta_i^j/Q^2$ coming from a matter propagator is replaced by $4\pi^2 C(R)_i{}^j \delta^4(Q)$. All expressions obtained after the replacements of one propagator should be summed.

3. After calculating the integrals of the $\delta$-functions the results are compared with the corresponding contributions to the ghost (or matter) anomalous dimensions. These contributions are given by the sums of all two-point supergraphs which are produced by all possible cuts of ghost (or matter) propagators in the considered supergraph, see Fig. \ref{Figure_3Loop_Ghosts}.

We have done this for all three-loop vacuum supergraphs presented in Fig. \ref{Figure_3Loop_Ghosts}. As a result, we have obtained

\begin{eqnarray}
&& \Delta_{\mbox{\scriptsize B}i} \Big(\frac{\beta}{\alpha_0^2}\Big)\ \to\ \frac{C_2}{\pi} \Delta_{\mbox{\scriptsize B}i} \gamma_c,\qquad \mbox{for}\qquad i=12,\ldots,21;\\
&& \Delta_{\mbox{\scriptsize B}22} \Big(\frac{\beta}{\alpha_0^2}\Big) + \Delta_{\mbox{\scriptsize B}23} \Big(\frac{\beta}{\alpha_0^2}\Big)\ \to\ \frac{C_2}{\pi} \Delta_{\mbox{\scriptsize B}22 + \mbox{\scriptsize B}23} \gamma_c - \frac{1}{2\pi r} C(R)_i{}^j (\Delta_{\mbox{\scriptsize B}22 + \mbox{\scriptsize B}23}\gamma_\phi)_j{}^i,\qquad
\end{eqnarray}

\noindent
in agreement with Eqs. (\ref{Delta_Matter_Final}) and (\ref{Delta_Ghost_Final}). Note that this verification is different from the one made in Ref. \cite{Kuzmichev:2019ywn}, because the algorithm used for constructing the contribution to the function (\ref{Delta_Beta}) is different.

\subsection{${\cal N}=1$ SQED in the three-loop approximation}
\hspace*{\parindent}

As one more example for checking the method proposed in section \ref{Subsection_Graphs} we can consider ${\cal N}=1$ SQED with $N_f$ flavors. In this case the gauge group is $U(1)$, $r=1$, $C_2 = 0$, and $C(R)_i{}^j \to \delta_{\alpha\beta} \cdot 1_2$, where $\alpha,\beta=1,\ldots, N_f$ and $1_2$ denotes an identity matrix of the size $2\times 2$. This theory does not contain Yukawa terms triple in chiral superfields. Therefore, the calculations in this case have been done for $F(x)=1$. Although ${\cal N}=1$ SQED is a particular case of ${\cal N}=1$ supersymmetric gauge theories considered earlier, this example is not trivial, because for this theory the {\it complete} expression for the three-loop $\beta$-function has explicitly been calculated with the higher derivative regularization. That is why it is possible to perform one more nontrivial test of the method proposed in this paper.

Using the results of Ref. \cite{Aleshin:2020gec} the sum of two- and three-loop vacuum supergraphs modified by an insertion of $\theta^4 (v^B)^2$ and multiplied by $-2\pi/{\cal V}_4\cdot d/d\ln\Lambda$ in the general $\xi_0$-gauge is written as

\begin{eqnarray}\label{SQED_Vacuum}
&&\hspace*{-5mm} 4\pi N_f \frac{d}{d\ln\Lambda} \int \frac{d^4Q}{(2\pi)^4} \frac{d^4K}{(2\pi)^4} \frac{e_0^2}{K^2 R_K}\Big(\frac{1}{Q^2 (Q+K)^2} - \frac{1}{(Q^2+M^2)((Q+K)^2+M^2)}\Big)\nonumber\\
&&\hspace*{-5mm} - 4\pi N_f^2 \frac{d}{d\ln\Lambda} \int \frac{d^4Q}{(2\pi)^4} \frac{d^4K}{(2\pi)^4} \frac{d^4L}{(2\pi)^4} \frac{e_0^4}{K^2 R_K^2} \Big(\frac{1}{Q^2 (Q+K)^2} - \frac{1}{(Q^2+M^2)((Q+K)^2+M^2)}\Big)\nonumber\\
&&\hspace*{-5mm} \times \Big(\frac{1}{L^2 (L+K)^2} - \frac{1}{(L^2+M^2)((L+K)^2+M^2)}\Big) + 8\pi N_f \frac{d}{d\ln\Lambda} \int \frac{d^4Q}{(2\pi)^4} \frac{d^4K}{(2\pi)^4} \frac{d^4L}{(2\pi)^4}\nonumber\\
&&\hspace*{-5mm} \times \frac{e_0^4}{K^2 R_K L^2 R_L} \bigg\{\frac{1}{Q^2 (Q+K)^2 (Q+L)^2} - \frac{K^2}{Q^2 (Q+K)^2 (Q+L)^2 (Q+K+L)^2} \\
&&\hspace*{-5mm} + \frac{K^2+M^2}{(Q^2+M^2)((Q+K)^2+M^2)((Q+L)^2+M^2)((Q+K+L)^2+M^2)} - \frac{1}{(Q^2+M^2)}\nonumber\\
&&\hspace*{-5mm} \times \frac{1}{((Q+K)^2+M^2)((Q+L)^2+M^2)} + \frac{2M^2}{(Q^2+M^2)^2((Q+K)^2+M^2)((Q+L)^2+M^2)}\bigg\}, \nonumber
\end{eqnarray}

\noindent
and does not contain inverse momenta to the fourth power. According to the method discussed in section \ref{Subsection_Graphs}, to find a contribution to the function (\ref{Delta_Beta}), we should sum the expressions obtained from Eq. (\ref{SQED_Vacuum}) by replacing one of $1/P^2$ by $4\pi^2 \delta^4(P)$, where $P_\mu$ is a momentum of the (massless) matter superfields. Note that it is not necessary to take into account the gauge superfield propagators, because now $C_2=0$, and nonsingular propagators of the massive Pauli--Villars superfields. In the first two terms of Eq. (\ref{SQED_Vacuum}) the momentum of the quantum gauge superfield is denoted by $K_\mu$, while in the last term the momenta of the quantum gauge superfield propagators are $K_\mu$ and $L_\mu$. Then the above described procedure gives

\begin{eqnarray}
&& \frac{\beta(\alpha_0)}{\alpha_0^2} - \frac{\beta_{\mbox{\scriptsize 1-loop}}(\alpha_0)}{\alpha_0^2} = \frac{2 N_f}{\pi} \frac{d}{d\ln\Lambda} \int \frac{d^4K}{(2\pi)^4} \frac{e_0^2}{K^4 R_K} - \frac{4N_f^2}{\pi} \frac{d}{d\ln\Lambda} \int \frac{d^4K}{(2\pi)^4} \frac{d^4L}{(2\pi)^4} \frac{e_0^4}{K^4 R_K^2}\nonumber\\
&& \times \Big(\frac{1}{L^2 (L+K)^2} - \frac{1}{(L^2+M^2)((L+K)^2+M^2)}\Big) - \frac{N_f}{\pi} \frac{d}{d\ln\Lambda} \int \frac{d^4K}{(2\pi)^4} \frac{d^4L}{(2\pi)^4} \frac{e_0^4}{R_K R_L}\qquad\nonumber\\
&& \times \Big(\frac{4}{K^2 L^4 (K+L)^2} - \frac{2}{K^4 L^4}\Big) + O(e_0^6).
\end{eqnarray}

\noindent
Again, this expression correctly reproduces the result obtained earlier by different methods, see Refs. \cite{Stepanyantz:2011jy,Aleshin:2020gec}.

Therefore, for all considered supergraphs the results obtained with the help of the technique discussed in section \ref{Subsection_Graphs} (which is a certain modification of the one proposed in \cite{Stepanyantz:2019ihw}) coincided with the expressions found earlier by different methods. This confirms the correctness of the method which was used in this paper for the all-loop perturbative derivation of the exact NSVZ $\beta$-function.

\section{Conclusion}
\hspace*{\parindent}

In this paper we have finished the all-order perturbative derivation of the exact NSVZ $\beta$-function (\ref{NSVZ_Exact_Beta_Function}) for non-Abelian ${\cal N}=1$ supersymmetric gauge theories which was started in Refs. \cite{Stepanyantz:2016gtk,Stepanyantz:2019ihw,Stepanyantz:2019lfm}. Its main ingredient is the higher covariant derivative regularization, which allows revealing some interesting features of quantum corrections in these theories. For instance, according to Ref. \cite{Stepanyantz:2019lfm}, with this regularization all loop integrals giving the $\beta$-function defined in terms of the bare couplings are integrals of double total derivatives with respect to loop momenta. These integrals do not vanish due to singularities, which appear when double total derivatives act on massless propagators. The sum of the singularities produces all contributions to the $\beta$-function starting from the two-loop approximation. (The one-loop quantum corrections should be considered separately. This has been done in Ref. \cite{Aleshin:2016yvj}.) It is possible to divide singular contributions into three groups depending on a propagator on which the double total derivatives act. Namely, they can act on the matter superfield propagators, on the propagators of the Faddeev--Popov ghosts, and on the propagators of the quantum gauge superfield. Qualitatively, this can be interpreted as a cutting of the corresponding internal line in a certain vacuum supergraph. All such cuts give a set of superdiagrams contributing to the anomalous dimension of the corresponding quantum superfield. From the other side, attaching two external gauge lines of the background gauge superfield to the considered supergraph we obtain a set of superdiagrams contributing to the $\beta$-function. The NSVZ equation in the form (\ref{NSVZ_Equivalent_Form}) relates this contribution to the $\beta$-function to the parts of the anomalous dimensions of quantum superfields coming from the superdiagrams produced by cuts of internal lines. Note that the factorization into double total derivatives allows to calculate analytically one of loop integrals, so that the $\beta$-function in a certain order is really related to the anomalous dimensions in the previous order. The proof of this fact has been done in this paper for arbitrary higher-derivative regulator functions $R$, $F$, and $K$ and arbitrary values of the regularization parameters $a=M/\Lambda$ and $a_\varphi=M_\varphi/\Lambda$ independent of couplings. We have demonstrated that the sums of singularities coming from the cuts of certain propagators are really equal to the corresponding terms in Eq. (\ref{NSVZ_Equivalent_Form}). (Note that in Ref. \cite{Stepanyantz:2019lfm} this has been done for the matter and Faddeev--Popov ghost superfields. The results of this paper obtained by a different method are the same. However, in the present paper we have also calculated the all-loop sum of singularities produced by cuts of the gauge propagators.) Thus, Eq. (\ref{NSVZ_Equivalent_Form}) for RGFs defined in terms of the bare couplings is proved in all orders in the case of using the higher covariant derivative regularization. The original NSVZ $\beta$-function (\ref{NSVZ_Exact_Beta_Function}) for these RGFs can be obtained with the help of the non-renormalization theorem for the triple gauge-ghost vertices proved in \cite{Stepanyantz:2016gtk}. Note that RGFs defined in terms of the bare couplings are scheme-independent for a fixed regularization, so that both these equations are valid for any renormalization prescription supplementing an arbitrary version of the higher covariant derivative regularization.

Taking into account that in the HD+MSL scheme RGFs defined in terms of the renormalized couplings coincide with the ones defined in terms of the bare couplings up to the renaming of arguments, we conclude that for standardly defined RGFs one of the NSVZ schemes is given by the HD+MSL prescription in all orders. By other words, to obtain the NSVZ equation in all loops, one should regularize a theory by higher covariant derivatives and include into renormalization constants only powers of $\ln\Lambda/\mu$ (or, equivalently, set all finite constants to 0). Note that the HD+MSL schemes corresponding to different versions of the higher covariant derivative regularization constitute a continuous set of the NSVZ schemes, which are in general different.

As we have already mentioned, explicit calculations exactly confirm the statements discussed above even in the approximations, where the dependence on a regularization and a renormalization prescription becomes essential, see, e.g., \cite{Shakhmanov:2017soc,Kazantsev:2018nbl,Kuzmichev:2019ywn,Aleshin:2020gec}. Also the results of this paper confirm the correctness of the expression for the three-loop $\beta$-function of a general ${\cal N}=1$ supersymmetric gauge theory with matter superfields and a simple gauge group derived in Ref. \cite{Kazantsev:2020kfl} from the NSVZ equation.

\vspace{5mm}

\section*{Acknowledgments}
\hspace*{\parindent}

The author would like to express the gratitude to S.~S.~Aleshin, A.~E.~Kazantsev, M.~D.~Kuzmichev, N.~P.~Meshcheriakov, S.~V.~Novgorodtsev, and I.~E.~Shirokov for numerous useful discussions and valuable comments on the manuscript.

The research was supported by the Foundation for the Advancement of Theoretical Physics and Mathematics `BASIS', grant No. 19-1-1-45-1.

\end{document}